\documentclass{article}
\usepackage{arxiv}

\usepackage[normalem]{ulem}

\usepackage{tikz}
\usepackage{amsmath}
\usepackage{tikz}
\usepackage{amsmath}
\usepackage{amssymb}
\usepackage{footmisc}

\usepackage{enumitem}
\usepackage{csquotes}
\usepackage{booktabs}
\usepackage{float}
\usepackage{tabularx}
\usepackage{amsmath}
\usepackage{makecell}
\usepackage{multirow}
\usepackage[hidelinks]{hyperref}
\RequirePackage[capitalise]{cleveref} %
\usepackage{cleveref}
\usepackage{caption}
\usepackage{subcaption}
\usepackage{epsfig,endnotes, graphicx,booktabs, multirow, array}
\renewcommand{\arraystretch}{1.2}
\definecolor{aliceblue}{rgb}{0.94, 0.97, 1.0}
\definecolor{maroon}{cmyk}{0,0.87,0.68,0.32}
\usepackage{tcolorbox}
\usepackage{listings}
\newtcolorbox{mybox}[1]{colback=aliceblue,colframe=black,fonttitle=\bfseries,title=#1}
\usepackage{pgfplots}

\usepackage{tikz}
\usepackage{amsmath}
\usepackage{xspace}
\usepackage{colortbl}
\usepackage{tabularx}
\usepackage[title]{appendix}
\usepackage[utf8]{inputenc}
\usepackage{mdframed}
\usepackage{url}

\definecolor{aliceblue}{rgb}{0.94, 0.97, 1.0}
\definecolor{beaublue}{rgb}{0.74, 0.83, 0.9}

\usepackage[utf8]{inputenc}
\usepackage{xcolor}
\usepackage{tcolorbox}
\tcbuselibrary{listings, skins, breakable}

\definecolor{sagegreen}{RGB}{235, 247, 235} 
\definecolor{darksage}{RGB}{60, 100, 70}
\definecolor{modblue}{RGB}{214, 230, 245}   %
\definecolor{borderblue}{RGB}{70, 130, 180} %
\definecolor{academicyellow}{RGB}{255, 250, 235} %
\definecolor{goldenborder}{RGB}{205, 170, 50}   %
\definecolor{lavenderbg}{RGB}{230, 230, 250}
\definecolor{darkviolet}{rgb}{0.58, 0.0, 0.83}

\usepackage{filecontents}

\newcommand{\benchmark}{\mbox{\textit{WebSP-Eval}}\xspace}
\newcommand{\numtasks}{200\xspace}
\newcommand{\numwebsites}{28\xspace}
\newcommand{\numwebsitecategories}{9\xspace}
\newcommand{\numtaskcategories}{9\xspace}

\newcommand{\taskname}{website security and privacy tasks\xspace}
\newcommand{\tasknameE}{\emph{website security and privacy tasks}\xspace}

\newcommand{\moduleone}{Task Curation\xspace}
\newcommand{\moduletwo}{Agent Instantiation\xspace}
\newcommand{\modulethree}{Automated Verification\xspace}
\newcommand{\moduleoneE}{\emph{Task Curation}\xspace}
\newcommand{\moduletwoE}{\emph{Agent Instantiation}\xspace}
\newcommand{\modulethreeE}{\emph{Automated Verification}\xspace}

\newcommand{\webspindexT}{\texttt{data-websp-index}\xspace}

\newcommand{\withnav}{WithNav\xspace}
\newcommand{\withnavE}{\emph{WithNav}\xspace}
\newcommand{\wonav}{W/oNav\xspace}
\newcommand{\wonavE}{\emph{W/oNav}\xspace}

\newcommand{\passatk}[1]{\ensuremath{\mathrm{pass@}#1}}
\newcommand{\passexpk}[1]{\ensuremath{\mathrm{pass}^{#1}}}

\usepackage{booktabs}

\begin{document}

\title{WebSP-Eval: Evaluating Web Agents on Website Security and Privacy Tasks}

\author{\hspace*{-1cm}
Guruprasad Viswanathan Ramesh, Asmit Nayak, Basieem Siddique, Kassem Fawaz\\
  University of Wisconsin-Madison \\
  \hspace*{-1cm}\url{viswanathanr@wisc.edu}, \url{anayak6@wisc.edu}, \url{bsiddique@wisc.edu}, \url{kfawaz@wisc.edu}
  }

\maketitle
\begingroup
\makeatletter
\renewcommand\@makefnmark{}
\makeatother
\footnotetext{Code and data available at \url{https://github.com/wi-pi/webspeval_code}}
\endgroup

\begin{abstract}
Web agents automate browser tasks, ranging from simple form completion to complex workflows like ordering groceries. While current benchmarks evaluate general-purpose performance~(e.g., WebArena) or safety against malicious actions~(e.g., SafeArena), no existing framework assesses an agent's ability to successfully execute user-facing website security and privacy tasks, such as managing cookie preferences, configuring privacy-sensitive account settings, or revoking inactive sessions.
  
To address this gap, we introduce \benchmark, an evaluation framework for measuring web agent performance on \taskname. \benchmark comprises 1) a manually crafted task dataset of 200 task instances across 28 websites; 2) a robust agentic system supporting account and initial state management across runs using a custom Google Chrome extension; and 3) an automated evaluator. We evaluate a total of 8 web agent instantiations using state-of-the-art multimodal large language models, conducting a fine-grained analysis across websites, task categories, and UI elements. Our evaluation reveals that current models suffer from limited autonomous exploration capabilities to reliably solve \taskname, and struggle with specific task categories and websites. Crucially, we identify stateful UI elements are a primary reason for agent failure, with toggles causing more than 45\% task failure across many models. 

\end{abstract}

\section{Introduction}

Web agents~\cite{ning2025survey, google2024mariner,openai2025operator,perplexity2025comet} are powerful tools that help in automating mundane tasks on the web. Recently, LLM-powered web agents~(also referred to as browser-use agents) have gained significant traction offering flexible web interactions on the web~\cite{zhou2023webarena, he2024webvoyager}. Modern browsers such as Perplexity's Comet~\cite{perplexity2025comet} and Open AI's Atlas~\cite{openai2025atlas} incorporate web agents into their user-experience. These agents allow users to delegate repetitive tasks such as finding the cheapest flight, ordering groceries, or listing in a code repository with a specific label~\cite{zhou2023webarena}. Web agents act on textual user instructions and leverage contextual information of the operating environment such as screenshots and~(or) DOM trees to understand the page's current state and execute actions using web automation frameworks like Selenium~\cite{selenium}, Playwright~\cite{playwright}, etc.  

While the utility of these agents is rapidly expanding, their autonomous nature necessitates rigorous evaluations, including security- and privacy-based ones. Standard web agent benchmarks are primarily general-purpose, focusing heavily on information-retrieval tasks~(e.g., WebArena~\cite{zhou2023webarena} and WebVoyager~\cite{he2024webvoyager}). Given the sensitive data these agents process, recent benchmarks also evaluate their safety and security~(SafeArena~\cite{tur2025safearena}, ST-WebAgentBench~\cite{levy2024st}) and assess their propensity to leak sensitive information during routine tasks~(AgentDam~\cite{zharmagambetov2026agentdam}, EIA~\cite{liao2025eia}).

However, existing benchmarks do not account for whether web agents can emulate web users in making security and privacy decisions. Examples include managing cookies, updating data sharing permissions, and revoking older sessions. We refer to these routine actions performed by users to keep their data private and accounts safe and secure as \tasknameE. It is critical to assess web agents on these tasks as they might not only be explicitly prompted by users, but also might encounter them during live exploration of websites while performing other tasks~(refer to \Cref{sec:mid-flow-appendix}). As agents advance toward making persona-driven decisions on behalf of users, their ability to safely navigate these settings becomes paramount~\cite{rossi2026m}. Deploying web agents without rigorous evaluation on \taskname is risky, as even a single erroneous action by an agent could inadvertently weaken a user's account security or expose their private data.

Evaluating web agents faithfully on \taskname requires a consistent initial state that can be precisely controlled across different evaluation runs. But as these tasks are performed on live websites where states are maintained on the server-side, an evaluator would require a robust account and state management setup to faithfully evaluate their models. A simple solution would be to use a new sock puppet account for each run, but this does not scale. Addressing not just this infrastructural challenge but also the lack of a standard dataset, we introduce \textbf{\benchmark}, an evaluation framework to assess web agents on \tasknameE. It comprises three modules, corresponding to our three-fold contributions: 1) \textbf{A Novel Benchmark of} manually crafted dataset of 200 task instances across 28 websites; 2) \textbf{A Robust Agentic framework} built upon WebVoyager~\cite{he2024webvoyager}, along with an account state and session management setup using a custom Google Chrome extension; and 3) \textbf{Extensive Empirical analysis} of 8 state-of-the-art model performance using an automated ensemble of judges comprising three state-of-the-art models. Task instances in our dataset are paired with one or multiple initial states of the websites that allow consistent evaluation of different models.
Although we build our agentic implementation over WebVoyager, we make improvements from extending the action space to enhancing its features, thereby enabling the agentic framework to perform the tasks in our dataset tied to user accounts on live websites~(refer to \Cref{sec:voyage_differences}).

We instantiate our agentic system with eight state-of-the-art multimodal large language models~(MLLMs), evaluating trajectories with our automated judge and addresses three research questions: 1) \textbf{RQ1:} Can agents autonomously execute tasks, and how does explicit navigational instruction impact performance? 2) \textbf{RQ2:} How does performance vary across different websites and task categories? and 3) \textbf{RQ3:} How do specific UI elements and their initial states impact agent success? 4) \textbf{RQ4:} Are agents robust across independent trials~(runs) of the same task?

We find that while top-tier models like Gemini-3-Pro perform reliably well~(achieving an 84.5\% success rate with navigation and 82.5\% without), forcing autonomous exploration causes significant performance degradation across all models. Smaller models suffer the most, with Gemini-2.5-Flash experiencing an 16.5\% drop in success rate without navigation. Furthermore, our analysis reveals that agent performance is highly sensitive to website-specific UI layouts and task categories. Fine-grained breakdown of agent performance by UI elements related to a task reveals that while reliably navigate using standard links and buttons, they fail significantly on stateful elements, with Gemini-2.5-Flash explicitly failing in 50\% of toggle-related tasks. We also notice that there is a strong bias from the models to take actions even when the initial state matches the required state according to the task. These results are supported by our manual analysis of agent failures~(see \Cref{sec:human_failure_analysis}). During our experiments, we found a potential destructive failure, where Gemini-2.5-Flash deactivates Pinterest account upon being asked to sign out~(see \Cref{fig:pinterest_signout}). Lastly, our robustness evaluation shows that web agents cannot consistently solve tasks accurately.

\section{Background and Related Work}
\label{sec:background}
This section consists of a background on web agents, notation used in the paper, and an overview 
of the related work. We categorize the related work into three subsections, each describing a separate related area of research: 1) general LLM security and privacy 2) general web agent benchmarks and frameworks, 3) web agent security and privacy. Finally, we highlight our research contributions.

\subsection{Web Agents Background and Notation}
\label{sec:web_agents}
\begin{figure}[t]
  \begin{center}
    \includegraphics[width=1\columnwidth]{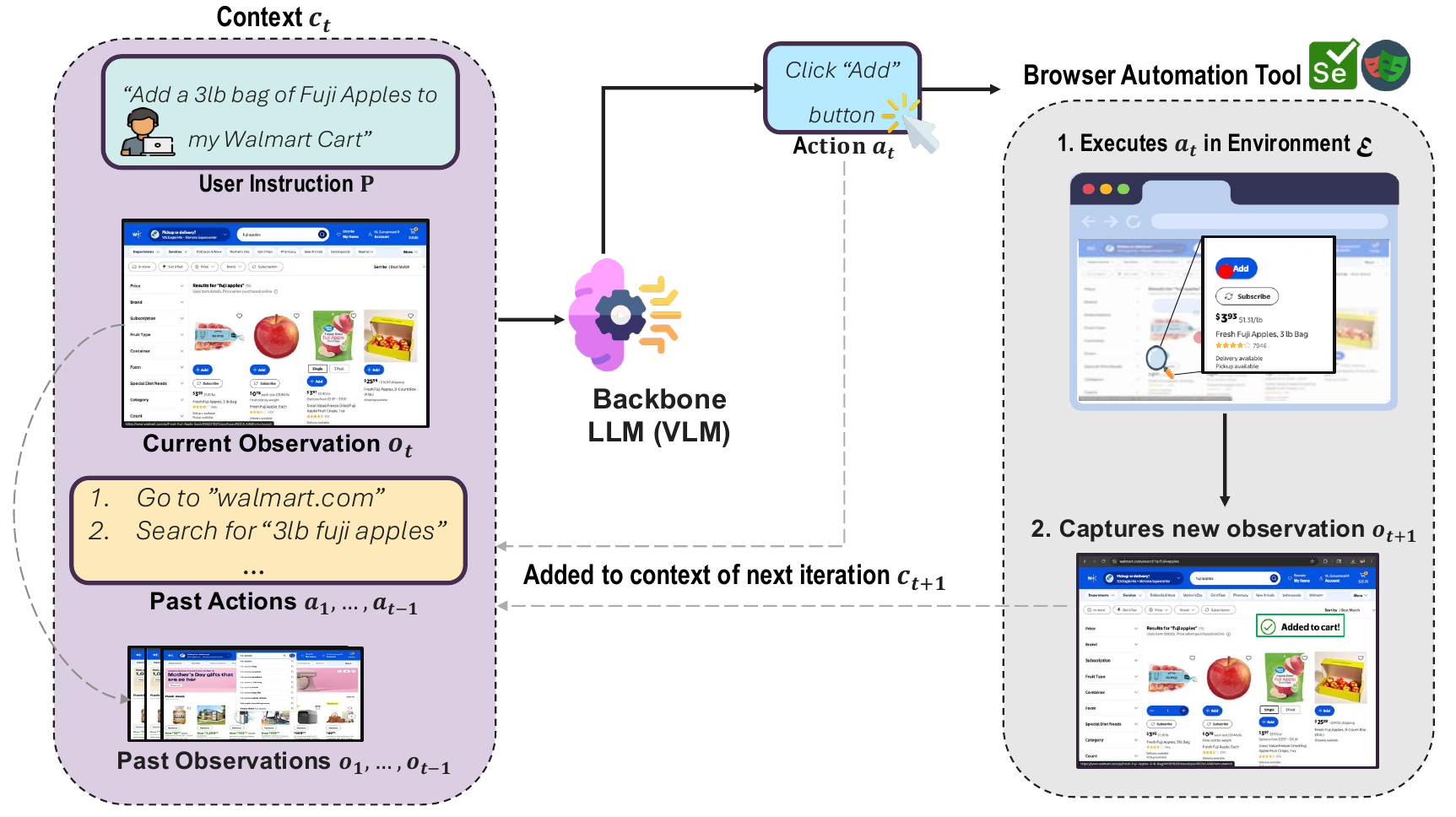}
  \end{center}
  \caption{A high-level overview of a web agent consisting of a backbone model and an automation framework to execute actions based on the input prompt, previous actions, and environmental feedback.}
  \label{fig:bua}
\end{figure}

Web agents~\cite{ning2025survey, zhou2023webarena, he2024webvoyager}, also commonly referred to Browser-use agents, enable dynamic and autonomous interaction within web environments. Unlike traditional web scrapers that are brittle to changing website structure, these agents attempt to mimic human behavior, i.e., they consider the current state of a website, reason about its content, and execute actions through the page's interactable elements. At a high level, web agents comprise three components (Fig.~\ref{fig:bua}): the user interface (UI), browser automation frameworks for actuation, and backbone models for reasoning.  %

Web agents rely on a natural language prompt $P$ to accept user instruction to perform a task $\mathcal{T}$ on a website $W$~(e.g., ``Disable all possible cookies for the website \texttt{shein.com}.'')~\cite{deng2023mind2web}. The agent accepts $P$, generates an execution plan, and executes the relevant actions. As the agent operates, the UI provides real-time transparency, either by keeping the browser instance in the foreground or by displaying execution logs, screenshots of the agent's current view, or a stream of ``thoughts'' explaining the agent's next move. The UI also allows the user to intervene to resolve ambiguities and make decisions that the agent cannot make on its own. 

The automation framework (e.g., Playwright~\cite{playwright}, Puppeteer~\cite{puppeteer}, Selenium~\cite{selenium}) acts as the interface to the environment $\mathcal{E}$, responsible for the agent's perception and actuation. At time step $t$, it captures a representation of the web page to generate an observation $o_t \in \mathcal{O}$. This observation space includes the Document Object Model (DOM), the Accessibility Tree, and visual screenshots, and is fed to the backbone model $M_A$. 

The backbone model $M_A$, typically an MMLM~\cite{singh2025openai, geminiteam2025gemini3, liu2023visual}, provides the agent's reasoning capability. At time step $t$, $M_A$ receives the context $c_t$, which consists of the user instruction $P$, the history of past actions and observations, and the current observation: $c_t = (P, o_1, a_1, \dots, a_{t-1},o_{t-1}, o_t)$. The model analyzes $c_t$ and produces an action $a_t \in \mathcal{A}$ such that $a_t = M(c_t)$. Once the model decides on an action $a_t$, the automation framework executes it within the environment $\mathcal{E}$. The action  $a_t$ yields the next subsequent observation $o_{t+1} = \mathcal{E}(o_t, a_t)$. This cycle continues until $M_A$ generates a termination action at step $T_{final}$ or a maximum step count $T_{max}$ is reached. The set of the actions along with the environmental changes is referred to as the trajectory of the agent. 

Actions $a_t$ correspond to specific browser events (e.g., \texttt{scroll to page footer}, \texttt{click("manage cookies")}, \texttt{click("marketing cookies switch")}) required to fulfill the task. Crucially, some actions change the underlying configuration of a website $W$, altering its state at step $t$, i.e., $S_t$. For example, if $S_t$ represents a configuration where marketing cookies are active, the action $a_t$ transitions the environment to $S_{t+1}$, where the setting is deactivated in accordance with the instruction $P$. Thus, ensuring a uniform initial state $S_0$ is essential for a faithful evaluation across runs.

\subsection{Related Work}
\label{sub:related_work}

\subsubsection{Web Agent Benchmarks and Frameworks}
\label{subsub:web_agent_bench}

Web Agents are predominantly evaluated on general-purpose, information-retrieval benchmarks. These include  WebArena~\cite{zhou2023webarena}, Mind2Web~\cite{deng2023mind2web}, WebShop~\cite{yao2022webshop}, WebVoyager~\cite{he2024webvoyager}, WorkArena~\cite{drouin2024workarena}, WorkArena++~\cite{boisvert2024workarena++}, and AssistantBench~\cite{yoran2024assistantbench}. These benchmarks also differ in using either sandboxed website snapshots~\cite{deng2023mind2web, zhou2023webarena, tur2025safearena, boisvert2024workarena++} or live websites~\cite{he2024webvoyager, yoran2024assistantbench} for evaluation.

Several LLM-powered web agent frameworks have been proposed, varying in their design and operation. WebVoyager~\cite{he2024webvoyager} uses an end-to-end multimodal architecture to process visual and textual inputs for real-world interactions. WebArena~\cite{zhou2023webarena} provides a self-hostable browser environment with fully functional websites across four domains. AgentLab, utilizing BrowserGym~\cite{chezelles2024browsergym},  unifies diverse benchmarks under a gym-like interface for scalable evaluation. Finally, CowPilot~\cite{huq2025cowpilot} allows human to collaborate during task-execution. Frameworks also differ in their web automation tool. Webvoyager uses Selenium~\cite{selenium}, while others use Playwright~\cite{playwright}.

We introduce \benchmark, a novel evaluation framework comprising a dataset to benchmark agents on solving \taskname. Compared to WebVoyager (643 tasks, 15 websites), WebArena (812 tasks, 4 websites), ST-WebAgentBench (375 tasks, 4 websites), and SafeArena (250 tasks, 4 websites), which repeat task templates across websites or configurations, our dataset covers more websites (28), comprising 200 tasks that represent different security and privacy options across these websites.

To the best of our knowledge, ours is also the first benchmark to evaluate web agents on live websites tied to user accounts. While Amazon-Bench~\cite{zhang2025functionality} involves a subset of user-account-tied tasks, it is restricted to a single website (Amazon) and its methodology lacks any state-management mechanism for consistent evaluation.

\subsubsection{Security and Privacy of LLMs}
\label{subsub:llm_secpriv}
LLMs integrated into agentic frameworks with tools add further vulnerabilities, typically comprising: (1) tool-execution risks, such as data leakage~\cite{shao2024privacylens} or destructive actions~\cite{ruan2024identifying}; (2) indirect prompt injection (e.g., adversarial injection in environment)~\cite{zhan2024injecagent, debenedetti2024agentdojo}; and (3) direct jailbreaks by malicious users bypassing safety guardrails~\cite{andriushchenko2025agentharm}. These threats have been examined in domain-specific settings, including coding agents~\cite{lee2026sec, guo2024redcode}, web agents~\cite{tang2026dark, ying2025securewebarena}, multi-agent systems~\cite{yagoubi2026agentleak, zhang2024psysafe}, and general tool-calling environments~\cite{fu2025ras, ye2024toolsword}. Many benchmarks like InjecAgent~\cite{zhan2024injecagent}, OS-Harm~\cite{kuntz2025harm}, PrivaCI-Bench~\cite{li2025privaci}, ASB~\cite{zhang2025agent} also evaluate the safety, security, and privacy of LLM-based agents.

\subsubsection{Web Agent Security and Privacy}
\label{subsub:webagent_secpriv}
Web agents often process sensitive user data. SafeArena~\cite{tur2025safearena}, ST-WebAgentBench~\cite{levy2024st}, and SecureWebArena~\cite{ying2025securewebarena} include safety- and security-oriented tasks. AgentDAM~\cite{zharmagambetov2026agentdam} evaluates whether web agents leak sensitive information while completing WebArena and VisualWebArena~\cite{koh2024visualwebarena} tasks. Indirect prompt injection attacks have also been shown to cause web agents to leak sensitive data~\cite{liao2025eia, evtimov2026wasp}. Web agents are likewise vulnerable to manipulative and deceptive UI such as dark patterns, which steer them away from the user's request~\cite{ersoy2025investigating, tang2026dark, guo2026susbench, cuvin2026how}, and to adversarial manipulations of the UI that divert them toward attacker-chosen goals~\cite{wu2025dissecting, zhang-etal-2025-attacking}. To mitigate such risks, PrivAgentFlow~\cite{ma2026privagentflow} proposes a distributed policy-driven agentic workflow to enforce data minimization.

Existing safety and privacy evaluations of web agents comprise agent response safety~\cite{tur2025safearena, levy2024st} and data leakage~\cite{zharmagambetov2026agentdam, liao2025eia}. In contrast, we evaluate web agent reasoning on direct prompts referring to \taskname. Our evaluation is on live websites in their native state, without adversarial manipulation~\cite{zhang-etal-2025-attacking} or injected dark patterns~\cite{cuvin2026how}.

\section{\benchmark Framework}
\label{sec:framework}
We introduce an evaluation framework, \benchmark, to analyze the performance of web agents on \taskname. Our framework comprises three modules~(as shown in Fig.~\ref{fig:phase_1}) -- \moduleoneE, \moduletwoE, and \modulethreeE. 

\begin{figure*}[t]
    \centering
    \includegraphics[width=\linewidth]{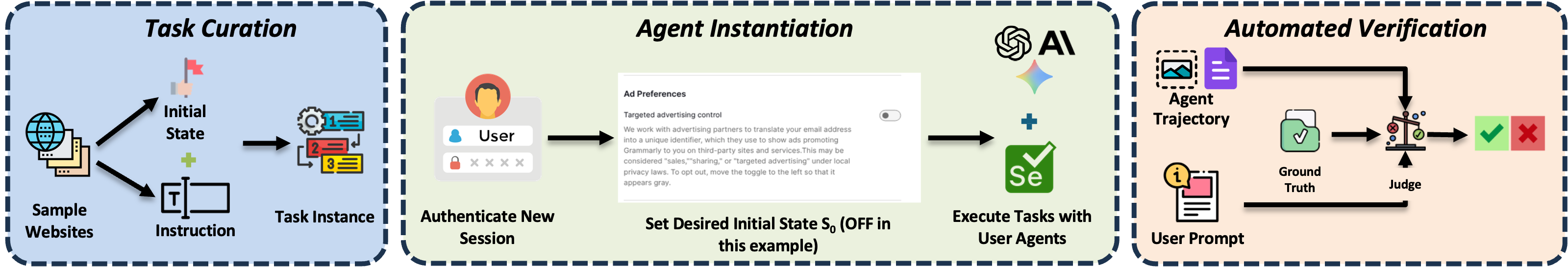}
    \caption{Modules of the \benchmark evaluation framework: 1) \moduleoneE \space -- Curation of a dataset consisting of \taskname across websites. 2) \moduletwoE \space -- A novel web agent deployment supporting account and state management, utilizing an MLLM and a Selenium driven backbone to execute actions 3) \modulethreeE \space -- An automated Vision Language Model-based judge to assess agent failure across five categories.}
    \label{fig:phase_1}
\end{figure*}

\subsection{\moduleone}
\label{sec:task_curation}
Modern website design is influenced by a variety of components, including evolving frontend libraries, CSS specifications,  and developer-specific implementation preferences~\cite{ivory2001empirically, vanderheijden2020structural}. Website interfaces vary significantly in design and menu structure even for standard security and privacy controls associated with user accounts. A representative example is a cookie notice, which although standardized, has varied implementations. To capture this heterogeneity, we manually curate a dataset of \taskname consisting of diverse popular websites as explained below.

First, we start with the top 50 websites from the Tranco list~\cite{pochat2018tranco} and augment this set with websites from WebVoyager~\cite{he2024webvoyager} to incorporate standard web agent benchmarks. Next, we categorize these websites using the Trellix TrustedSource~\cite{trellixtrustedsource} service. We filter categories such as \emph{Finance/Banking} that require Personally Identifiable Information (PII) like SSNs and real phone numbers, and websites requiring complex account verification involving biometrics~(e.g., Instagram). We discard websites that require Multi-Factor Authentication~(MFA) for sign-in. Two authors analyze the filtered websites to identify representative \taskname. This selection is based on established cybersecurity guidelines issued by governmental agencies, including the National Cyber Security Centre (NCSC) in the UK~\cite{ncscadvice}, and the National Institute of Standards and Technology (NIST)~\cite{nistcsf2} and the Federal Trade Commission (FTC)~\cite{ftccollectuseinfo} in the US. 

For each identified task, the authors draft a textual instruction representing the task, and ensure the task is constrained to Selenium's functionalities~(for example, native browser prompts are not supported). To represent sparse categories lacking tasks, such as Sports, we expand our search to the Tranco top 5,000 and add popular websites with any representative tasks. The curated dataset includes tasks of varying complexity: simpler tasks that require navigation to the page footer to opt out of data collection using a consent banner, and multi-step tasks that involve navigating multiple account layers to toggle various checkboxes

Unlike the aforementioned existing benchmark involving web agents performing general-purpose tasks~\cite{zhou2023webarena, he2024webvoyager, tur2025safearena}, a typical \taskname involves modifying the website's state from  $S_0$, its initial state, to $S_{final}$, wherein $S_t$ represents the state of the website at step $t$.
Although the desired final state $S_{final}$ for a task $\mathcal{T}$ is always the same, the actions to achieve $S_{final}$ are different, depending on $S_0$. For example, let's consider a task requiring disabling marketing cookies; an agent should take no action if the cookies are already inactive in $S_0$, however, the agent must execute specific clicks to disable these cookies if active.

To account for such variability, we rigorously define $S_0$ for every task. Each task in our task dataset $D_T$ is composed of -- 1) a user query $P$ representing a task $\mathcal{T}$, and 2) a consistent initial State $S_0$ of $W$ with respect to $\mathcal{T}$. For tasks with binary settings~(such as toggles), we include two possible states: ON and OFF. For multi-choice settings (such as radio buttons), we initialize $S_0$ to the least-private or least-secure setting by default. Therefore, some tasks in our dataset are paired with multiple initial states $S_0$. We refer to these prompt-state pair in our dataset as an \emph{instance}. 

This process yields a total of \numtasks instances, representing 138 distinct \taskname across \numwebsites websites~(7 from WebVoyager). The task instances span \numwebsitecategories website categories~(\Cref{tab:websites}) and \numtaskcategories task categories~(\cref{tab:suc_by_tasktype_and_model_wonav}). Despite this wide coverage, a few categories remain underrepresented. For example, the Data \& Asset Management and UI/UX Preferences categories contain only 6 and 5 task instances, respectively. Furthermore, a handful of websites, such as Al Jazeera and IKEA, are limited to cookie notice tasks, and feature only 1 or 2 task instances.

To evaluate each agent fairly, we must ensure consistent $S_0$. A trivial solution to ensure consistency would be to create a fresh user account for each evaluation run of an agent. However, this approach is not scalable. Thus, we enable a consistent initial state $S_0$ with the help of a browser extension developed in-house~(as detailed in \Cref{sec:agent_architecture}).

\subsection{\moduletwo} 
\label{sec:agent_architecture}

The second module in \benchmark is \moduletwo. Here, we develop a system to execute instances from our dataset by building upon  the base implementation logic and action space of WebVoyager~\cite{he2024webvoyager}. 
WebVoyager takes a user prompt $P$ as input, along with context $c_t$ comprising 1) current and past visual screenshots grounded with Set of Marks~(SoM)~\cite{yang2023set} (numbered bounding boxes over interactable elements), 2) textual information of the interactive elements on page, and 3) text-based action history of the agent. The LLM decides an action $a_t$ based on $P$ and $c_t$ that is executed based on a Selenium-based browser automation framework~\cite{selenium}. The action space $\mathcal{A}$ comprises actions like `\texttt{CLICK}', `\texttt{SCROLL}', `\texttt{TYPE}', and `\texttt{ANSWER}'~(indicates model's completion).

We specifically enhance the action space to cover more actions typically performed by a real user and add \emph{Account and State Management} controls to create a system that executes instances in our dataset in a fully automated manner in the following steps:

1) \textbf{Authentication:} If the task requires an authenticated session, our system automatically logs in to the website.  

2) \textbf{State Initialization:} The system then sets the desired initial state $S_0$ for the instance.

3) \textbf{Task Execution:} The agent interacts with the environment using Selenium as the interface and performs actions necessary to execute the task.

\subsubsection{Account and State Management}
\label{sec:acc_state_management}
The WebVoyager benchmark~\cite{he2024webvoyager} comprises information general-purpose tasks within stateless, unauthenticated sessions. However, all tasks in our dataset require a consistent initial state, and most require authenticated user accounts on a website. To address this requirement, we develop an account and state management component that operates independent of the LLM backbone within the agent. These components handle the first two steps of our instance execution system.

Our system integrates persistent Chrome profiles into Selenium sessions, using a manual, pre-authenticated Google account to perform cross-site logins (via OAuth). We create a copy of this base profile to execute the three-step process described above.
Furthermore, to enforce initial states ($S_0$) and support automated authentication without Google OAuth, we implement a record-and-replay mechanism~\cite{barman2016ringer}. This approach prevents account logouts across evaluation runs and ensures reproducible $S_0$. Using our in-house browser extension, we record execution traces—sequences of user actions and DOM attributes (e.g., XPaths)—which are subsequently replayed via Selenium. This setup ensures consistent state management across agent evaluations. We detail the implementation of the record-and-replay mechanism and state management components below.

\paragraph{Record-and-Replay Mechanism:}

There exist many implementations of the record-and-replay mechanism~\cite{barman2016ringer,nass2024improving,leotta2016robula+}, allowing automated execution of recorded user actions. These tools typically capture user interactions on a website by extracting web element locators (e.g., XPath, CSS selectors, etc.) during recording, and later replicate~(replay) the same interactions by matching the locators to the elements in the DOM. One such implementation is the Chrome DevTools Recorder~\cite{chromedevtools2024recorder}, which we initially use to set the initial state $S_0$ of our system. However, we observe that the tool failed on dynamically rendered, React-based web applications~(e.g., Grammarly), where DOM structures are frequently re-generated, attributes may be non-deterministic, and component re-rendering alters element identities. As a result, the recorded locators were often not found during replay. We also observe that this tool does not always support elements present within the Shadow DOM~\cite{mdn2025shadowdom} and iframes. Similarly, another tool, Ringer~\cite{barman2016ringer}, was developed to capture user interactions and replay them later; however, it relied on stable user interfaces and has not been updated to be usable with the dynamic nature of modern user interface designs. These limitations motivate the design of our own record-and-replay tool.

Our tool, similar to Ringer~\cite{barman2016ringer}, is implemented as a browser extension to record execution traces for setting $S_0$, and as a Selenium script to replay the recording with the desired $S_0$. The browser extension contains three major components: (i) a content script injected into every page~(including iframes) using Google Manifest V3~\cite{chrome2026manifestv3}, (ii) a background service worker that manages the recording session, and (iii) a popup interface through which a user can start, stop, and name a recording session.

When a user begins recording a session, the content script dynamically overrides the \texttt{Event.\allowbreak prototype} methods during page load. This helps reliably capture user interactions even on websites that block extensions or injected scripts from recording them. As the user interacts with the page~(\texttt{click}, \texttt{mousedown}, or \texttt{pointerdown} events), the extension observes the event and determines the target element using the event's \texttt{composedPath()}, preserving traversal information across Shadow DOM boundaries. For each event, the extension captures a comprehensive list of web element locators and semantic metadata, ensuring replay robustness for websites that render elements dynamically. This includes basic locators like the CSS selector path~(annotated with \texttt{::shadow} markers at shadow-root boundaries) and XPath; standard attributes like \texttt{id}, \texttt{name}, \texttt{data-testid}, and \texttt{href}; native interactive tag names~(e.g., \texttt{button}, \texttt{input}) and the element's \texttt{outerHTML}; ARIA attributes\footnote{\url{https://developer.mozilla.org/en-US/docs/Web/Accessibility/ARIA}}
; the element's label text; and sibling and parent element text.

However, we found that relying solely on these generic attributes is insufficient for highly dynamic pages, where the attributes change upon page reload (e.g., Grammarly or Reddit).  To address this, we implement a novel deterministic indexing mechanism that we refer to as \webspindexT. This index is generated whenever the rendered DOM of the page changes, monitored using \texttt{MutationObserver}, using the \texttt{TreeWalker} API to traverse the DOM (including shadow DOMs if applicable) and identify all focusable and interactive elements. Each element is assigned an index, stored in the custom \webspindexT, based on its rendered DOM order. This ensures that the \webspindexT is stable even for dynamically loaded elements. 

To summarize, the content script captures event information and sends it to the background service worker, which exports it as a structured JSON file that is used by the Selenium-based replay script later. Every recorded event in this file comprehensively details the event type, frame path, event state, generic locators, semantic metadata~(ARIA labels and nearby text), and our deterministic \webspindexT. Additionally, the extension captures screenshots for all interactions, enabling visual inspection of the recorded session. 

The replay component of our tool is a Selenium-based script that uses the extension's exported JSON session to replay the events required to reach the desired state $S_0$. The script uses a cascading fallback strategy to re-identify the respective elements across both simple and dynamically rendered DOM structures. It first attempts shadow DOM-aware lookups for the stable attributes such as \texttt{data-testid}, \texttt{id}, \texttt{name}, and \texttt{aria-label}. If these attributes fail, the script moves on to identifying the elements using label text, nearby sibling text, CSS selector paths, and XPaths. The last fallback option is the \webspindexT. 

Once a target element is identified, the script reliably executes the intended action by attempting standard Selenium clicks, JavaScript-based click injections, and \texttt{ActionChains} mouse simulations using a fallback stratefy until the target element successfully registers the interaction. The script also dynamically manages execution contexts to support complex authentication flows~(e.g., Google or Microsoft OAuth), automatically detecting and switching the Selenium WebDriver focus to cross-origin iframes or pop-up windows prior to event execution. 

The replay script also accepts a configuration parameter to explicitly set stateful elements~(e.g., toggles and checkboxes) to either an ON or OFF state. Before interacting with a target element, the script determines the existing state based on ARIA attributes~(\texttt{aria-checked}, \texttt{aria-pressed}, \texttt{aria-selected}) or the native \texttt{checked} property. In some cases, it also incorporates domain-specific heuristics, such as evaluating CSS classes~(e.g., \texttt{a-switch-active}, \texttt{a-disabled}) to infer switch states. It skips the interaction if the element already matches the desired configuration. Otherwise, the script performs the interaction and subsequently verifies whether the desired state is achieved.
This conditional replay mechanism removes the need for the user to record the interactions for different desired initial states, as the script automatically handles both desired states `ON' and `OFF' with a single recorded session capturing the necessary events and elements.

\paragraph{Account Management:}
We create a primary Google sock puppet account that we use to create sock puppet accounts on other websites in our dataset either through Single Sign-On~(SSO) or with the email address and password. We initialize accounts with random attributes, such as name, age, interests, and demographics for websites that require profile details. The primary google account is linked to the Chrome profile tagged with the Selenium automation during the agent runs. Furthermore, some tasks in our dataset require the agent to operate on artifacts present in the created sock puppet accounts. For example, making a repository on GitHub or HuggingFace as private. To facilitate such task, we programmatically create empty artifacts~(e.g., HuggingFace and GitHub repositories). 

As mentioned above, it is possible that some websites might logout authenticated sessions due to inactivity between evaluation runs. Thus, we record execution traces using our browser extension for automatically logging in to the websites with tasks requiring authentication. The login trace captures a typical sign-in workflow, involving either Google SSO or standard email-password login, starting from the login page.

\paragraph{State Management:} We achieve a consistent $S_0$ for a majority of the required tasks in our dataset using our in-house record-and-replay tool. As mentioned in \Cref{sec:task_curation}, we decide the initial state for these tasks based on the type of element that is involved in the task. For elements that exhibit binary state~(ON and OFF) in isolation like toggle switches and checkboxes, we consider two initial states $S_0$: 1) an \textit{All-ON} state, where the target and adjacent elements are active; and 2) an All-OFF state, where they are inactive. For elements like radio buttons or dropdowns, where only one element in a group can be active at a time, we initialize $S_0$ with the least-private or least-secure setting (e.g., 'Send me daily email notifications' over 'Do not send me any email notifications') as active. 

Apart from ensuring consistent evaluation across runs, the dual initialization also helps evaluate models' ability to interpret the required state from the prompt $P$ and the existing state $S_t$ of relevant elements, and to avoid unnecessary actions. For example, if $P$ instructs model to disable all email notifications, and $S_0$ already reflects this, then the model's ideal action should be to navigate to the settings page and terminate without interacting the elements. Two authors recorded the necessary actions to set the desired $S_0$ for all tasks using the browser extension and the recorded actions are replayed using the Selenium replay script before an agent attempts the task.

For a handful of tasks, we set $S_0$ using alternative methods. For some Hugging Face and GitHub tasks, we use the respective official APIs to set $S_0$, e.g., repository visibility tasks. For tasks involving revoking inactive sessions, we define $S_0$ by creating five Selenium sessions that log in to the target website. Lastly, for cookie-related tasks, $S_0$ is always the default cookie setting from the website, which mostly is all cookies being active. This sets by default as we use a copy of the base Chrome profile for each run. 

\subsubsection{WebVoyager Enhancements and Differences}
\label{sec:voyage_differences}

The action space of WebVoyager includes basic actions `\texttt{CLICK}', `\texttt{TYPE}', `\texttt{SCROLL}', `\texttt{GOBACK}', `\texttt{GOOGLE}', `\texttt{WAIT}', and `\texttt{ANSWER}'. We significantly expand WebVoyager’s action space to accommodate the websites and tasks in our dataset. Webvoyager utilizes Selenium’s \texttt{ActionChains} to focus on elements and perform keyboard inputs. This approach would fail to scroll selected elements for a variety of reasons, including but not limited to non-focusable containers (e.g., \texttt{<div>}), non-scrollable elements (e.g. \texttt{overflow: hidden}), and sticky overlays that frequently intercept input events. To address this, we develop a JavaScript-based injection strategy that bypasses the limitations of simulated keyboard events. By querying the DOM stack at specific coordinates via \texttt{elementsFromPoint}, our framework identifies the highest-priority scrollable candidate. Next, the framework validates these candidates based on their computed styles (e.g., \texttt{overflow}) and properties (e.g., \texttt{scrollHeight}) to apply programmatic offsets directly to the scrollable DOM node. 

Furthermore, we extend the scrolling functionality to include 1) `\texttt{SCROLL\_TO\_END}' that allows rapid scrolling to page footer in long pages, 2) `\texttt{SCROLL\_WITHIN\_POPUP}`for scrolling inside modals 3) Horizontal scrolling on pages. We also add the action `\texttt{SWITCH\_TAB}' that allows the agent to switch between tabs seamlessly. These navigational enhancements are essential for the agent to perform many tasks in our dataset such as disabling cookies. 

We modify the JavaScript-based interactive element detection tool used in WebVoyager~(GPT-4V-Act~\cite{dupont2023gpt4vact}) to include elements in the Shadow DOM and modal/popup elements. The `\texttt{TYPE}' action of WebVoyager includes an automatic "ENTER" keypress, which we remove allowing the agent to perform tasks such as access token creation on websites like Docker without unintentionally hitting a `Submit' button on the page. We use randomized multi-color bounding boxes to improve element visibility on websites that render in dark mode by default. Additionally, we use the undetected chromedriver~\cite{undetected_chromedriver} and automatic captcha solvers~\cite{avci2024googlerecaptchabypass} in the setup to help avoid bot detection when interacting with websites. Lastly, we adopt the system prompt of WebVoyager with modifications to include the latest practices to system prompt writing and customizing it for evaluating \taskname. We include the full system prompt in \Cref{sec:agent_system_prompt}.

\subsection{Automated Verification}
\label{sec:automated_verification}

The third module of \benchmark is the fully automated evaluation of the agents on our dataset instances with an MLLM-as-a-Judge~\cite{chen2024mllm, lee2024prometheus}. This approach aligns with established Web Agent evaluation~\cite{he2024webvoyager, zhou2023webarena, lu2025agentrewardbench}. As prior benchmarks deal mostly with information retrieval tasks, their automated evaluators consisted mostly screenshots of the few steps of the agent trajectory. In contrast, we provide the entire trajectory as tasks in our benchmark also involve intermediate actions that are relevant to the task. Thus, the input to the judge, $M_J$ comprises: 1) the user prompt $P$, 2) the agent's entire task trajectory comprising the actions and environment snapshots~(screenshots) provided sequentially, and 3) a manually annotated ground truth action sequence~($\mathcal{G} = (g_0, g_1, \dots, g_m)$). The annotations were performed by an author, who is a web design expert and vetted by two authors. Each step $g_t$ consists of the action along with the target element. The target elements are based on standard convention followed by UI libraries~\cite{shadcnui}. The ground truth sequence $\mathcal{G}$ grounds the judge with necessary actions required to achieve desired final state, including when $S_0 = S_f$~(no action needed as initial state and final state are the same). 

We prompt $M_J$ to evaluate for successful task completion and answer with a binary \texttt{CORRECT} or \texttt{INCORRECT} classification along with a reasoning for its choice. We use this reasoning to assist us with our manual inspection of the trajectories and also use them in some of the analyses to derive stronger conclusions~\Cref{sec:rq3}. In addition to the task success and failure, we also track exceeding task maximum time limit~(timeout) and maximum iteration count, exceeding both results in automatic failure. We detail the judge's configuration and its performance on a human annotated subset in the upcoming section~(\Cref{sec:judge_validation}).

\section{MLLM Judge Development}
\label{sec:judge_validation}

In this section, we describe the manual annotation process used to curate a ground-truth subset of agent trajectories, and detail our automated judge ($M_J$) along with its performance on the annotated data. 

\smallskip 

\noindent \paragraph{Human Annotated Judge Evaluation Dataset Curation:}
We sample 200 agent trajectories across different models and datasets variants~(refer to \Cref{sec:eval_setup}). To ensure a balanced initial distribution, we use predictions from an early version of our automated judge based on Google's \emph{Gemini-2.5-Pro}~\cite{comanici2025gemini} to select 100 \texttt{CORRECT} and 100 \texttt{INCORRECT} instances. Next, one author manually inspected these agent trajectories against task instruction $P$ using the ground truth actions $\mathcal{G}$ as a reference, and assigned a label to them. Two authors combinedly vetted the annotations and corrected inaccuracies in the initial version. The Cohen's Kappa score between the initial and vetted version is $0.928$. The final dataset following the annotation process is slightly imbalanced towards the \texttt{CORRECT}~(115 instances). We then use this curated dataset to iteratively tune and evaluate our automated judge $M_J$. 

\smallskip 
\noindent \paragraph{Iterative Judge Development:} We consider four state-of-the-art candidates: Google's \emph{Gemini-3.1-Pro}~\cite{google2026gemini31pro} and \emph{Gemini-3-Pro}~\cite{geminiteam2025gemini3}, OpenAI's GPT-5.2~\cite{openai2025gpt52}, and Anthropic's \emph{Claude-Opus-4.6}~\cite{anthropic2026opus46}. We use a random subset of 10 examples from the evaluation dataset to tune the system prompt, temperature, reasoning budget, and ordering of the input on Google's AI Studio~\cite{googleaistudio} for the Gemini-3-Pro model. We slowly test improvements obtained in Gemini-3-Pro's judgment on the 10 examples and replicate the same setup across all other models on the whole evaluation dataset through the respective API endpoints. All four models operate with an almost similar system prompt that is available in \Cref{sec:judge_system_prompt}, a temperature of 1.0~(recommended/default setting for reasoning models) with a `high' or `dynamic' thinking budget. Finally, we implement a majority-vote ensemble of Gemini-3.1-Pro, Claude-Opus-4.6, and GPT-5.2 as our judge as this combination gives a higher F1 score than having Gemini-3-Pro in place of GPT-5.2.
 
\smallskip

\noindent \paragraph{Judge performance:}
The final evaluation results at the end of our iterative design process are in \Cref{tab:judge_table}. While, Gemini-3.1-Pro is more precise than the other three models, and Claude-Opus-4.6 achieves the best recall and F1 score of 93.91\% and 95.2\%, respectively. Although GPT-5.2 is less precise than other three models, its recall is as good as Claude-Opus-4.6. Overall, our majority-vote ensemble achieves an F1 score of 96.04\% and 95.5\% human agreement. Even though, Gemini-3-Pro has a higher F1 than GPT-5.2, the ensemble with GPT-5.2 does slightly better. Despite the standalone capabilities of Opus-4.6 and Gemini-3.1-Pro, we opt for the ensemble as our final automated judge $M_J$ for assessing the agents on the whole dataset as it not just achieves highest F1 scores but also mitigates a single-model's randomness.  

An analysis of the ten failure cases reveals that eight involve tasks targeting moderately or small-sized UI elements (e.g., checkboxes, toggles, or radio buttons), while two others relate to access token creation. One of the UI element failure is a cookie task where the judge is not able to identify the colors pertaining to ON and OFF states. Nevertheless, the overall accuracy of the ensemble approach remains sufficiently high to guarantee a sufficiently faithful evaluation of the agents across different backbone LLMs.

\begin{table*}[h]
\centering
\begin{minipage}[c]{0.35\textwidth}
\centering
\renewcommand{\arraystretch}{1.3}
\resizebox{\linewidth}{!}{%
\begin{tabular}{l c c}
\toprule
Total Instances & \#\texttt{CORRECT} & \#\texttt{INCORRECT} \\
\midrule
200 & 115 & 85 \\
\bottomrule
\end{tabular}
}

\vspace{0.8em}
\textbf{(a) Annotated Data Statistics}
\end{minipage}\hspace{1em}%
\begin{minipage}[c]{0.60\textwidth}
\centering
\resizebox{\linewidth}{!}{%
\renewcommand{\arraystretch}{1.4}%
\begin{tabular}{l c c c c}
\toprule
\textbf{Judge Model} & \textbf{Precision} & \textbf{Recall} & \textbf{F1-Score} & \textbf{Accuracy} \\
\midrule
Gemini-3-Pro & 93.0 & 92.2 & 92.6 & 91.5 \\
\rowcolor{aliceblue} 2. Claude-Opus-4.6 & 96.4 & \textbf{93.91} & \textbf{95.2} & 94.5 \\
Gemini-3.1-Pro & \textbf{98.15} & 92.17 & 95.07 & 94.5 \\
\rowcolor{aliceblue} 4. GPT-5.2 & 89.91 & 93.04 & 91.45 & 90.0 \\
\midrule
Majority Ensemble 1,2,3 & 97.30 & 93.91 & 95.57 & 95 \\
\rowcolor{aliceblue} Majority Ensemble 2,3,4 & \textbf{97.32} & \textbf{94.782} & \textbf{96.04} & \textbf{95.50} \\
\bottomrule
\end{tabular}%
}

\vspace{0.8em}
\textbf{(b) Automated Judge Performance (\%)}
\end{minipage}
\vspace{0.5em}
\caption{Evaluation Data Distribution and Judge Performance}
\label{tab:judge_table}
\end{table*}

\section{Experimental Results}
\label{sec:results}
In this section, first we describe the evaluation setup and introduce the three research questions that guide our analyses. Next, in each subsequent subsection, we report inferences on analyses addressing each research question individually. 

\subsection{Evaluation Setup}
\label{sec:eval_setup}

We run our \moduletwo with seven backbone MLLMs on the \numtasks instances in our dataset. The models comprise six proprietary models: Google's \emph{Gemini-3-Pro}, \emph{Gemini-2.5-Pro} \& \emph{Gemini-2.5-Flash}~\cite{comanici2025gemini}, Anthropic's \emph{Claude-Sonnet-4.5} \& \emph{Claude-Haiku-4.5}~\cite{anthropic2025claude45}, and OpenAI's \emph{GPT-5.1} \& \emph{GPT-5-mini}~\cite{singh2025openai}, and one open-weight model Google's \emph{Gemma-3-27B}~\cite{team2025gemma}. The proprietary models are all reasoning models and among the flagship models offered by the providers, while Gemma-3-27B is a leading open-weight non-reasoning model. Evaluating open models is crucial, as they enable privacy-preserving, on-device execution, and including Gemma-3-27B allows understanding the state of open-weight models for complex, non-retrieval web agent tasks.  We set the temperature to 1.0, use `dynamic' thinking mode for all reasoning models, and set maximum output tokens to 8192 for Claude models and 10,000 for the other models~(much higher than what agents require to describe its thoughts and action).

We integrate these models with our \moduletwo configuring maximum number of non-scroll and non-wait iterations per run to 20, and maximum runtime to 10 minutes. This setting is supported by the statistics from the ground truth: 1) Average number of non-scroll and non-wait actions across the instances in our dataset is 5.16~(56 instances with 5 actions), 2) Maximum number of actions is 13~(1 instance), and 3) Minimum number of actions is 2~(5 instances). We evaluate the agents on live websites in a `light mode' enforcing it through the Selenium ChromeDriver\footnote{Works except when `dark mode' is enforced by the website} and use randomized bounding box colors for the SoMs to ensure website background color does not inhibit the models from `observing' the interactive elements due to similar color bounding box colors. Please refer to \Cref{sec:eval_setup_details} for further experiment details, pipeline overhead, and maximum runtime setting of 10 minutes.

We pose the following four research questions: 
\begin{tcolorbox}[
    enhanced,
    colback=academicyellow,
    colframe=goldenborder,
    title=\textbf{Research Questions},
    fonttitle=\bfseries\sffamily,
    arc=0.5mm,
    boxrule=0.8pt,
    left=2mm, right=2mm, top=2mm, bottom=2mm
]
\begin{itemize}[leftmargin=2.5em, labelsep=0.5em]
    \item[\textbf{RQ1:}] Can web agents autonomously execute tasks, and how does their performance differ with and without explicit navigational instructions?
    \item[\textbf{RQ2:}] How does agent performance vary across different websites and task categories?
    \item[\textbf{RQ3:}] How does agent performance vary across different UI elements and their initial states?
    \item[\textbf{RQ4:}] How robust is agent performance across multiple independent trials of the task?
    
\end{itemize}
\end{tcolorbox}

We assess each agent trajectory's success and failure using our automated judge~(\Cref{sec:judge_validation}) and answer the above questions predominantly quantitatively, measuring success rate~(percentage of successful instances) and failure rate for each model across the dataset variants. The failure count is a total of the explicit mistakes by the model~(as predicted by the judge) and the number of instances that hit time or iteration limit~(automatically resulting in a failure). We also discuss representative failures qualitatively for each question. For RQ1-RQ3, we run a single trial per task instance for each model. For RQ4, we select a subset of both failed and succeeded task instances from the initial trial and run two additional trials, reporting the resulting $\passatk{k}$ and $\passexpk{k}$ scores~(see \Cref{sec:eval_setup_details} for definitions). $\passatk{k}$ measures if an agent solves a task at least once across `k' trials, whereas $\passexpk{k}$ calculates successful completion across all `k' trials. Due to the sensitivity of \taskname, $\passexpk{k}$ is the more critical of the two, as it captures agent's reliability.

\subsection{RQ1: Analyzing Exploration Capabilities of Backbone Agents} 
\label{sec:rq1}
As security and privacy options are nested within a website, an agent has to locate the correct page before executing actions. This requires the agent to \emph{plan and reason} about the task from the user prompt, \emph{explore} the website, and perform actions to solve the task. Thus, we construct two variants of our dataset to understand if agents can autonomously navigate an account to find the specific page relevant to a task. In the first variant, hereafter referred to as \emph{WithNav}, we construct user prompts that also includes navigation. For example, ``Navigate to my account settings and then privacy settings, and ensure my trip type is enabled as viewable while my name and location are disabled for my reviews.'' In the second variant, hereafter referred to as \wonavE, the user prompt $P$ is just the task instruction. For example, `Ensure my trip type is enabled as viewable while my name and location are disabled for my reviews.'' 

Table~\ref{tab:suc_fai_by_model_and_data_variant} contains the overall performance of the eight models across the 200 instances in both the \withnavE and \wonavE dataset variants. Gemini-3-Pro-Preview is the best performing model~(84.5\% on \withnavE and 82.5\% on \wonavE). The open-weight model Gemma-3-27b is the worst performing, achieving 25\% and 20\% on the \withnavE and \wonavE variants. Models typically show higher success rates for \withnavE compared to \wonavE, with the only anomaly being Claude-Sonnet-4.5 failing 5 more tasks in the \withnavE variant. The biggest performance drop from \withnavE to \wonavE is for Gemini-2.5-Flash, with a relative difference of 16.5\%~(21 instances). 

Comparing the models from the same provider, we generally observe that the `bigger' or 'more expressive' model show a higher success rate than its `smaller' or `less expressive' counterparts when navigation is not a part for the instruction. This could potentially be due to the stronger reasoning and planning capabilities. For instance, Claude-Haiku-4.5 fails 16 more tasks compared to Claude-Sonnet-4.5. The same applies to the models from Google, with Gemini-2.5-Pro performing better than Gemini-2.5-Flash. GPT's models are the only ones with similar performance~(1 more successful task for GPT-5.1).

Overall trends show that models perform better on average, when provided with the navigation needed to execute the task.

\begin{table}[ht!]
\centering
\resizebox{0.8\linewidth}{!}{
\begin{tabular}{l  ccc  ccc}
\toprule
\multirow{2}{*}{\textbf{Model}} & \multicolumn{3}{c|}{\textbf{\emph{WithNav} Variant}} & \multicolumn{3}{c}{\textbf{W/oNav Variant}} \\
\cmidrule(lr){2-4} \cmidrule(lr){5-7}
& \textbf{Success} & \textbf{Failure} & \textbf{T${\text{out}}$}
& \textbf{Success} & \textbf{Failure} & \textbf{T${\text{out}}$} \\
\midrule
Gemini-2.5-Flash     & 127 & 66 & 7 & 106 & 83 & 11 \\
\rowcolor{aliceblue} Gemini-2.5-Pro       & 131 & 59 & 10 & 122 & 74 & 4 \\
Gemini-3-Pro-Preview & 169 & 24 & 7 & 165 & 26 & 9 \\
\rowcolor{aliceblue} Claude-Haiku-4.5     & 117 & 27 & 56 & 106 & 25 & 69 \\
Claude-Sonnet-4.5    & 117 & 32 & 51 & 122 & 29 & 49 \\
\rowcolor{aliceblue} GPT-5-Mini           & 91 & 23 & 86 & 87 & 32 & 81 \\
GPT-5.1              & 108 & 65 & 27 & 88 & 85 & 27 \\
\rowcolor{aliceblue} Gemma-3-27b          & 50 & 128 & 22 & 40 & 128 & 32 \\
\bottomrule
\end{tabular}
}
\caption{Overall performance of the evaluated backbone models on the \withnavE and \wonavE dataset variants. The \textbf{Success} column indicates the number of successfully completed instances. The \textbf{Error} column represents explicit failures from the model. \textbf{T${\text{out}}$} denotes tasks that terminated due to a maximum task time of 600 seconds or a 20 iteration limit on non-scroll and non-wait actions. (Success + Error + T${\text{out}}$ equals the 200 total evaluation instances for both variants).}
\label{tab:suc_fai_by_model_and_data_variant}
\end{table}

\begin{table}[htbp]
\centering
\resizebox{0.8\linewidth}{!}{
\begin{tabular}{lrrrr}
\toprule
Model & Both Correct & Only WithNav & Only W/oNav & Both Failed \\
\midrule
Gemini-2.5-Flash & 89 & 38 & 17 & 56 \\
\rowcolor{aliceblue} Gemini-2.5-Pro & 101 & 30 & 21 & 48 \\
Gemini-3-Pro-Preview & 147 & 22 & 18 & 13 \\
\rowcolor{aliceblue} Claude-Haiku-4.5 & 83 & 34 & 23 & 60 \\
Claude-Sonnet-4.5 & 90 & 27 & 32 & 51 \\
\rowcolor{aliceblue} GPT-5-Mini & 64 & 27 & 23 & 86 \\
GPT-5.1 & 64 & 44 & 24 & 68 \\
\rowcolor{aliceblue} Gemma-3-27b & 23 & 27 & 17 & 133 \\
\bottomrule
\end{tabular}
}
\caption{Instance-level performance comparison across the \withnav and \wonav dataset variants. The columns indicate the number of instances where a model succeeded in both variants~(Both Correct), or exclusively in one variant~(Only \withnav or Only \wonav), or failed in both~(Both Failed).}
\label{tab:with_and_wo_nav_performance}
\end{table}

To further understand if navigational information in the prompt improves model performance, we present another instance-level analysis in \Cref{tab:with_and_wo_nav_performance}, detailing instances where models succeed with and without navigational information, succeed only with or only without it, or fail in both cases. The results again reinforce that models generally benefit from explicit navigation details. Gemini-3-Pro-Preview is the most consistent and capable model across the two variants. It solves 147 out of the 200 instances regardless of whether navigation is included in its prompt, while failing in both variants only 13 times. Gemma-3-27b fails in both variants on 133 instances, while succeeding in both on just 23 instances.   

The results also reveal that when a model succeeds on only one variant, it is almost always the \withnavE variant. For example, there is a wide gap in the performance when comparing instances that were only solved in \withnavE and \wonavE: 38 vs 17 instances for Gemini-2.5-Flash and 44 vs 24 instances for GPT-5.1. Claude-Sonnet-4.5 alone solves more tasks only for \wonavE variant~(32 tasks) than only for \withnavE variant~(27 tasks).

Apart from the overall success rate, \cref{tab:suc_fai_by_model_and_data_variant} also includes task failures due to task timeout or iteration limit on non-scroll and non-wait instances, revealing differences in how models fail across different tasks. The failures of all Gemini models are more due to explicit errors while completing the task than to timing out~(maximum only 11 instances). GPT-5-Mini's failures are significantly more due to timeout or iteration limit across both \withnavE and \wonavE, highlighting potential limitations in exploring websites and deciding actions in the given time even when provided with navigation. Claude-Sonnet-4.5 and Claude-Haiku-4.5 also show more failures due to timeout than explicit mistakes for both variants. Gemma-3-27b's failures are dominated by explicit mistakes in completing the task rather than timeouts.

We present examples of two successful task completions on the \wonavE variant by models Gemma-3-27b and Claude-Haiku-4.5 in \Cref{fig:reddit_success}, and \Cref{fig:grammarly_success}. We also present a specific case in \Cref{fig:twitch_combined}, where Gemini-3-Pro successfully completes a Twitch task to disable story mentions only when provided with explicit navigational instruction.

\subsection{RQ2: Analyzing Performance Across Websites and Task Categories}
\label{sec:rq2}
In this subsection we address RQ2 by breakdowning success rate by website and task category. 

\begin{table*}[ht!]
\centering
\begin{minipage}[t]{0.5\textwidth}
\vspace{0pt} 
\resizebox{\linewidth}{!}{
\begin{tabular}{p{2.6cm} cccccccc c} %
\toprule
\multirow{2}{*}{\textbf{Website}} & \multicolumn{8}{c}{\textbf{\wonavE Variant}} & \multirow{2}{*}{\textbf{\# Inst.}} \\
\cmidrule(lr){2-9}
 & \textbf{2.5F} & \textbf{2.5P} & \textbf{3P} & \textbf{H4.5} & \textbf{S4.5} & \textbf{5m} & \textbf{5.1} & \textbf{3Ge} & \\
\midrule
Airbnb & 4 & 6 & \textbf{9} & 2 & 7 & 1 & 4 & 5 & 9 \\
\rowcolor{aliceblue}
AlJazeera & 0 & 0 & 0 & \textbf{1} & 0 & 0 & 0 & \textbf{1} & 1 \\
AllRecipes & \textbf{1} & 0 & \textbf{1} & \textbf{1} & \textbf{1} & 0 & \textbf{1} & 0 & 1 \\
\rowcolor{aliceblue} Amazon & 6 & 5 & \textbf{8} & 4 & 5 & 5 & 4 & 2 & 8 \\
BBC & \textbf{3} & \textbf{3} & \textbf{3} & \textbf{3} & \textbf{3 }& \textbf{3} & 2 & 0 & 3 \\
\rowcolor{aliceblue} Coursera & 2 & 3 & 4 & \textbf{6} & 3 & 2 & 1 & 0 & 6 \\
Docker & 4 & 2 & \textbf{5} & 2 & \textbf{5} & 3 & 1 & 0 & 8 \\
\rowcolor{aliceblue} Duolingo & 4 & 5 & \textbf{7} & 4 & 5 & 5 & 3 & 3 & 7 \\
GitHub & 12 & 11 & \textbf{13} & 9 & 5 & 9 & 9 & 4 & 14 \\
\rowcolor{aliceblue} Goal & 2 & 1 & \textbf{5} & \textbf{5} & \textbf{5} & 1 & 3 & 1 & 6 \\
Goodreads & 1 & 5 & 6 & 1 & 1 & 1 & 2 & 1 & 6 \\
\rowcolor{aliceblue} GoogleAdCenter & 5 & 5 & \textbf{7} & 6 & 6 & 2 & 3 & 1 & 7 \\
Grammarly & 3 & 8 & \textbf{10} & 7 & 6 & 9 & 5 & 3 & 10 \\
\rowcolor{aliceblue} HuggingFace & 6 & 6 & 7 & 5 & 7 & 6 & \textbf{8} & 2 & 9 \\
\bottomrule
\end{tabular}}
\end{minipage}\hfill
\begin{minipage}[t]{0.5\textwidth}
\vspace{0pt} 
\resizebox{\linewidth}{!}{
\begin{tabular}{p{2.6cm} cccccccc c} %
\toprule
\multirow{2}{*}{\textbf{Website}} & \multicolumn{8}{c}{\textbf{\wonavE Variant}} & \multirow{2}{*}{\textbf{\# Inst.}} \\
\cmidrule(lr){2-9}
 & \textbf{2.5F} & \textbf{2.5P} & \textbf{3P} & \textbf{H4.5} & \textbf{S4.5} & \textbf{5m} & \textbf{5.1} & \textbf{3Ge} & \\
\midrule
IKEA & 1 & 1 & \textbf{2} & \textbf{2} & \textbf{2} & 1 & 1 & 0 & 2 \\
\rowcolor{aliceblue} Moodle & 2 & 2 & \textbf{4} & 0 & 2 & 0 & 0 & 0 & 5 \\
NVIDIA & 0 & 2 & 2 & \textbf{3} & \textbf{3} & \textbf{3} & 1 & 0 & 3 \\
\rowcolor{aliceblue} OldReddit & 4 & 6 & \textbf{7} & 3 & 6 & 2 & 3 & 4 & 8 \\
OpenStreetMap & 1 & 1 & \textbf{2} & 1 & 1 & 1 & 1 & 0 & 2 \\
\rowcolor{aliceblue} Pinterest & 6 & 12 & \textbf{13} & 7 & 10 & 5 & 4 & 1 & 17 \\
Quora & 6 & 7 & \textbf{8} & 0 & 1 & 1 & 2 & 3 & 9 \\
\rowcolor{aliceblue} Reddit & 7 & 6 & \textbf{10} & 7 & 5 & \textbf{10} & 8 & 2 & 10 \\
Shein & 1 & 2 & 1 & 1 & \textbf{3} & 0 & 2 & 1 & 3 \\
\rowcolor{aliceblue} Steam & \textbf{9} & 5 & 8 & 7 & 7 & 2 & 6 & 1 & 17 \\
Twitch & 6 & 7 & \textbf{8} & 3 & 8 & 3 & 2 & 2 & 11 \\
\rowcolor{aliceblue} USAToday & 0 & 1 & 2 & \textbf{3} & \textbf{3} & 2 & 2 & 0 & 4 \\
Wattpad & 5 & 4 & \textbf{6} & \textbf{6} & \textbf{6} & 5 & \textbf{6} & 3 & 6 \\
\rowcolor{aliceblue} Wolfram & 5 & 6 & \textbf{7} & \textbf{7} & 6 & 5 & 4 & 0 & 8 \\
\bottomrule
\end{tabular}}
\end{minipage}
\caption{Agent performance breakdown by website for the \wonavE dataset variant\protect\footref{note:shorthand}. The best or join-best performing model are in bold.}
\label{tab:suc_by_web_and_model_wonav}
\end{table*}

\footnotetext{Shorthands 2.5F, 2.5P, 3P, H4.5, S4.5, 5m, 5.1, and 3Ge refer to Gemini-2.5-Flash, Gemini-2.5-Pro, Gemini-3-Pro, Claude-Haiku-4.5, Claude-Sonnet-4.5, GPT-5-mini, GPT-5.1, and Gemma-3-27B, respectively.\label{note:shorthand}}

\begin{table}[ht!]
\centering
\resizebox{0.8\linewidth}{!}{
\begin{tabular}{l cccccccc cc}
\toprule
\textbf{Task Category} & \textbf{2.5F} & \textbf{2.5P} & \textbf{3P} & \textbf{H4.5} & \textbf{S4.5} & \textbf{5m} & \textbf{5.1} & \textbf{3Ge} & \textbf{\# Inst.} & \textbf{\# Websites}\\
\midrule
Account Security \& Access Control & 16 & 16 & \textbf{20} & 10 & 16 & 11 & 13 & 4 & 22 & 10\\
\rowcolor{aliceblue} Advertising \& Personalization Control & 10 & 13 & \textbf{19} & 13 & 14 & 10 & 9 & 6  & 19 & 7\\
Cookie \& Tracking Consent Management & 11 & 14 & 16 & 18 & \textbf{19} & 11 & 10 & 2 & 24 & 8\\
\rowcolor{aliceblue} Data \& Asset Management & 4 & 4 & 5 & 3 & 5 & 4 & \textbf{6} & 2 & 6  & 2 \\
Notification \& Communication Preferences & 30 & 26 & \textbf{41} & 32 & 31 & 24 & 24 & 10 & 51 & 13 \\
\rowcolor{aliceblue} Profile Visibility \& Customization & 10 & 14 & \textbf{21} & 5 & 10 & 6 & 10 & 9 & 22 & 9\\
Social Safety \& Content Moderation & 16 & 20 & \textbf{25} & 12 & 16 & 9 & 10 & 3 & 31 & 8\\
\rowcolor{aliceblue} UI/UX Preferences & 1 & 2 & \textbf{3} & 1 & 2 & 2 & 2 & 0 & 5 & 3\\
User Privacy \& Data Rights & 8 & 13 & \textbf{15} & 12 & 9 & 10 & 4 & 4 & 20 & 8\\
\bottomrule
\end{tabular}}
\caption{Agent success rates broken down by task category for the \wonavE dataset variant\protect\footnotemark. The best or join-best performing models are in bold.}
\label{tab:suc_by_tasktype_and_model_wonav}
\end{table}

\noindent \paragraph{\textbf{Performance breakdown by websites:}}
Although websites share underlying structural principles~(e.g., settings accessible through profile icon on top right corner), their specific user interfaces are unique. To understand if models struggle with any distinct website environments, we analyze the performance breakdown across individual websites in the \wonavE variant~(refer to \Cref{tab:suc_by_web_and_model_wonav}). As one would expect from earlier results, Gemini-3-Pro is the best or joint-best performing model for 21 out of the 28 websites. 

However, the overall results convey some interesting anomalies where lower-ranked models from \Cref{sec:rq1} perform better. For example, Claude-Haiku-4.5 achieves a perfect success rate on Coursera~(6 out of 6 instances), outperforming both Gemini-3-Pro~(4 instances) and its more capable counterpart Claude-Sonnet-4.5~(3 instances). Similarly, GPT-5-mini matches Gemini-3-Pro on Reddit by solving all 10 instances, and nearly solves all Grammarly instances~(9 out of 10). For both these websites, GPT-5-mini also outperforms GPT-5.1~(8 and 5 instances, respectively). 

The results also reveal that even the best performing models can struggle with certain websites that are harder to understand and interact with. For example, it can be noted that seven out of the eight models fail to achieve 50\% success rate on Steam~(17 instances), and even Gemini-3-Pro can only get 8 out of 17 tasks right. The same is also observable with Docker and Goal, where five of the eight models can only solve less than or equal to 50\% of the instances. Note that all 6 Goal instances belong to the same task category~(\Cref{tab:task_categories}), Notification \& Communication Preferences. This supports the hypothesis that models indeed struggle with specific UI patterns and layouts that are unique across websites.

We provide two examples of website specific design elements that confuse models. The first is that of Steam~(\Cref{fig:phase_1}), where the task is to disable two settings options in communication preferences. The model, Gemini-3-Pro, correctly disables these options but fails to scroll down to click on `Save Changes' button. Thus, the models changes never get stored properly. The next example is that of Duolingo~(\Cref{fig:duolingo_rq2}), where Gemini-2.5-Pro, drifts away from solving the task of making the user's profile private and instead solves an introductory French lesson. 

\noindent \paragraph{\textbf{Performance breakdown by task categories:}}
Task categories also rely on similar structural principles with unique, website-specific implementations. For example, while cookie notices are mostly present in the footer through either `Privacy Choices' or `Cookies' textual elements, the elements used to enable or disable cookies differ, ranging from toggle switches~(NVIDIA) to radio buttons~(BBC). Thus, we also breakdown the performance based on the task category in \Cref{tab:suc_by_tasktype_and_model_wonav} to analyze if models can successfully perform similar tasks across different websites. Consistent with earlier trends, Gemini-3-Pro is the best model, achieving best or join-best performance in 7 out of 9 categories, including task categories like Notifications \& Communication Preferences~(41 out of 51), and Social Safety \& Content Moderation~(25 out of 31). 

Similar to performance breakdown on websites, here too we notice that specific models perform better than Gemini-3-Pro on some task categories. For example, Claude-Sonnet-4.5 achieves the highest success rate on Cookie \& Tracking Consent Management instances, solving 19 out of 24, outperforming both Claude-Haiku-4.5~(18 instances) and Gemini-3-Pro~(16 instances). GPT-5.1 solves all 6 Data \& Asset Management instances belonging to the websites HuggingFace and GitHub. And we also notice GPT-5-mini successfully completes 10 out of 20 instances in User Privacy \& Data Rights category, while GPT-5.1 only does 4 tasks. Gemma-3-27b shows extremely limited performance across most categories, with a notable case of Cookie \& Tracking Consent Management~(2 out of 24 instances) tasks highlighting the gap that exists between open-weight and proprietary reasoning models. 

Our results also reveal that models struggle mostly with three categories: 1) UI/UX Preferences~(where all models except Gemini-3-Pro have less than 50\% success rate); 2) Profile Visibility \& Customization~(where 6 models have below 50\% success across 22 instances); and 3) Social Safety \& Content Moderation~(where 4 models show less than 50\% success rate across 31 instances).

We present an example in \Cref{fig:docker_rq2}, where GPT-5-Mini keeps opening and closing the cookie notice even after successfully setting the cookies as requested in the instruction.

\subsection{RQ3: Analyzing Performance Across UI elements and Initial States}
\label{sec:rq3}
\begin{table}[ht!]
\centering
\resizebox{0.8\linewidth}{!}{
\begin{tabular}{l cccccccc c}
\toprule
\textbf{UI Element} & \textbf{2.5F} & \textbf{2.5P} & \textbf{3P} & \textbf{H4.5} & \textbf{S4.5} & \textbf{5m} & \textbf{5.1} & \textbf{3Ge} & \textbf{\# Inst.} \\
\midrule
Button & 54 (26) & 70 (15) & 92 (3) & 61 (0) & 76 (10) & 48 (12) & 49 (31) & 19 (40) & 111 \\
\rowcolor{aliceblue} Checkbox & 16 (19) & 22 (15) & 30 (6) & 19 (0) & 18 (9) & 10 (9) & 13 (17) & 6 (23) & 40 \\
Dropdown & 54 (9) & 50 (7) & 72 (2) & 42 (0) & 50 (2) & 38 (8) & 41 (25) & 16 (32) & 93 \\
\rowcolor{aliceblue} Icon & 29 (2) & 34 (1) & 44 (0) & 26 (0) & 28 (0) & 22 (2) & 24 (11) & 9 (11) & 52 \\
Link & 93 (20) & 101 (16) & 138 (7) & 90 (0) & 103 (3) & 73 (14) & 73 (41) & 32 (47) & 172 \\
\rowcolor{aliceblue} Menu & 2 (2) & 5 (0) & 7 (0) & 1 (0) & 5 (0) & 0 (1) & 2 (2) & 3 (2) & 7 \\
Option & 41 (5) & 48 (1) & 64 (0) & 35 (0) & 43 (0) & 34 (2) & 34 (16) & 16 (20) & 77 \\
\rowcolor{aliceblue} Radio Button & 15 (2) & 14 (6) & 13 (5) & 9 (0) & 13 (4) & 7 (5) & 10 (7) & 4 (10) & 20 \\
Text Input & 8 (3) & 5 (3) & 11 (0) & 4 (0) & 7 (0) & 3 (3) & 6 (4) & 1 (5) & 14 \\
\rowcolor{aliceblue} Toggle & 45 (49) & 56 (45) & 80 (18) & 54 (0) & 58 (20) & 47 (9) & 39 (49) & 21 (64) & 98 \\
\bottomrule
\end{tabular}}
\caption{Performance breakdown by target UI element for the \wonavE dataset variant\protect\footref{note:shorthand}. The \# Inst. column indicates the total number of unique instances where the corresponding UI element is part of the task's solution. The model columns report the successful task count involving the UI element, with task failures directly attributed to that specific element shown in parentheses.}
\label{tab:ui_element_without_nav}
\end{table}

\begin{table}[h!]
\centering
\resizebox{0.8\linewidth}{!}{
\begin{tabular}{l  cccc}
\toprule
\textbf{Model} & \textbf{Both Correct} & \textbf{Only ON} & \textbf{Only OFF} & \textbf{Both Failed} \\
\midrule
Gemini-2.5-Flash & 18 & 17 & 9 & 18 \\
\rowcolor{aliceblue} Gemini-2.5-Pro & 22 & 19 & 8 & 13 \\
Gemini-3-Pro-Preview & \textbf{47} & 4 & 6 & 5 \\
\rowcolor{aliceblue} Claude-Haiku-4.5 & 22 & 10 & 8 & 22 \\
Claude-Sonnet-4.5 & 20 & 17 & 9 & 16 \\
\rowcolor{aliceblue} GPT-5-Mini & 18 & 7 & 5 & 32 \\
GPT-5.1 & 10 & 12 & 15 & 25 \\
\rowcolor{aliceblue} Gemma-3-27b & 4 & 8 & 9 & 41 \\
\bottomrule
\end{tabular}}
\caption{Task-level performance comparison across a total of 62 tasks with both initital states `ON' and `OFF'. The columns indicate the number of instances where a model succeeded in both states (Both Correct), or exclusively in one state (Only ON or Only OFF), or failed in both states (Both Failed).}
\label{tab:on_off_without_nav}

\paragraph{Performance}
\end{table}

In this subsection, we address RQ3 by analyzing how models comprehend different UI elements and their state through two separate analyses. 

\noindent
\paragraph{\textbf{Performance breakdown by UI elements:}}
To determine if models are sensitive to specific interaction modalities, we evaluate performance based on the target UI element types required to execute each task~(refer to \Cref{tab:ui_element_without_nav}), meaning the model must correctly interact with every designated element type to successfully complete the instance. We extract these target elements for each instance from the manually annotated ground truth actions~(see \Cref{sec:automated_verification}). Additionally, we use the reasoning output from Gemini-3.1-Pro, the most precise model in our ensemble judge, to identify if explicit model failures~(without considering timeout) are directly due to one or more specific UI elements in ground truth elements. To perform this, we pass Gemini-3.1-Pro's reason along with the ground truth actions and UI elements to Gemini-3.1-Pro in a separate evaluator. We include the system prompt for this evaluator in \Cref{sec:system_prompt_ui_failure}. We manually check a sample to ensure it is correct. In \Cref{tab:ui_element_without_nav}, these element-specific failure counts are denoted in parentheses.

Gemini-3-Pro achieves the highest success rate across all elements except \texttt{Radio Button}s. For the other models, tasks involving the UI elements \texttt{Text Input}~(14 instances) and \texttt{Menu}~(7 instances) lead to uniformly low success rates. For instance, GPT-5-mini is able to complete only 3 and 0 tasks involving these elements, respectively, explicitly failing due to the \texttt{Text Input} in 3 instances and the \texttt{Menu} in 1 instance. Furthermore, stateful elements like \texttt{Toggle}~(98 instances) and \texttt{Radio Button}~(20 instances), which form a major portion of our dataset, lead to lower success rates and high direct failure rates across all models.

Similarly, Gemma-3-27B successfully completes tasks involving toggles in only 21 instances. %
We also observe that within the Claude and GPT families, performance on toggles is relatively consistent between variants, whereas the larger models~(Claude-Sonnet-4.5 and GPT-5.1) demonstrate better performance on radio buttons. Lastly, we notice that the UI elements \texttt{Link} and \texttt{Button} show considerably lower direct failure rates relative to their frequency across all models. This suggests that while models reliably complete tasks involving standard navigational elements like links and buttons, their capabilities degrade significantly on elements that require conditional interaction. We explore this state-dependency further in the subsequent analysis. For example, both Gemini-2.5-Flash and Gemini-2.5-Pro struggle significantly with toggles: 2.5-Flash has 49 instances of direct failures~(50.0\% of toggle tasks), and Gemini-2.5-Pro shows 45 direct failures.

\paragraph{\textbf{Performance across tasks with dual initial state:}}
As mentioned earlier in \Cref{sec:task_curation}, our dataset includes tasks evaluated under dual initial states~(`ON' and `OFF'), requiring different solutions for the same user instruction. For example, when instructed to disable a toggle that is already `OFF', the agent must correctly perceive this initial state $S_0$ of OFF and not interact with the elements unnecessarily. We analyze these paired instances representing 62 tasks in \wonavE~(mostly involving UI elements toggle and checkboxes), to assess how frequently models successfully solve both configurations of the exact same task~( see \Cref{tab:on_off_without_nav}). Gemini-3-Pro exhibits best state-awareness, successfully completing 47 tasks for both initial states and only for the `ON' and `OFF' states, in 4 and 6 instances respectively. We observe that most models succeed more frequently when the initial state $S_0$ is `ON' rather than `OFF', showing a strong dependency on $S_0$. For instance, Gemini-2.5 Pro exclusively solves more tasks when $S_0$ is `ON'~(19 tasks) than when it is `OFF'~(8). It is to be noted that a majority of these 62 tasks involve disabling options or a combination of enabling and disabling options. This discrepancy along with results indicate that models frequently fail to accurately perceive the initial element state, often executing incorrect action sequences when the setting is already `OFF'~(more apparent in smaller models). This shows the current web agents have to be improved on comprehending element state, with respect to the instruction, before deployments to solve \taskname. We present qualitative examples of such stateful element releated failures in \Cref{sec:failure_figures}.

\subsection{RQ4: Analyzing Agent Robustness across multiple runs}
\label{sec:rq4}
To assess agent robustness in performing \taskname, we curate a subset of the \wonavE variant for each model, comprising all failures from trial 1 along with a sample of 2 successful tasks instances per website (whenever available). We perform two additional trials on this subset and report the $\passatk{1}$, $\passatk{3}$ and $\passexpk{3}$ metrics in \Cref{tab:passatk}. As Gemini-3-Pro was discontinued, we use Gemini-3.1-Pro to analyze robustness across multiple runs as it comes from the same family~(refer to \Cref{sec:eval_setup_details}). While $\passatk{k}$~\cite{chen2021evaluating} captures whether a model can solve a task at least once, $\passexpk{k}$~\cite{yao2024tau} is the more important metric for \taskname, since any single failure can potentially lead to additional data collection or reduced security, making consistency across trials essential. The $\passatk{k}$ and $\passexpk{k}$ scores show expected trends, i.e., $\passatk{k}$ for all models increases and $\passexpk{k}$ decreases as as k increases.  

\begin{table}[ht]
\centering
\resizebox{0.8\linewidth}{!}{
\begin{tabular}{lrcccccc}
\toprule
Model & $N$ & pass@1 & pass@2 & pass@3 & pass$^{2}$ & pass$^{3}$ \\
\midrule
Gemini-2.5-Flash  & 145 & 0.41 & 0.55 & 0.61 & 0.27 & 0.19 \\
\rowcolor{aliceblue} Gemini-2.5-Pro    & 128 & 0.46 & 0.62 & 0.70 & 0.30 & 0.23 \\
Gemini-3.1-Pro    & 98 & \textbf{0.71} & \textbf{0.84} & \textbf{0.90} & \textbf{0.58} & \textbf{0.51} \\
\rowcolor{aliceblue} Claude-Haiku-4.5  & 141 & 0.39 & 0.53 & 0.60 & 0.25 & 0.18 \\
Claude-Sonnet-4.5 & 129 & 0.46 & 0.62 & 0.70 & 0.30 & 0.22 \\
\rowcolor{aliceblue} GPT-5-Mini        & 151 & 0.38 & 0.51 & 0.59 & 0.24 & 0.17 \\
GPT-5.1           & 157 & 0.33 & 0.47 & 0.55 & 0.19 & 0.13 \\
\rowcolor{aliceblue} Gemma-3-27b       & 188 & 0.12 & 0.20 & 0.27 & 0.04 & 0.02 \\

\bottomrule
\end{tabular}}
\caption{$\passatk{k}$, $\passexpk{k}$ scores for k=1,2,3 across three trials for a subset of \wonavE variant. $N$ is the number of tasks judged across three trials.}
\label{tab:passatk}
\end{table}

Across three trials, Gemini-3.1-Pro is the strongest model on both \passatk{k}
and $\passexpk{k}$, solving 90\% of the selected tasks at least once across the three trials. Gemini-2.5-Pro and Claude-Sonnet-4.5 are the next strongest models, with each solving close to 70\% of the selected tasks at least once. In contrast, Gemma-3-27B, the only open-weight model, achieves a $\passatk{3}$ of just 0.27. On $\passexpk{3}$, the best-performing model is Gemini-3.1-Pro with a score of 0.51. The next best models Claude-Sonnet-4.5 and Gemini-2.5-Pro are only able to solve 22\% and 23\% of tasks across all three trials respectively, while Gemma-3-27B reliably solves just 2\% of tasks across all three trials. This highlights the significant difficulty web agents face in solving \taskname consistently.

\section{Human Analysis of Agent Failure} \label{sec:human_failure_analysis}

To better understand agent failures on \taskname, we manually analyze a random sample of 40 failed trajectories~(35 for Gemini-3-Pro) per backbone model from trial 1 of the \wonav variant. We categorize them into five types: 1) \emph{Incorrect Navigation:} The agent fails to reach the relevant task page or the specific section within it; 2) \emph{Hallucinated Success}: The agent takes actions contrary to its thought and misinterprets its own action sequence as a successful completion of the given task; 3) \emph{State Misunderstanding:} The agent performs the opposite of the required task when interacting with stateful UI elements; 4) \emph{Repeated Actions:} The agent repeats a set of actions in a loop without making progress towards the task; 5) \emph{Partial Completion:} The agent successfully completes only a subset of the task requirements. A single failure can be attributed to more than one category. Our taxonomy follows AgentRewardBench~\cite{lu2025agentrewardbench}, refining its categories to characterize \taskname. Most importantly, we introduce \emph{State Misunderstanding}, representing an important failure mode in our dataset.

\begin{table}[t]
\centering
\begin{tabular}{lrrrrr}
\toprule
Model & IN & SM & HS & RA & PC \\
\midrule
Claude-Haiku-4.5      & 23 &  5 &  8 & 3 &  8 \\
\rowcolor{aliceblue} Claude-Sonnet-4.5     & 23 & 13 & 10 & 1 &  1 \\
GPT-5-Mini            & 18 & 13 &  6 & 8 &  8 \\
\rowcolor{aliceblue} GPT-5.1               & 22 &  7 & 19 & 1 &  2 \\
Gemini-2.5-Flash      & 19 & 16 & 26 & 0 & 11 \\
\rowcolor{aliceblue} Gemini-2.5-Pro        & 11 & 23 & 33 & 0 & 13 \\
Gemini-3-Pro-Preview  & 14 &  8 & 17 & 2 &  4 \\
\rowcolor{aliceblue} Gemma-3-27b           & 27 &  9 & 16 & 1 &  2 \\
\bottomrule
\end{tabular}%
\caption{Human Analysis of agent failures per model. The abbreviations IN, SM, HS, RA, PC indicate the failure types incorrect navigation, state misunderstanding, hallucinated success, repeated actions, and partial completion.}
\label{tab:failure_categorization}
\end{table}

Two authors independently analyzed disjoint subsets of the sampled failures. \Cref{tab:failure_categorization} reports the results after discarding 27 instances where the judge incorrectly predicted a failure. We observe distinct failure profiles across models. Navigational issues are frequent across all models, including Gemini-3-Pro-Preview. Both Gemini-2.5-Flash and Gemini-2.5-Pro struggle to interpret stateful elements, with 16 and 23 failures attributable to a misunderstanding of state, and both models tend to complete tasks only partially. GPT-5-Mini, in contrast, most frequently exhibits repetitive actions. Finally, both Claude models fail predominantly due to navigational issues, with comparatively fewer occurrences of other failure types. 

As noted in RQ1, failures of the Claude and GPT model families stem more from timeout issues than explicit task errors, and the model error profiles from our manual analysis reveal that these models fail more often due to navigational issues leading to timeout. Gemma-3-27b's incorrect navigation failures also consist of examples like \Cref{fig:qual_analysis_fig_1}, where the model searches on Google for instructions and quickly answers that the task does not exist without exploring the website properly.

Across the analyzed samples, we identify 76 instances where models failed due to a combination of state misunderstanding and hallucinated success, i.e., models' thoughts and actions did not match, and they mistakenly clicked the wrong buttons and proceeded to the next step without properly verifying the change. We present examples of these errors in the appendix and encourage the reader to refer to them.  \Cref{fig:twitch_combined,fig:qual_analysis_fig_10} illustrate navigational difficulties of Gemini-3-Pro. \Cref{fig:qual_analysis_fig_19,fig:qual_analysis_fig_15,fig:qual_analysis_fig_14,fig:qual_analysis_fig_9} are examples of model misunderstanding stateful elements and performing an action opposite to the task. In addition to these failure modes, we occasionally observed models making potentially destructive actions. Gemini-2.5-Flash \textbf{deactivates Pinterest account} when asked to sign out~(see \Cref{fig:pinterest_signout}), and Gemma-3-27B completes steps before payment for Huggingface PRO subscription~(see \Cref{fig:qual_analysis_fig_11}).

\section{Discussion}
In this section, we discuss the challenges and limitations of our work. We also outline the key takeaways from our evaluations and propose approaches to address them in future work. 

\subsection{Limitations and Challenges}

\paragraph{\textbf{Website and Task Restrictions:}}
In this work, we specifically select websites and tasks that can be automatically performed by an agent on a site. As such, we do not include sites that enforce 2-Factor Authentication (e.g., ESPN), as this would require human input. Furthermore, for the websites on our list, we create sock puppet accounts with random, unverifiable personal information, such as names and dates of birth. As such, our dataset does not include websites, like banking and financial websites, that rely on verifiable personal information, like Social Security Number. Lastly, we exclude tasks requiring actions outside the DOM, i.e., parts Selenium cannot control such as native browser overlays.
.

\paragraph{\textbf{Initial Manual Setup and Replication Overhead:}}
Creating the dataset required significant manual effort, including registering sock puppet accounts across numerous websites and identifying relevant \taskname. Due to ethical constraints, researchers using our benchmark should create their own accounts and manually record login traces. To facilitate this, our repository contains detailed setup instructions, specifying platforms requiring Google SSO and standard email authentication.

\paragraph{\textbf{State Management Susceptibility:}}
We evaluate web agents on live websites. Thus, the state management step in \moduletwo, is susceptible to structural and UI updates of the websites. For example, during our experiments, Steam modified the sidebar cookie window label on the 'Preferences' page from `Cookies and Browsing' to `Data and Browsing'. To manage this issue, we open-source our recording extension. On average, it takes around 2 minutes to record and test the execution trace.

We further note that a network latency of > 500ms may impact the stability of our custom \webspindexT since the index is calculated after a page load delay of 500ms. However, our replay mechanism relies primarily on identifying elements via DOM selectors, using \webspindexT only as a fallback~(\Cref{sec:acc_state_management}). Our experiments were run over stable network connections, and we did not observe any failures in our replay mechanism due to network latency.

\paragraph{\textbf{Result Replication:}} Exact replication of our results may vary depending on the locations as privacy and security settings on websites are governed by guidelines like GDPR, CCPA, etc. Thus, cross-region deployment of the same website's interfaces often differs. For example, cookie notices can vary across countries~\cite{safna2026cookie}.

\paragraph{\textbf{Scope of Agent Evaluation:}} Our goal is to evaluate model reasoning for \taskname. Thus, we measure task success on the frontend using the agent's trajectory. This mirrors existing web agent research~\cite{he2024webvoyager, zhou2023webarena, lu2025agentrewardbench}, checking if an agent can emulate a web user. While it is possible that websites do not honor privacy setting changes~\cite{matte2020cookie,trevisan20194, kancherla2025johnny}, verifying backend compliance requires information flow tracking such as reverse-engineering obfuscated JavaScript and analyzing encrypted network traffic, and is orthogonal to our goal.

\subsection{Evaluation Takeaways and Future Work}
Based on our evaluation of different backbone LLMs, we identify the following takeaways from our work and propose future directions the research community can improve upon: 

\paragraph{\textbf{Local Deployments need Stronger Open-Weight Models:}}
A privacy-preserving deployment of web agents is crucial as sensitive user information is often used while executing tasks, especially when the tasks are \taskname. In our evaluation, we observed a substantial performance gap between the open-weight model (Gemma-3-27B) and proprietary models. Bridging this gap is critical for deployments without compromising user data. In addition to using open-weight models locally, alternate approaches to manage settings, such as dedicated web APIs~\cite{song2025beyond}, can reduce privacy risks for the user.

\paragraph{\textbf{Improving Stateful Element Understanding:}} Our evaluation reveals backbone models struggle to comprehend the correct state of UI elements like \texttt{Toggle}, \texttt{Radio Button} and a mistake in a single step while performing the task can not just affect the task success but also compromise user data. So, state comprehension of backbone models have to be improved for overall integrity of web agents. This can be achieved by improving the harness for the backbone model~\cite{young2025harnesses} and enhancing the inherent UI understanding of backbone models.

\paragraph{\textbf{Overcoming Site-Specific UI Variations:}} RQ2's results indicate agent failure due to website and task-specific patterns. Examples include popups interrupting execution~(\Cref{fig:duolingo_rq2}), and isolated save buttons to confirm setting changes~(\Cref{fig:qual_analysis_fig_1}). This limitation can be mitigated by including agent policies based on common website designs, ensuring the model systematically verifies relevant criteria before terminating the task. Additionally, web developers can adopt agent-friendly UI for higher success rate~\cite{lu2025build}. For example, incorporating a keyword based search bar on a complex settings pages will provide agents with a direct, alternative route to locate a specific option in a page instead of relying on excessive scrolling~(refer to example in \Cref{sec:agent-friendly-design-appendix}). Sanboxed-deployments of websites with such variations is a useful dataset to finetune the backbone models to understand different website designs in the wild. 

\paragraph{\textbf{Mid-flow Security and Privacy Decisions:}} In-the-wild, web agents often need to make security and privacy decisions mid-flow during general task execution. For example, handling a cookie banner before accessing a website~(see \Cref{sec:mid-flow-appendix}). Existing agentic browsers manage these scenarios by controlling the agent's choice using predefined policies or delegating them to the user. Thus, it is necessary to disentangle the agent's inherent understanding of the \taskname from its policy adherence. In this paper, we establish a baseline for the former by directly prompting agents to solve \taskname. In future work, to understand mid-flow decisions, a systematic evaluation should include policies representing diverse privacy personas to measure the agent's policy adherence separately from its security and privacy reasoning and overall task success.

\paragraph{\textbf{Robustness to Deceptive UIs:}} Real websites often contain deceptive patterns, and prior research has highlighted web agent vulnerabilities to deceptive patterns and other deceptive UI changes, such as adversarial injections. Our evaluation uses live websites without explicitly introducing any structural or visual changes. While some websites in our dataset might inherently contain deceptive patterns, we do not isolate this as a cause for agent failure, instead measuring success based on whether agents can emulate web users. Given the critical nature of these vulnerabilities, we urge future work to investigate the effect of deceptive UI changes on \taskname that handle sensitive user data.  

\paragraph{\textbf{Improving Agent Instantiation Context:}} Our agent instantiation includes our novel state and account management components, and predominantly uses screenshots with SoM grounding as context to the backbone model. This approach has been the most used across many other established benchmarks~\cite{koh2024visualwebarena, he2024webvoyager, yang2023set}. We leave the evaluation of alternate harness~\cite{young2025harnesses}~(context) to future work, due to compute budget restrictions.

\section{Conclusion}
In this paper, we introduce \textbf{\benchmark}, a comprehensive evaluation framework designed to assess web agent evaluation on a new dataset of 200 \emph{website security and privacy} task instances that are paired with an initial state. We develop a robust a account and state management tool based on a custom Google Chrome extension and use it to build our agentic system for executing website tasks. We instantiate this systems with 8 MLLMs and perform fine-grained evaluation across websites, task categories, and UI element types. Our analyses reveal that models experience a performance drop when explicit navigational details are not part of the instruction,  and struggle extensively with stateful UI elements, often demonstrating a bias towards altering already correct initial states. 

These vulnerabilities highlight critical barriers for safe deployment of web agents. If an agent is trusted with sensitive account settings, making these kinds of basic execution errors could easily compromise a user's security or leak private data. Our robustness analysis in \Cref{sec:rq4} illustrates the severity of these risks. Overall, \textbf{\benchmark} presents the research community a dataset, a system with state management plus agent implementation, and an automated judge—to rigorously to test performance.

\section{Acknowledgements}

This work is partially supported by the NSF through
awards CNS-1942014, CNS-2247381, and TI-2533192. It is also supported by a grant from the Google PSS Privacy Faculty Award program.

\clearpage

\bibliographystyle{plain}
\bibliography{main}

\clearpage

\appendix
\section{Open Science}

Our dataset is available under a CC-BY-NC 4.0 license and the code is Apache 2.0 licensed. Both the code and dataset are available in in \url{https://github.com/wi-pi/webspeval_code}. It includes the following:

\begin{itemize}
    \item Both the \withnavE and \wonavE variants of our dataset, including manual annotations for task categories and ground truth actions.
    \item The custom Google Chrome extension used to record login and state reset traces along with a Python script to test the replay of the recorded trace.
    \item The Python scripts to run agent instantiation taking care of session and state management, and automated evaluation.
    \item Comprehensive documentation and all necessary artifacts required to execute, record, and replay the tasks.
\end{itemize}

\section{Generative AI Usage Statement}
We hereby confirm that large language models were used to assist in improving grammar, checking spelling, and polishing the text written by the authors. No large language model text was directly pasted into this document. We also used them for analyzing results. However, the initial code was fully written by the authors and all other subsequent generations were vetted carefully before proceeding. 

\section{Task Details}
\label{sec:more_task_details}

\begin{table*}[h!]
\centering
\resizebox{0.9\linewidth}{!}{
\begin{tabular}{l | p{12cm}}
\toprule
\textbf{Task Category} & \textbf{Task IDs} \\
\midrule
Account Security \& Access Control & Airbnb\_task-111, Airbnb\_task-219, Amazon\_task-221, Docker\_task-2, Docker\_task-3, Docker\_task-4, Duolingo\_task-222, GitHub\_task-132, GitHub\_task-133, GitHub\_task-134 (2), GitHub\_task-135 (2), Goodreads\_task-100, Grammarly\_task-24, HuggingFace\_task-120, HuggingFace\_task-121, HuggingFace\_task-122, HuggingFace\_task-220, Moodle\_task-209, Pinterest\_task-218, Pinterest\_task-68 \\
\midrule
Advertising \& Personalization Control & Amazon\_task-190, Amazon\_task-88, Duolingo\_task-104 (2), Goodreads\_task-75, GoogleAdCenter\_task-138 (2), GoogleAdCenter\_task-139, GoogleAdCenter\_task-140, GoogleAdCenter\_task-82 (2), Grammarly\_task-16 (2), Pinterest\_task-61 (2), Pinterest\_task-64 (2), Reddit\_task-50 (2) \\
\midrule
Cookie \& Tracking Consent Management & AllRecipes\_task-154, BBC\_task-167, BBC\_task-168, BBC\_task-169, Coursera\_task-155, Coursera\_task-156, Coursera\_task-157, Docker\_task-142, Docker\_task-143, Docker\_task-144, IKEA\_task-158, IKEA\_task-159, NVIDIA\_task-145, NVIDIA\_task-146, NVIDIA\_task-147, Shein\_task-161, Shein\_task-162, Shein\_task-163, Steam\_task-191 (2), Steam\_task-192 (2), Steam\_task-193 (2) \\
\midrule
Data \& Asset Management & GitHub\_task-126, HuggingFace\_task-113, HuggingFace\_task-114, HuggingFace\_task-116, HuggingFace\_task-117, HuggingFace\_task-118 \\
\midrule
Notification \& Communication Preferences & Amazon\_task-86 (2), Amazon\_task-87 (2), Coursera\_task-182, Coursera\_task-203, Duolingo\_task-105 (2), GitHub\_task-129 (2), GitHub\_task-130 (2), GitHub\_task-131, Goal\_task-93 (2), Goal\_task-94 (2), Goal\_task-95 (2), Moodle\_task-206 (2), OldReddit\_task-56 (2), Quora\_task-78 (2), Quora\_task-79, Reddit\_task-54, Reddit\_task-55, Steam\_task-198 (2), Steam\_task-199, Steam\_task-200 (2), USAToday\_task-36 (2), USAToday\_task-37 (2), Wattpad\_task-223 (2), Wattpad\_task-224 (2), Wattpad\_task-225 (2), Wolfram\_task-10 (2), Wolfram\_task-7 (2), Wolfram\_task-8 (2), Wolfram\_task-9 (2) \\
\midrule
Profile Visibility \& Customization & Airbnb\_task-107 (2), Airbnb\_task-108 (2), Duolingo\_task-103 (2), GitHub\_task-127 (2), Goodreads\_task-101, Goodreads\_task-99, OldReddit\_task-57 (2), OldReddit\_task-58 (2), OpenStreetMap\_task-91, OpenStreetMap\_task-92, Pinterest\_task-65 (2), Quora\_task-76 (2), Reddit\_task-52 (2) \\
\midrule
Social Safety \& Content Moderation & Airbnb\_task-106 (2), Goodreads\_task-102 (2), Moodle\_task-205 (2), Pinterest\_task-67, Pinterest\_task-69, Quora\_task-77, Quora\_task-80, Reddit\_task-51 (2), Reddit\_task-53 (2), Steam\_task-194, Steam\_task-195, Steam\_task-196 (2), Steam\_task-197 (2), Twitch\_task-226, Twitch\_task-227 (2), Twitch\_task-228 (2), Twitch\_task-230, Twitch\_task-231, Twitch\_task-232 (2), Twitch\_task-233 (2) \\
\midrule
UI/UX Preferences & Amazon\_task-89, OldReddit\_task-59, OldReddit\_task-60, Pinterest\_task-70 (2) \\
\midrule
User Privacy \& Data Rights & Airbnb\_task-180, AlJazeera\_task-179, Coursera\_task-183, Docker\_task-1 (2), GoogleAdCenter\_task-141, Grammarly\_task-17 (2), Grammarly\_task-18 (2), Grammarly\_task-19 (2), Grammarly\_task-20, Pinterest\_task-62 (2), Pinterest\_task-63, Pinterest\_task-66 (2), Quora\_task-81 (2) \\
\bottomrule
\end{tabular}}
\caption{Task Categories and Associated Task IDs. Tasks with (2) have both ON and OFF state variants.}
\label{tab:task_categories}
\end{table*}

Our benchmark comprises 138 unique \taskname distributed across nine types. The nine types and their counts are: Notification \& Communication Preferences (29 tasks), Cookie \& Tracking Consent Management (21 tasks), Account Security \& Access Control (21 tasks) Social Safety \& Content Moderation (20 tasks), Profile Visibility (13), User Privacy \& Data Rights (13 tasks), Advertising \& Personalization Control (12 tasks), Data \& Asset Management (6 tasks), and UI/UX Preferences (4 tasks). The full list of categories and their respective task identifiers are present in \Cref{tab:task_categories}.  

We obtain the broad categories based on guidelines from NCSC~\cite{ncscadvice}, NIST~\cite{nistcsf2} and FTC~\cite{ftccollectuseinfo}. We also add tasks that we feel are relevant website privacy and security tasks such as Data \& Asset Management tasks that deal with setting the visibility of repositories on HuggingFace and GitHub.

\Cref{tab:task_categories} also consists of taskIDs that have dual initial state~(both `ON' and `OFF') in our dataset. Out of 138 unique tasks, we have 62 tasks with dual initial state, 52 tasks with single initial state, and lastly 24 that are not state dependent~(such as logging out of a website or adding an access token with specific conditions). 

The websites in our dataset and the categories are in \Cref{tab:websites}.

\begin{table}[t]
\centering
\resizebox{0.8\linewidth}{!}{
\begin{tabular}{@{}lp{0.65\columnwidth}@{}}
\toprule
\textbf{Category} & \textbf{Websites} \\ \midrule
Education/Reference & Coursera, Duolingo, Goodreads, Moodle, Wolfram \\ \addlinespace
Entertainment \& Games & Steam, Twitch, Wattpad \\ \addlinespace
Travel & Airbnb, AllRecipes, OpenStreetMap \\ \addlinespace
Sports & Goal \\ \addlinespace
General News & AlJazeera, BBC, USAToday \\ \addlinespace
Online Shopping & Amazon, IKEA, Shein \\ \addlinespace
Social Networking & Pinterest, Quora, Reddit, OldReddit \\ \addlinespace
Interactive Web Applications & Grammarly \\ \addlinespace
Technology \& Business & Docker, GitHub, GoogleAdCenter, Grammarly, HuggingFace, NVIDIA \\ \bottomrule
\end{tabular}%
}
\caption{The list of websites in the dataset of \benchmark categorized according to Trellix TrustedSource~\cite{trellixtrustedsource}}
\label{tab:websites}
\end{table}

\section{Evaluation Setup Details}
\label{sec:eval_setup_details}

\paragraph{API Endpoints:} We use the AI studio API endpoint for Gemini-3.1-Pro, Gemini-3-Pro, Gemini-2.5-Pro, and Gemini-2.5-Flash. For Claude-Opus-4.6, Claude-Sonnet-4.5, and Claude-Haiku-4.5, we use the Google Cloud Platform API endpoints. For GPT-5.1, GPT-5-mini, we utilize Microsoft Azure endpoints. For Gemma-3-27B, we use the AI Studio endpoints for trial 1, and OpenRouter endpoints for trials 2 and 3. The change to OpenRouter was necessary as Google discontinued Gemma-3-27B from all its platforms~\cite{gemma3discontinued}. Similarly, Gemini-3-Pro was also discontinued during trials 2 and 3, and is no longer available on any platform~\cite{gemini3prodeprecation}. Thus, we use Gemini-3.1-Pro to report results for RQ4. We reran trial 1 for the corresponding subset of tasks using Gemini-3.1-Pro to ensure consistency in RQ4 results for robustness across different trials. 

\paragraph{Experiment Details:} We set temperature for both the web agent backbone models and the judge to $1.0$. The temperature choice for agent is same as established web agent research~\cite{he2024webvoyager, tur2025safearena}. A higher temperature makes the agent more explorational due to the less deterministic nature. The choice for the judge is based on Google's temperature guidelines for reasoning models, that allows the model to explore more options during the thinking phase.

\paragraph{Compute and Pipeline Overhead:} We host our web agents and execute tasks on a GMKtec Intel N150 mini-PC. A full pipeline run takes between 15 and 18 hours, depending on the model. 

\paragraph{Maximum Runtime to 10 minutes:} During our initial tests with a 3-minute timeout, we made two observations: 1) tasks finished prematurely before the agent was just about to solve them, and 2) agents often performed repetitive scroll, wait, and click~(e.g., on the same icon) actions without progressing. Because we do not penalize scroll and wait actions due to varying webpage lengths, we decided to increase the timeout to 10 minutes to account for 10–30 second API request times. We also set a maximum limit of 20 non-scroll and non-wait iterations, which is approximately 1.5$\times$ the maximum number of non-scroll iterations in our ground truth (13). This aligns with existing benchmarks like WebVoyager and WebArena, which enforce limits of 15 and 30 iterations including scroll action, respectively. During our initial test run with 3 minute limit, the Haiku-4.5 and Sonnet-4.5 74 and 92 on the \withnav variant. On increasing the time limit to 10 minutes~(our final setting), their performance improved to 118 and 121, respectively. We perform a detailed analysis of their failures in \Cref{sec:human_failure_analysis}.

\paragraph{Metrics for Measuring Agent Robustness:} As backbone models are probabilistic, evaluating robustness across trials~(runs) is a common practice and we perform that in \Cref{sec:rq4}. We select a subset of the tasks that failed and succeeded in trial 1 and run them for two more trials~(3 in total). To measure robustness across trials, we use the metrics $\passatk{k}$~\cite{chen2021evaluating} and $\passexpk{k}$~\cite{yao2024tau}. $\passatk{k}$ measures if the agent can solve a task in at least once in `k' trials, whereas $\passexpk{k}$ measures if the agent can solve a task reliably in all `k' trials. They are defined as follows:
\begin{equation}
    \passatk{k} := \mathbb{E}_{\text{tasks}}\left[\,1 - \frac{\binom{n-c}{k}}{\binom{n}{k}}\,\right], \qquad \passexpk{k} := \mathbb{E}_{\text{tasks}}\left[\,\frac{\binom{c}{k}}{\binom{n}{k}}\,\right],
\end{equation}
where $n$ is the total number of trials per task, $c$ is the number of successful trials, and $k \leq n$. In our setup, we set $n=3$ and report results for $\passatk{k}$ and $\passexpk{k}$, for $k=1, 2, 3$.

\section{System Prompts}
In this section, we include the system prompts for our agent instantiation, models in the ensemble judge, and failure due to UI element detection.
\subsection{System Prompt of our \moduletwo}
\label{sec:agent_system_prompt}
We adopt the system prompt of WebVoyager~\cite{he2024webvoyager} and modify it to suit the tasks in our dataset while also adding the new actions we introduce in our agent instanation. Our system prompt is 1079 tokens according to GPT-5 tokenizer\footnote{\url{https://platform.openai.com/tokenizer}}, while WebVoyager is 273 tokens. Our system prompt is as follows: 

\begin{tcolorbox}[
    enhanced,
    breakable,
    colback=sagegreen, %
    colframe=darksage, 
    title=\textbf{Agent System Prompt},
    fonttitle=\bfseries\sffamily,
    fontupper=\small
]

Imagine you are a robot browsing the web, just like humans. Now you need to complete a task. In each iteration, you will receive an Observation that includes a screenshot of a webpage and some texts. This screenshot will feature Numerical Labels placed in the TOP LEFT corner of each Web Element.

You will also see information about ALL OPEN TABS in each observation. When you click on links, new tabs may open automatically, and you will switch to them. You can also manually switch between tabs when needed.

Carefully analyze the visual information to identify the Numerical Label corresponding to the Web Element that requires interaction, then follow the guidelines and choose one of the following actions:

\begin{enumerate}
    \item \textbf{Click} a Web Element. (Note: If clicking opens a new tab, you will automatically switch to it)
    \item \textbf{Hover} over a Web Element. Use this to trigger hover effects, reveal hidden elements, or display tooltips without clicking.
    \item \textbf{Type}: Delete existing content in a textbox and then type content. 
    \item \textbf{Scroll} up, down, left, or right. Multiple scrolls are allowed to browse the webpage. Pay attention!! The default scroll is 2/3rd of the whole window. If the scroll widget is located in a certain area of the webpage, then you have to specify a Web Element in that area.
    \item \textbf{Scroll to end}. Use this when you need to reach the bottom of the page quickly without multiple scroll actions. Be smart about this action, you will use it only when it is absolutely useful. For example, you can use this to find the cookie notice.
    \item \textbf{Scroll within popup}. Use this when you need to scroll inside a modal, popup, dialog, or overlay element like a cookie notice, terms of service popup, or consent dialog. This action automatically detects the topmost popup and scrolls within it.
    \item \textbf{Switch tab}. Use this to switch between different browser tabs. You can see all open tabs with their titles and URLs in each observation.
    \item \textbf{Wait}. Typically used to wait for unfinished webpage processes, with a duration of 5 seconds.
    \item \textbf{Go back}, returning to the previous webpage.
    \item \textbf{Google}, directly jump to the Google search page. When you can't find information in some websites, try starting over with Google.
    \item \textbf{Answer}. This action should only be chosen when all questions in the task have been solved.
\end{enumerate}

\tcbline

\textbf{Correspondingly, Action should STRICTLY follow the format:}
\begin{itemize}
    \item \texttt{Click [Numerical\_Label]}
    \item \texttt{Hover [Numerical\_Label]}
    \item \texttt{Type [Numerical\_Label]; [Content]}
    \item \texttt{Scroll [Numerical\_Label or WINDOW]; [up or down or left or right]}
    \item \texttt{Scroll\_to\_end}
    \item \texttt{Scroll\_within\_popup; [up or down or left or right]}
    \item \texttt{Switch\_tab [URL]}
    \item \texttt{Wait}
    \item \texttt{GoBack}
    \item \texttt{Google}
    \item \texttt{ANSWER; [content]}
\end{itemize}

\tcbline

\textbf{Key Guidelines You MUST follow:}

\medskip
\textbf{Action guidelines}
\begin{enumerate}
    \item To input text, NO need to click textbox first, directly type content. After typing, the system does NOT automatically press Enter - you must explicitly click the search/submit button if needed. Try to use simple language when searching.  
    \item You must Distinguish between textbox and search button, don't type content into the button! If no textbox is found, you may need to click the search button first before the textbox is displayed. 
    \item Execute only one action per iteration. 
    \item STRICTLY Avoid repeating the same action if the webpage remains unchanged. You may have selected the wrong web element or numerical label. Continuous use of the Wait is also NOT allowed.
    \item When a complex Task involves multiple questions or steps, select \texttt{ANSWER} only at the very end, after addressing all of these questions (steps). Flexibly combine your own abilities with the information in the web page. Double check the formatting requirements in the task when \texttt{ANSWER}. 
    \item The \texttt{TYPE} action can also be used to just delete the content of a textbox without typing anything by giving an empty content.
\end{enumerate}

\medskip
\textbf{Web Browsing Guidelines}
\begin{enumerate}
    \item For tasks that require login and you do not find yourself already authenticated, you should try to login using default google account \texttt{\{DEFAULT\_GOOGLE\_ACCOUNT\}}.
    \item Don't interact with useless web elements like donation that appear in Webpages. Pay attention to Key Web Elements like search textbox and menu.
    \item Visit video websites like YouTube is allowed BUT you can't play videos. Clicking to download PDF is allowed and will be analyzed by the Assistant API.
    \item Focus on the numerical labels in the TOP LEFT corner of each rectangle (element). Ensure you don't mix them up with other numbers (e.g. Calendar) on the page.
    \item Focus on the date in task, you must look for results that match the date. It may be necessary to find the correct year, month and day at calendar.
    \item Pay attention to the filter and sort functions on the page, which, combined with scroll, can help you solve conditions like `highest', `cheapest', `lowest', `earliest', etc. Try your best to find the answer that best fits the task.
\end{enumerate}

\tcbline

\textbf{Your reply should strictly follow the format:}\\
\texttt{Thought: \{\{Your brief thoughts (briefly summarize the info that will help ANSWER)\}\}}\\
\texttt{Action: \{\{One Action format you choose\}\}}

\medskip
\textbf{Then the User will provide:}\\
\texttt{Observation: \{\{A labeled screenshot Given by User\}\}}

\end{tcolorbox}

\subsection{System Prompt of our \modulethree}
\label{sec:judge_system_prompt} 

Similar to the agent system prompt, we adopt the evaluation system prompt of WebVoyager, and customzie it to our evaluation needs, while also ensuring it is friendly to the Judge MLLM~(Gemini-2.5-Pro). Our prompt is 1107 tokens according to GPT-5 tokenizer.

\begin{tcolorbox}[
    enhanced,
    breakable,
    colback=modblue,      
    colframe=borderblue,
    title=\textbf{Judge System Prompt},
    fonttitle=\bfseries\sffamily,
    fontupper=\small %
]

\textbf{ROLE:} WebPrivSecEvaluator Bot

\medskip
\textbf{PRIMARY GOAL}\\
You are an expert evaluator tasked with analyzing the performance of a browser-automation agent (web agent). Your objective is to determine if the agent successfully completed a specific web privacy or security task based on the provided textual and visual evidence, compared against the Ground Truth steps.

\tcbline

\textbf{INPUT DATA}\\
For each evaluation instance, you will receive:
\begin{enumerate}
    \item \textbf{Task Query}: The natural language instruction given to the agent.
    \item \textbf{Ground Truth Steps}: The expected ideal sequence of actions.
    \item \textbf{Result Response}: The agent's textual log describing its thought process and actions.
    \item \textbf{Result Screenshots}: A sequence of images corresponding to the agent's actions.
\end{enumerate}

\textbf{Crucial Evidence Guidelines}
\begin{itemize}
    \item \textbf{Screenshot Trust}: \textbf{TRUST THE SCREENSHOTS OVER THE TEXT LOG.} The visual evidence is the ground truth for what actually happened on the page.
    \item \textbf{Screenshot-Thought Pairing}: Each thought in \texttt{thoughts.json} corresponds to a screenshot showing the page state AFTER that action was executed. Match iteration numbers to align screenshots with thoughts.
    \item \textbf{Element Identification}: Valid element IDs are typically two-digit numbers. Be vigilant for hallucinations where the model cites implausible element IDs (e.g., [1078], [678]).
    \item \textbf{Ground Truth Usage}: Use the Ground Truth steps as a guide for the \textit{correct solution}. It is possible that the agent does not follow the exact steps but still achieves the correct final outcome. So use the ground truth to understand what are a set of possible actions that are needed to achieve the goal and use that understanding to evaluate the agent's actions and whether they are correct or not.
\end{itemize}

\textbf{Agent Action Space}\\
The agent can perform the following actions. The final action is always an \texttt{ANSWER} acknowledging completion.
\texttt{Click}, \texttt{Type}, \texttt{Scroll}, \texttt{Scroll\_to\_end}, \texttt{Scroll\_within\_popup}, \texttt{Switch\_tab}, \texttt{Wait}, \texttt{GoBack}, \texttt{Google}, \texttt{ANSWER}, \texttt{Hover}

\tcbline

\textbf{CORE EVALUATION LOGIC}
\begin{enumerate}
    \item \textbf{Strict Evidence Adherence}: Do not assume actions were taken unless visible in screenshots or explicitly stated in the log (and supported by context).
    \item \textbf{Step Comparison}: Compare the agent's actions against Ground Truth steps to identify missing, redundant, or incorrect actions. Use the ground truth as a reference for the set of possible actions for achieving the goal. Agent can deviate but still achieve the goal. So check the final outcome based on the agent actions while using the ground truth solely as a reference.
    \item \textbf{Final State Verification}: Ensure all changes are finalized (e.g., `Save Preferences' button was actually clicked).
\end{enumerate}

\tcbline

\textbf{DETERMINING THE RESULT}

\medskip
\textbf{General Observations \& Edge Cases}\\
Before categorizing the result, consider these specific nuances:
\begin{itemize}
    \item \textbf{Data Request Tasks}: If the agent is unable to successfully complete a data request (e.g., recently requested data so optional not available currently), check the final \texttt{ANSWER} and the thoughts of the agent. If the agent correctly identifies the limitation and understands the task intent, this may still be considered valid depending on the context.
    \item \textbf{Element IDs (Click Actions)}: Valid element IDs are typically double-digit numbers. Be vigilant for hallucinations where the model cites implausible element IDs (e.g., \texttt{[1078]}, \texttt{[678]}). If the agent clicks a non-existent high-number ID, that specific action is a mistake from the agent and can impact the overall outcome of the task.
    \item \textbf{``Already Done'' States}: In some cases, the requested privacy or security setting may already be active. If the agent correctly identifies that the task is already completed and terminates without changing anything, this is a \textbf{CORRECT} result.
    \item \textbf{Toggle States}: Ensure to compare the state of elements like toggles with the desired state from the task instruction very carefully, as these elements form a major part of the tasks you will get.
\end{itemize}

\textbf{1. CORRECT}\\
The agent executed necessary steps and the final state reflects the desired outcome (matches the expected outcome from Ground Truth).

\medskip
\textbf{2. INCORRECT}\\
The agent failed to achieve the goal due to any reason (navigation errors, incomplete steps, hallucinated actions, semantic reversals, or post-completion destructive actions).

\tcbline

\textbf{REASONING GUIDELINES}\\
When writing the \texttt{reason} field, you must adhere to the following structure:
\begin{enumerate}
    \item \textbf{Summary vs. Expected}: Summarize the actual actions taken by the agent and directly compare them with the expected outcome (Ground Truth).
    \item \textbf{Why It Failed}: Clearly explain the specific reason(s) why the task was not completed successfully.
    \item \textbf{Destructive Action Check}: You must explicitly mention if the agent attempted any destructive or irrelevant high-risk actions.
    \begin{itemize}
        \item \textit{Examples of destructive actions include:} Deleting accounts, creating new affiliations/subscriptions, or interacting with unrelated external websites.
    \end{itemize}
\end{enumerate}

\tcbline

\textbf{REQUIRED OUTPUT FORMAT}\\
Provide your evaluation as a JSON object with exactly the following fields:

\medskip
\noindent\texttt{\{\\}
\texttt{\hspace*{1.5em}"result": "CORRECT" or "INCORRECT",\\}
\texttt{\hspace*{1.5em}"reason": "Detailed explanation of the actions taken by the agent, how they compare to the Ground Truth steps, and the reason for the final outcome."\\}
\texttt{\}}

\medskip
\noindent Output ONLY the JSON object with no additional text, markdown, or code fences.

\end{tcolorbox}

\subsection{System Prompt to detect UI element causing failure}
\label{sec:system_prompt_ui_failure}
Below, we present the system prompt we use to analyze the cases where the failure was due to the model not comprehending target elements involved in solving a task successfully. This pertains to analysis addressing RQ3~\Cref{sec:rq3}
\begin{tcolorbox}[
    enhanced,
    breakable,
    colback=lavenderbg, %
    colframe=darkviolet, %
    title=\textbf{UI Performance Analyzer System Prompt},
    fonttitle=\bfseries\sffamily,
    fontupper=\small
]

\textbf{ROLE:} Expert UI performance analyzer.

\medskip
The user will provide a JSON file containing multiple task evaluation entries. Each entry includes the following fields:
\begin{itemize}
    \item \textbf{\texttt{TaskID}}: A unique identifier for the task.
    \item \textbf{\texttt{Task\_Instruction}}: The instruction originally given to the web agent.
    \item \textbf{\texttt{Ground\_Truth\_UI\_Elements}}: The types of UI elements that must be interacted with to successfully complete the task (e.g., Link, Button, Checkbox, Toggle, Dropdown).
    \item \textbf{\texttt{Ground\_Truth\_Actions}}: The sequence of actions required to correctly complete the task. These are provided for reference only and should be used to understand the intended UI interactions.
    \item \textbf{\texttt{Gemini3.1\_Response}}: An analysis explaining how and why the agent failed the task.
\end{itemize}

\tcbline

\textbf{IMPORTANT CONTEXT}\\
All entries in the JSON represent model failures. You do not need to determine whether the model failed. Your only task is to identify which UI element type(s) caused the failure.

\medskip
\textbf{YOUR OBJECTIVE}\\
For each entry, determine which specific UI element type(s) were responsible for the failure. Use \texttt{Gemini3.1\_Response} as the primary source of reasoning. Refer to \texttt{Ground\_Truth\_UI\_Elements} and \texttt{Ground\_Truth\_Actions} only to understand what UI components were involved.

\tcbline

\textbf{INSTRUCTIONS}
\begin{enumerate}
    \item Carefully analyze the failure explanation in \texttt{Gemini3.1\_Response}.
    \item Identify which UI element type(s) from \texttt{Ground\_Truth\_UI\_Elements} caused the failure.
    \item Add a new key called \texttt{WHICH\_UI\_ELEMENT\_FAILED} to each entry.
    \item The value of \texttt{WHICH\_UI\_ELEMENT\_FAILED} must contain only the UI element type(s) that directly caused the failure.
    \item If multiple UI element types contributed to the failure, include all of them.
    \item Do not critique or evaluate the correctness of the ground truth. Every entry is already a confirmed failure.
\end{enumerate}

\tcbline

\textbf{OUTPUT REQUIREMENTS}
\begin{itemize}
    \item Return a JSON list of dictionaries.
    \item Each dictionary must contain exactly the following keys:
    \begin{itemize}
        \item \texttt{"TaskID"}
        \item \texttt{"WHICH\_UI\_ELEMENT\_FAILED"}
        \item \texttt{"Ground\_Truth\_UI\_Elements"}
    \end{itemize}
    \item Do not include explanations, commentary, or additional fields.
    \item Ensure the output is strictly valid JSON.
    \item Analyze the entire input file and provide results for all entries.
\end{itemize}

\tcbline

\textbf{EXAMPLE}\\
If the failure occurred because the agent did not click a required save button, and the relevant UI element type was a Button, the output should be:

\medskip
\noindent\texttt{\{\\}
\texttt{\hspace*{1.5em}"TaskID": "OldReddit\_task-59",\\}
\texttt{\hspace*{1.5em}"WHICH\_UI\_ELEMENT\_FAILED": "Button",\\}
\texttt{\hspace*{1.5em}"Ground\_Truth\_UI\_Elements": "Link, Checkbox, Button"\\}
\texttt{\}}

\medskip
\noindent If multiple entries are provided, return them as a JSON list of such objects.

\end{tcolorbox}

\section{\withnav variant results}

We present the \withnav variant results for RQs 2 and 3 in \Cref{tab:suc_by_web_and_model_withnav,tab:suc_by_tasktype_and_model_withnav,tab:ui_element_with_nav,tab:suc_by_web_and_model_withnav}. 

\begin{table}[ht!]

\centering
\resizebox{0.8\linewidth}{!}{
\begin{tabular}{l  cccccccc  c}
\toprule
\multirow{2}{*}{\textbf{Website}} & \multicolumn{8}{c|}{\textbf{\withnavE Variant}} & \multirow{2}{*}{\textbf{\# Instances}} \\
\cmidrule(lr){2-9}
 & \textbf{2.5F} & \textbf{2.5P} & \textbf{3P} & \textbf{H4.5} & \textbf{S4.5} & \textbf{5m} & \textbf{5.1} & \textbf{3Ge} & \\
\midrule
Airbnb & 7 & 8 & 9 & 5 & 7 & 1 & 5 & 6 & 9 \\
\rowcolor{aliceblue} AlJazeera & 0 & 0 & 0 & 1 & 0 & 0 & 0 & 1 & 1 \\
AllRecipes & 1 & 0 & 1 & 0 & 1 & 0 & 1 & 0 & 1 \\
\rowcolor{aliceblue} Amazon & 5 & 5 & 8 & 5 & 4 & 4 & 6 & 2 & 8 \\
BBC & 3 & 3 & 3 & 3 & 3 & 3 & 1 & 0 & 3 \\
\rowcolor{aliceblue} Coursera & 3 & 1 & 6 & 5 & 4 & 3 & 3 & 0 & 6 \\
Docker & 3 & 5 & 7 & 6 & 6 & 3 & 6 & 1 & 8 \\
\rowcolor{aliceblue} Duolingo & 5 & 1 & 7 & 6 & 4 & 3 & 1 & 4 & 7 \\
GitHub & 12 & 13 & 14 & 13 & 4 & 12 & 12 & 4 & 14 \\
\rowcolor{aliceblue} Goal & 2 & 2 & 4 & 4 & 5 & 1 & 3 & 1 & 6 \\
Goodreads & 3 & 4 & 4 & 2 & 3 & 3 & 2 & 0 & 6 \\
\rowcolor{aliceblue} GoogleAdCenter & 7 & 6 & 7 & 6 & 5 & 3 & 6 & 3 & 7 \\
Grammarly & 8 & 7 & 10 & 7 & 6 & 9 & 6 & 5 & 10 \\
\rowcolor{aliceblue} HuggingFace & 8 & 7 & 7 & 5 & 7 & 7 & 6 & 4 & 9 \\
IKEA & 1 & 0 & 2 & 2 & 2 & 0 & 1 & 1 & 2 \\
\rowcolor{aliceblue} Moodle & 2 & 5 & 5 & 1 & 0 & 0 & 1 & 0 & 5 \\
NVIDIA & 0 & 2 & 3 & 3 & 3 & 0 & 0 & 0 & 3 \\
\rowcolor{aliceblue} OldReddit & 5 & 5 & 6 & 2 & 6 & 2 & 1 & 0 & 8 \\
OpenStreetMap & 1 & 1 & 2 & 2 & 1 & 1 & 2 & 1 & 2 \\
\rowcolor{aliceblue} Pinterest & 10 & 12 & 15 & 6 & 12 & 6 & 7 & 4 & 17 \\
Quora & 8 & 7 & 6 & 0 & 3 & 0 & 6 & 2 & 9 \\
\rowcolor{aliceblue} Reddit & 7 & 7 & 9 & 3 & 3 & 9 & 8 & 3 & 10 \\
Shein & 2 & 2 & 1 & 3 & 1 & 0 & 0 & 0 & 3 \\
\rowcolor{aliceblue} Steam & 9 & 9 & 7 & 6 & 8 & 5 & 5 & 1 & 17 \\
Twitch & 5 & 6 & 10 & 7 & 7 & 5 & 6 & 3 & 11 \\
\rowcolor{aliceblue} USAToday & 2 & 2 & 4 & 2 & 2 & 2 & 3 & 1 & 4 \\
Wattpad & 2 & 5 & 5 & 6 & 4 & 4 & 5 & 3 & 6 \\
\rowcolor{aliceblue} Wolfram & 6 & 6 & 7 & 6 & 6 & 5 & 5 & 0 & 8 \\
\bottomrule
\end{tabular}}
\caption{Agent performance breakdown by website for the \withnavE dataset variant. Shorthands 2.5F, 2.5P, 3P, H4.5, S4.5, 5m, 5.1, and 3Ge refer to Gemini-2.5-Flash, Gemini-2.5-Pro, Gemini-3-Pro, Claude-Haiku-4.5, Claude-Sonnet-4.5, GPT-5-mini, GPT-5.1, and Gemma-3-27B, respectively.}
\label{tab:suc_by_web_and_model_withnav}
\end{table}

\begin{table}[ht!]
\centering
\resizebox{0.8\linewidth}{!}{
\begin{tabular}{l cccccccc c}
\toprule
\textbf{Task Category} & \textbf{2.5F} & \textbf{2.5P} & \textbf{3P} & \textbf{H4.5} & \textbf{S4.5} & \textbf{5m} & \textbf{5.1} & \textbf{3Ge} & \textbf{\# Tasks} \\
\midrule
Account Security \& Access Control & 17 & 18 & 20 & 13 & 12 & 14 & 17 & 6 & 22 \\
\rowcolor{aliceblue} Advertising \& Personalization Control & 16 & 14 & 17 & 12 & 11 & 11 & 10 & 8 & 19 \\
Cookie \& Tracking Consent Management & 13 & 14 & 18 & 18 & 14 & 3 & 7 & 2 & 24 \\
\rowcolor{aliceblue} Data \& Asset Management & 5 & 5 & 5 & 4 & 5 & 5 & 5 & 3 & 6 \\
Notification \& Communication Preferences & 29 & 32 & 42 & 33 & 27 & 21 & 30 & 10 & 51 \\
\rowcolor{aliceblue} Profile Visibility \& Customization & 12 & 13 & 18 & 7 & 11 & 8 & 9 & 6 & 22 \\
Social Safety \& Content Moderation & 17 & 18 & 27 & 13 & 18 & 14 & 15 & 7 & 31 \\
\rowcolor{aliceblue} UI/UX Preferences & 4 & 4 & 4 & 4 & 5 & 3 & 2 & 0 & 5 \\
User Privacy \& Data Rights & 14 & 13 & 18 & 13 & 14 & 12 & 13 & 8 & 20 \\
\bottomrule
\end{tabular}}
\caption{Agent success rates broken down by task category for the \withnavE dataset variant. Shorthands 2.5F, 2.5P, 3P, H4.5, S4.5, 5m, 5.1, and 3Ge refer to Gemini-2.5-Flash, Gemini-2.5-Pro, Gemini-3-Pro, Claude-Haiku-4.5, Claude-Sonnet-4.5, GPT-5-mini, GPT-5.1, and Gemma-3-27B, respectively.}
\label{tab:suc_by_tasktype_and_model_withnav}
\end{table}

\begin{table}[ht!]
\centering
\resizebox{0.8\linewidth}{!}{
\begin{tabular}{l cccccccc c}
\toprule
\textbf{UI Element} & \textbf{2.5F} & \textbf{2.5P} & \textbf{3P} & \textbf{H4.5} & \textbf{S4.5} & \textbf{5m} & \textbf{5.1} & \textbf{3Ge} & \textbf{\# Tasks} \\
\midrule
Button & 73 & 73 & 93 & 69 & 71 & 52 & 64 & 25 & 111 \\
\rowcolor{aliceblue} Checkbox & 20 & 22 & 30 & 16 & 24 & 13 & 14 & 6 & 40 \\
Dropdown & 53 & 58 & 74 & 44 & 45 & 40 & 51 & 25 & 93 \\
\rowcolor{aliceblue} Icon & 37 & 36 & 45 & 32 & 31 & 29 & 31 & 13 & 52 \\
Link & 107 & 111 & 142 & 100 & 97 & 73 & 93 & 40 & 172 \\
\rowcolor{aliceblue} Menu & 5 & 6 & 7 & 3 & 5 & 0 & 3 & 4 & 7 \\
Option & 43 & 46 & 65 & 34 & 42 & 35 & 40 & 21 & 77 \\
\rowcolor{aliceblue} Radio Button & 16 & 16 & 15 & 12 & 15 & 11 & 10 & 4 & 20 \\
Text Input & 8 & 7 & 11 & 5 & 9 & 7 & 9 & 3 & 14 \\
\rowcolor{aliceblue} Toggle & 54 & 55 & 82 & 57 & 49 & 36 & 46 & 24 & 98 \\
\bottomrule
\end{tabular}}
\caption{Performance breakdown by target UI element for the \withnavE dataset variant\protect\footref{note:shorthand}. The best or join-best performing model are in bold. The \#Inst. column indicates the number of unique instances in which the respective UI element is part of the solution to the task.}
\label{tab:ui_element_with_nav}
\end{table}

\begin{table}[h!]
\centering
\resizebox{0.8\linewidth}{!}{
\begin{tabular}{l  cccc}
\toprule
\textbf{Model} & \textbf{Both Correct} & \textbf{Only ON} & \textbf{Only OFF} & \textbf{Both Failed} \\
\midrule
Gemini-2.5-Flash & 24 & 21 & 4 & 13 \\
\rowcolor{aliceblue} Gemini-2.5-Pro & 26 & 15 & 11 & 10 \\
Gemini-3-Pro-Preview & 45 & 10 & 3 & 4 \\
\rowcolor{aliceblue} Claude-Haiku-4.5 & 21 & 12 & 9 & 20 \\
Claude-Sonnet-4.5 & 22 & 14 & 4 & 22 \\
\rowcolor{aliceblue} GPT-5-Mini & 19 & 7 & 8 & 28 \\
GPT-5.1 & 16 & 15 & 11 & 20 \\
\rowcolor{aliceblue} Gemma-3-27b & 3 & 14 & 8 & 37 \\
\bottomrule
\end{tabular}}
\caption{ON/OFF State Comparison: WithNav Variant}
\label{tab:on_off_with_nav}
\end{table}

\section{Qualitative Sample of Task Trajectories}
\label{sec:failure_figures}

\begin{figure*}[t]
    \centering
    \includegraphics[width=0.8\linewidth]{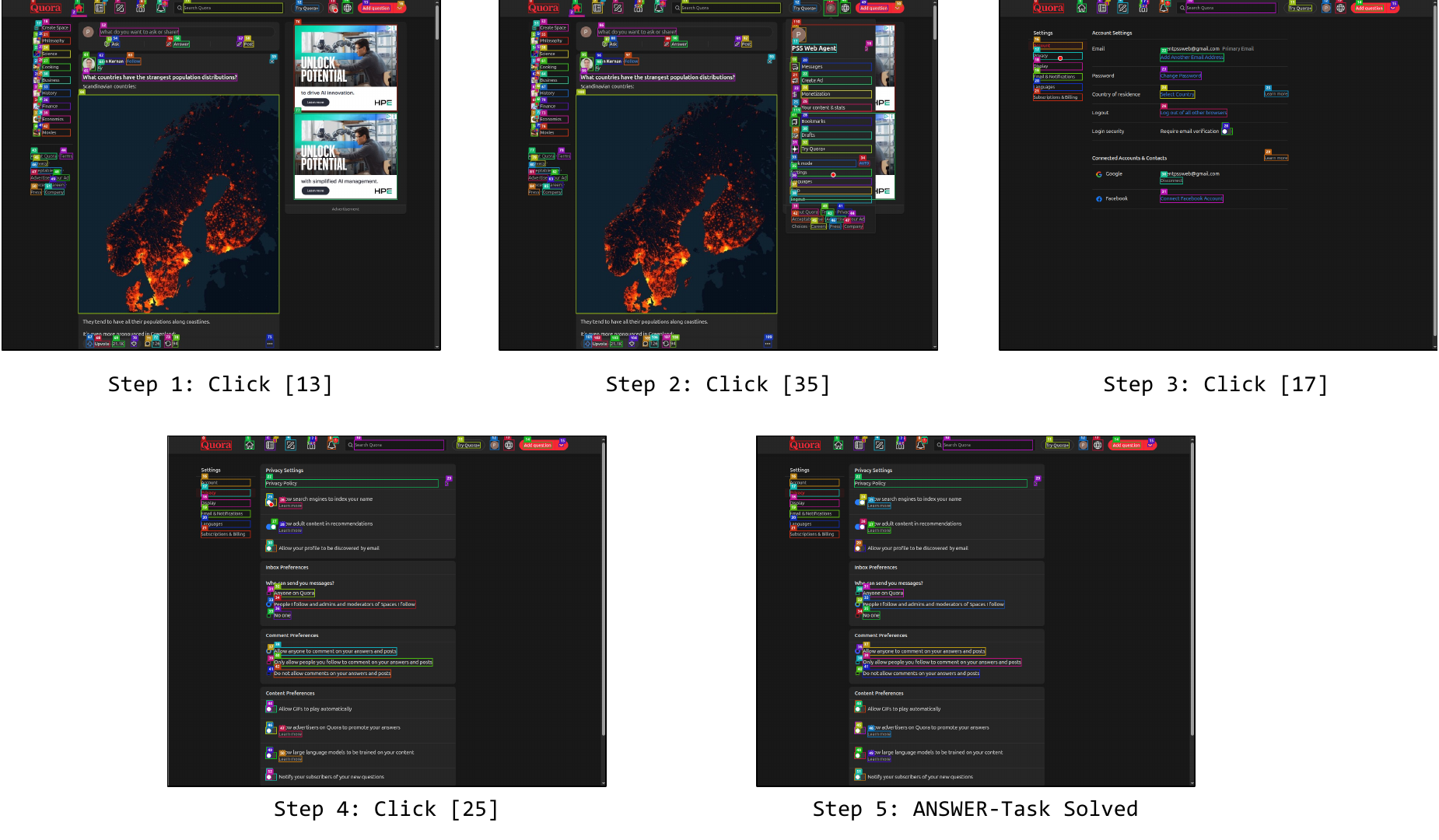}
    \caption{Failure highlighting misunderstanding and enabling a Quora option~(Gemini-2.5-Flash, \wonavE variant). Given the instruction ``Disable the option that allows search engines to index my name.'', the agent enables the option to allow search engines to index the name, instead of disabling it.}
    \label{fig:qual_analysis_fig_19}
\end{figure*}

Through \Cref{fig:phase_1} to \cref{fig:qual_analysis_fig_19}, we present a qualitative sample of task trajectories that illustrate agent successes and failures encountered across websites and tasks in our dataset. The examples span multiple backbone models
and mostly contain the \wonavE variant, the main dataset we use in this paper. Our examples showcase different failure modes discussed in \Cref{sec:human_failure_analysis}, including state misunderstanding on toggles, partial completion of a task, hallucinated success, and navigational difficulties. Each caption details the specific instruction provided to the agent and the underlying reason for the observed success or failure.

\begin{figure*}[t]
    \centering
    \includegraphics[width=\linewidth]{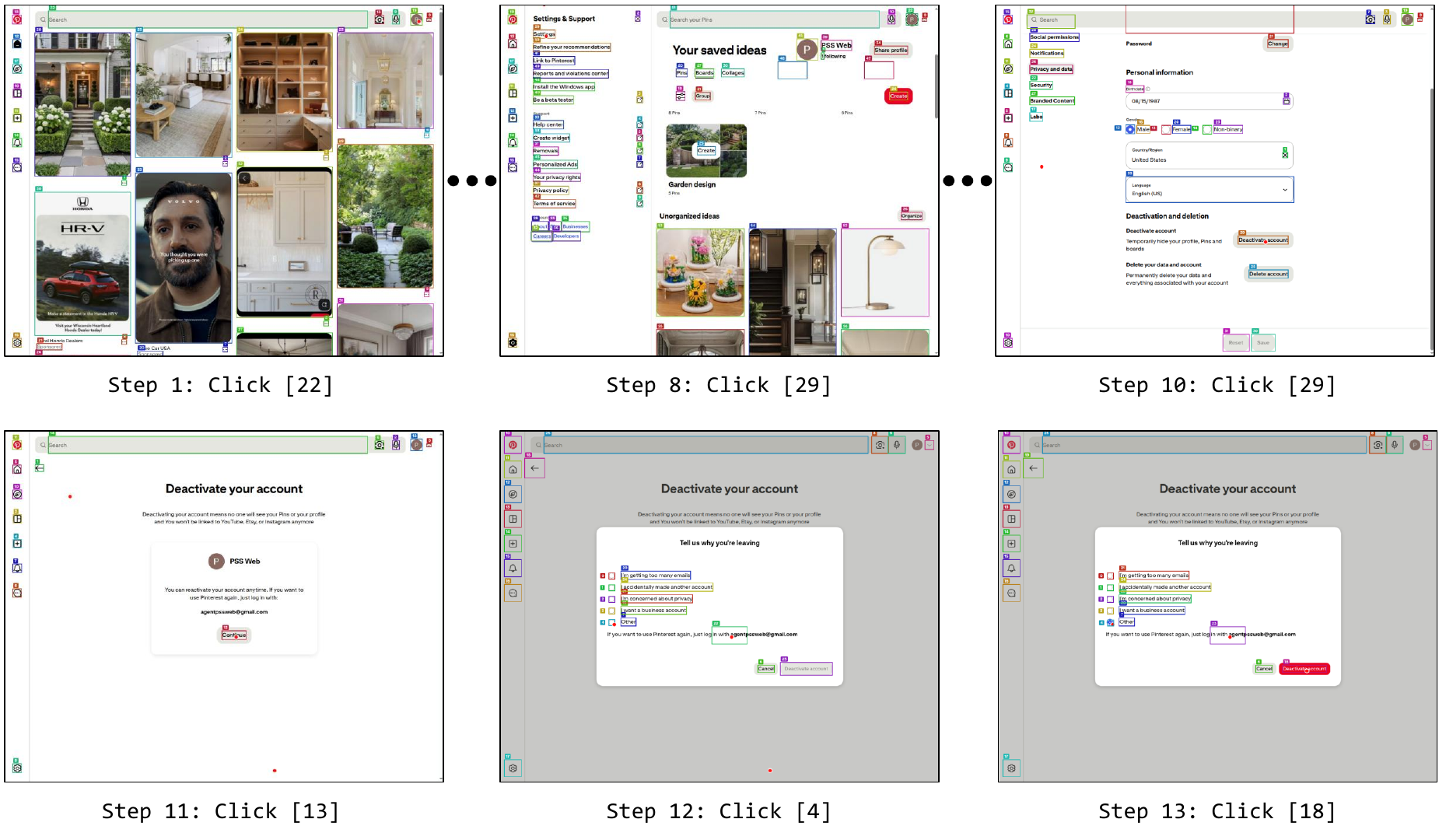}
    \caption{Failure example highlighting agent deactivating Pinterest account on being asked to sign out~(Gemini-2.5-Flash, \wonavE variant). Given the instruction ``Sign out of my Pinterest account.'' the model struggles to locate the sign out option and eventually deactivates the account altogether to sign out. The model described in its thoughts: ``I'm on the `Account management' page. I still haven't found a `Sign out' option. I see `Deactivation and deletion' at the bottom of the page. This sounds like it could lead to an option to sign out, or at least deactivate the account which would sign me out ...''}
    \label{fig:pinterest_signout}
\end{figure*}

\begin{figure*}[t]
    \centering
    \includegraphics[width=\linewidth]{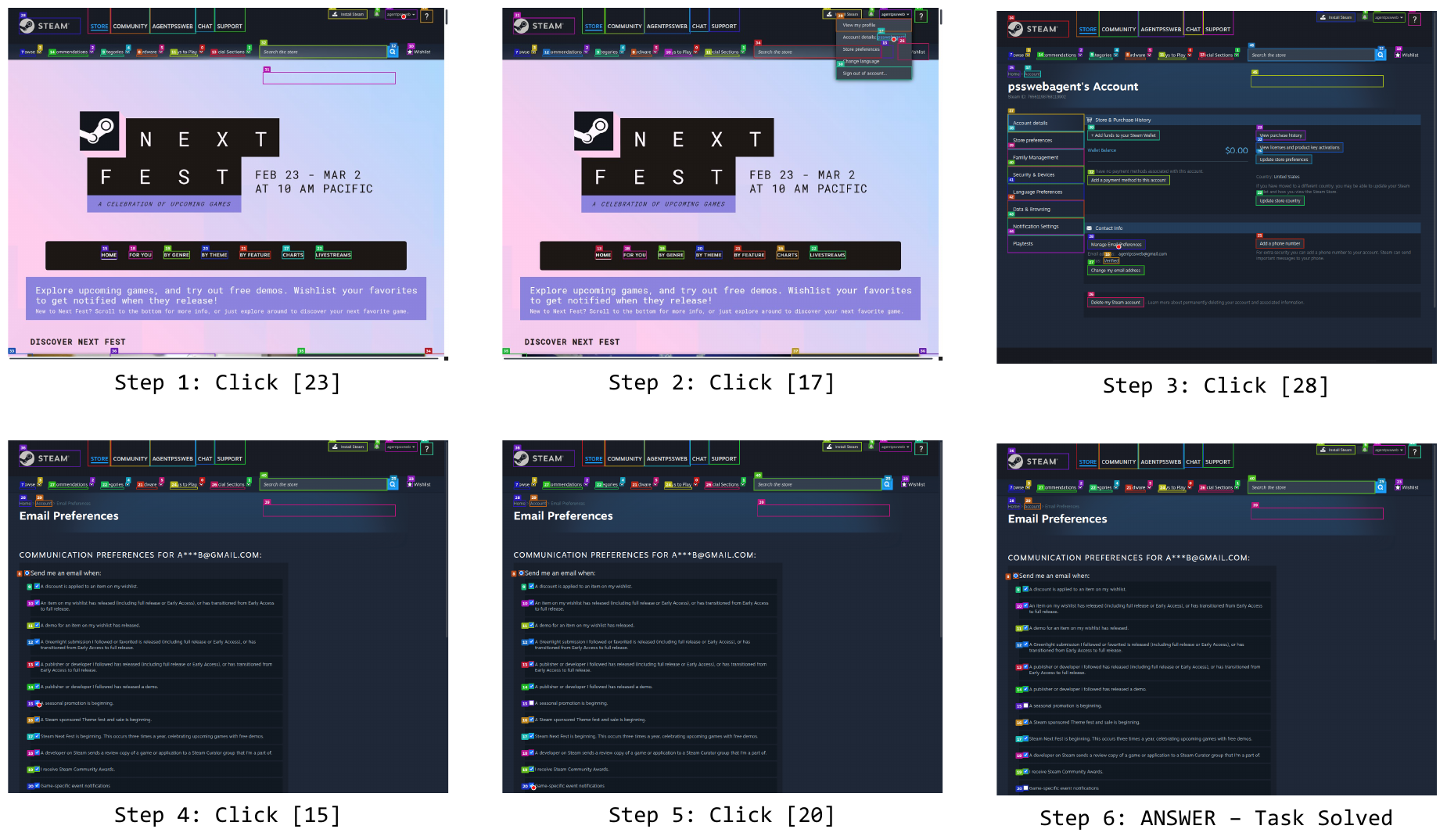}
    \caption{Failure example highlighting website specific design on Steam~(Gemini-3-Pro, \wonavE variant). Given the instruction ``Disable email notifications for `A seasonal promotion is beginning' and `Game-specific event notifications','' the model correctly disables the options but fails to scroll down to click on `Save Changes' button, meaning the changes are never stored.}
    \label{fig:phase_1}
\end{figure*}

\begin{figure*}[t]
    \centering
    \includegraphics[width=0.8\linewidth]{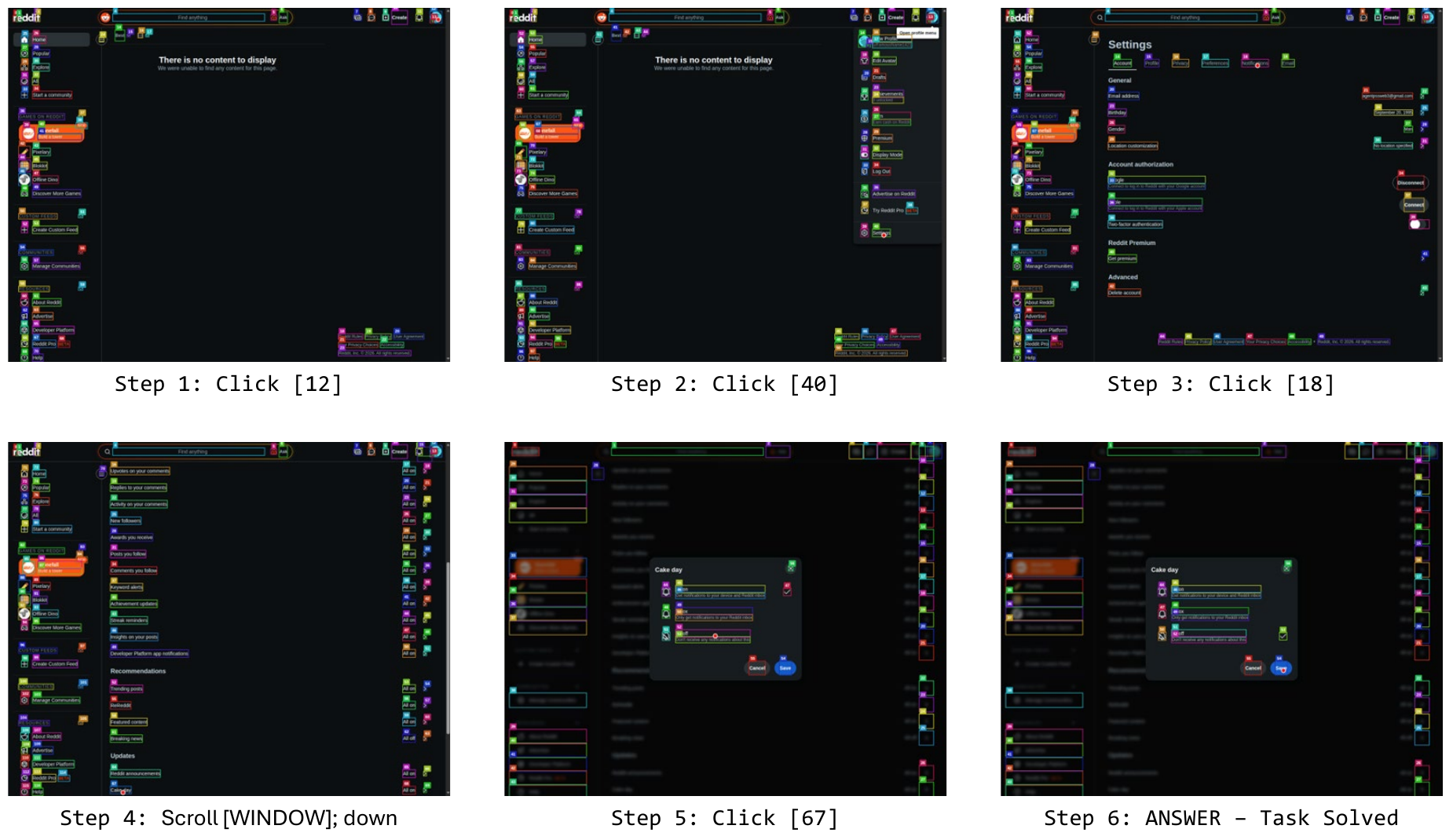}
    \caption{Success example for Gemma-3-27b on the \wonavE variant. Given the instruction ``Disable the notifications for cake day updates.'', the model successfully navigates to the page and clicks on the Cake Day updates option and disables notifications.}
    \label{fig:reddit_success}
\end{figure*}

\begin{figure*}[t]
    \centering
    \includegraphics[width=0.75\linewidth]{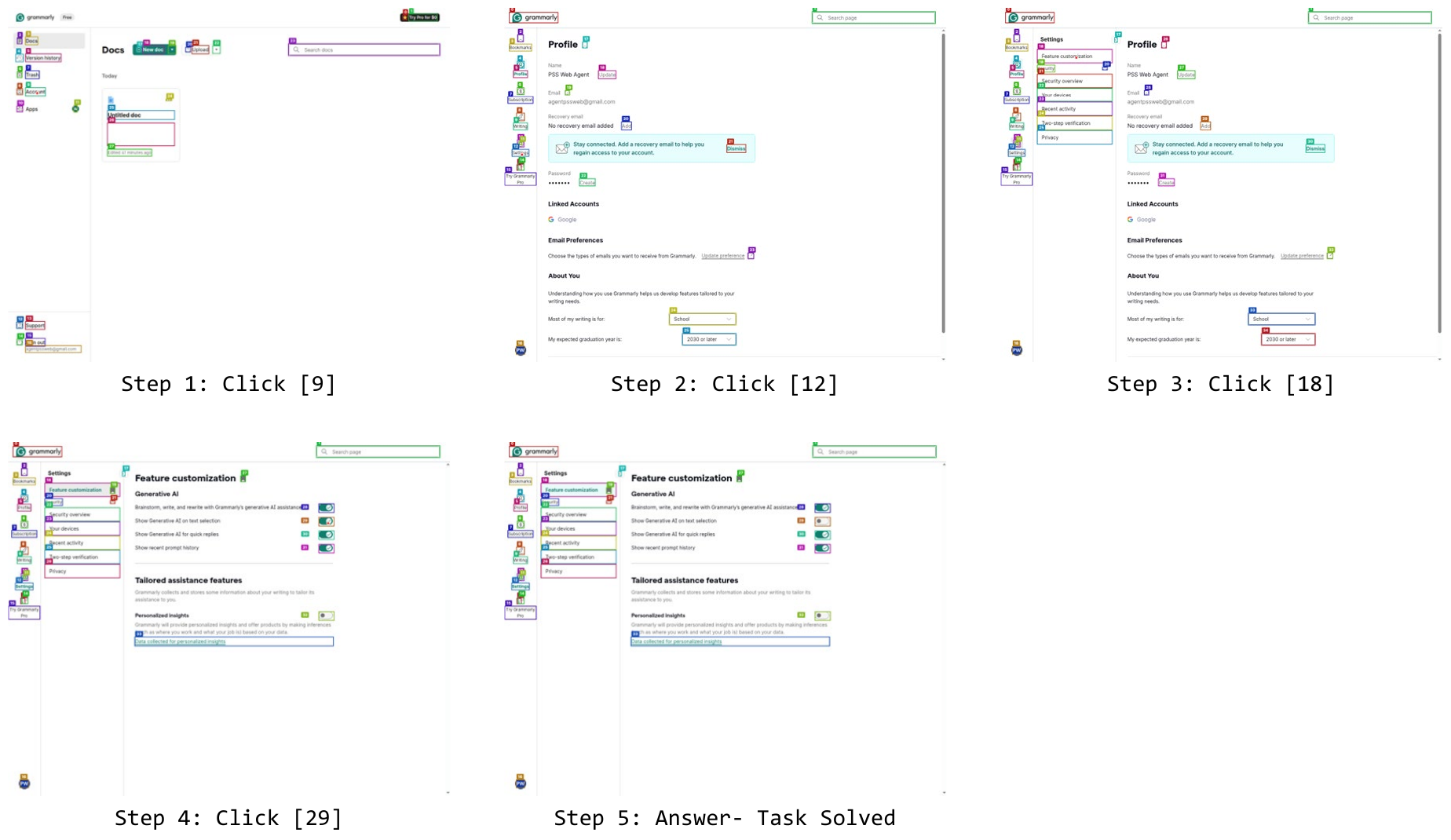}
    \caption{Success example for Claude-Haiku-4.5 on the \wonavE variant. Given the instruction ``Disable 'show generative AI on text selection' in feature customization section.'', the model successfully navigates to the page and toggle for generative AI.}
    \label{fig:grammarly_success}
\end{figure*}

\begin{figure*}[t]
    \centering
    \includegraphics[width=\linewidth]{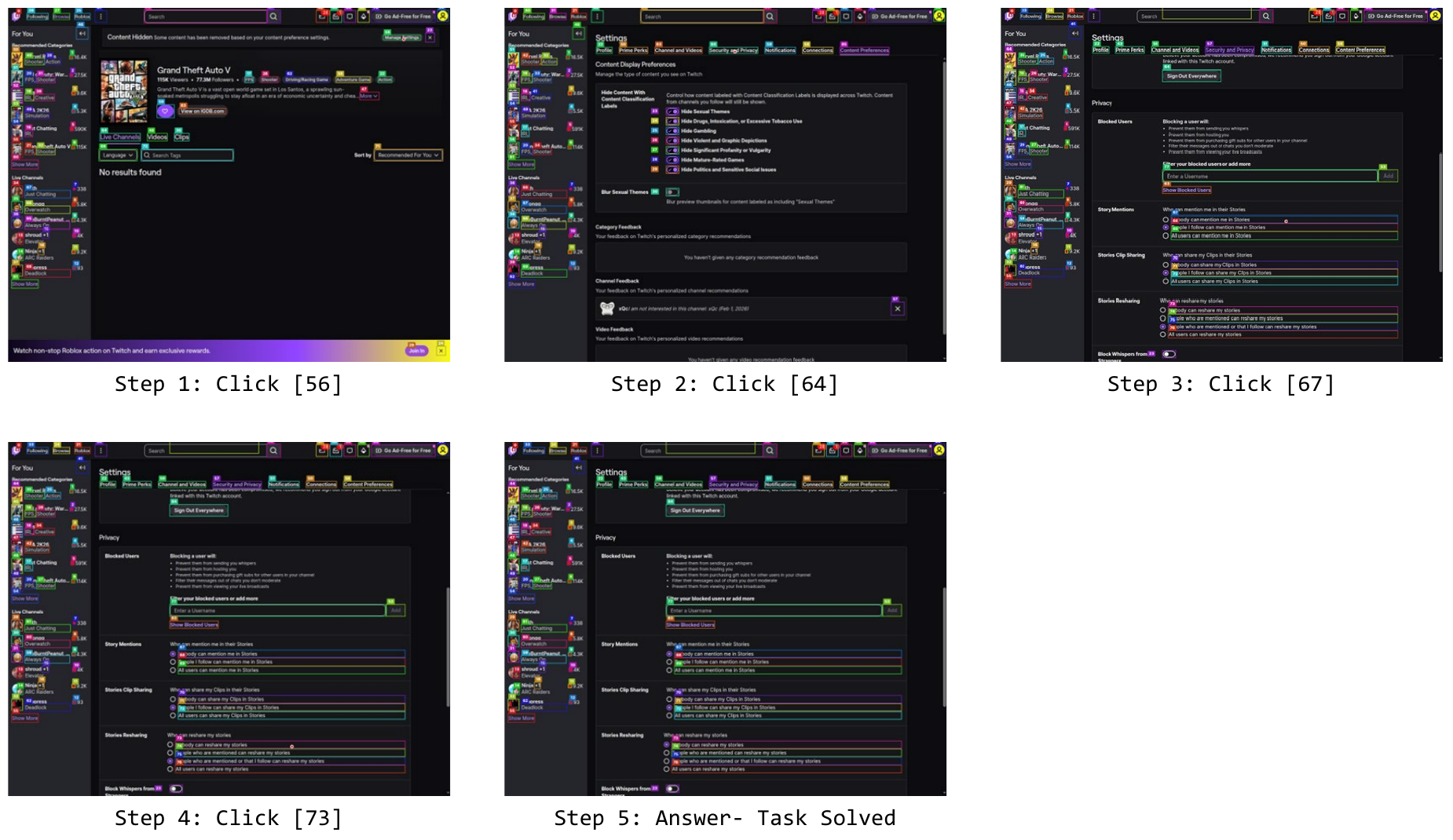}
    
    \vspace{0.5cm} %
    
    \includegraphics[width=\linewidth]{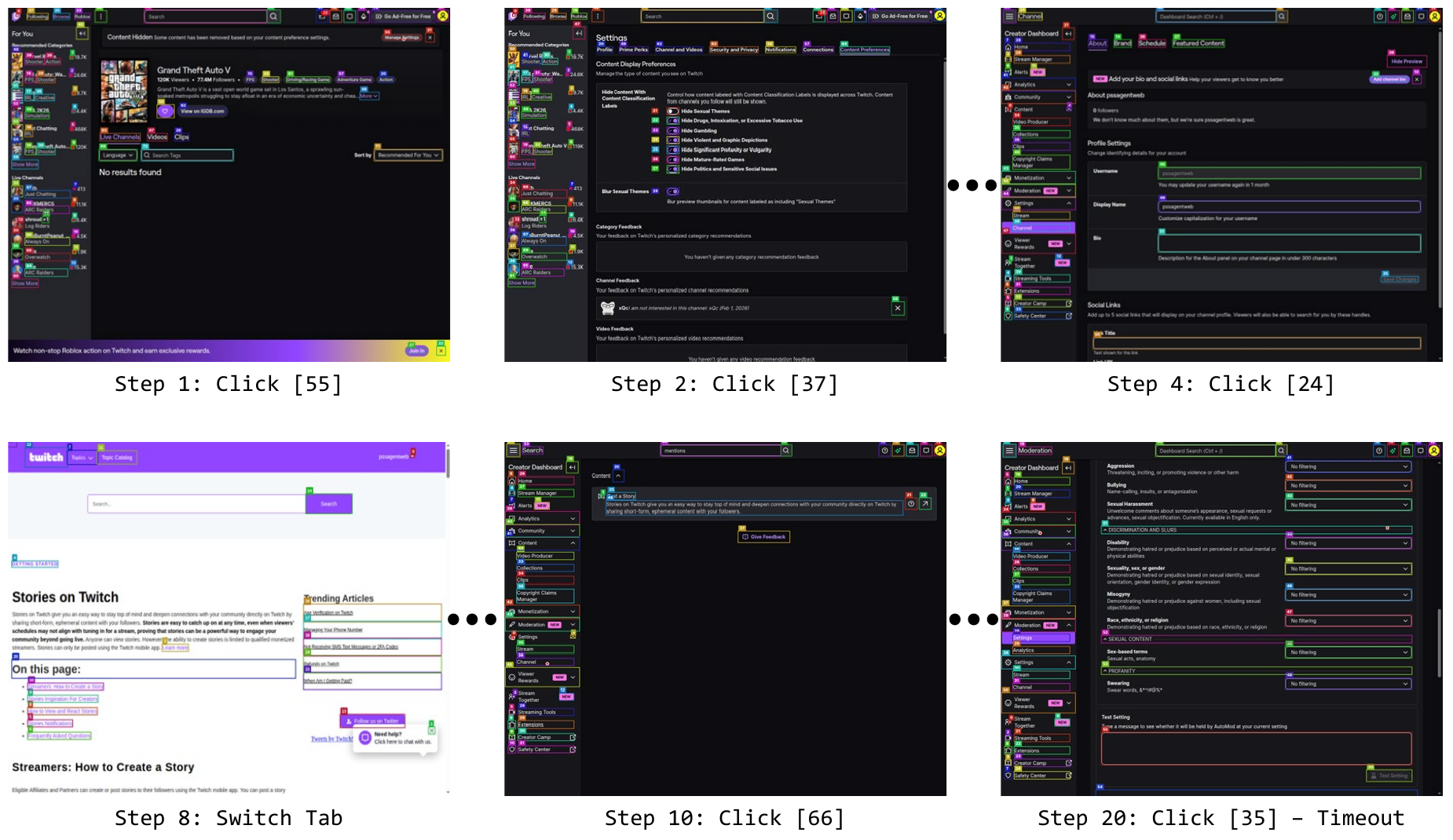}
    
    \caption{Comparison of Gemini-3-Pro trajectories on Twitch with~(top) and without~(bottom) explicit navigational instructions. Given the prompt, ``Navigate to `Security and Privacy' in my account settings and turn on options to ensure nobody can mention me on their stories and reshare my stories,'' the agent successfully completes the task when guided by navigational details but fails during autonomous exploration.}
    \label{fig:twitch_combined}
\end{figure*}

\begin{figure*}[t]
    \centering
    \includegraphics[width=0.8\linewidth]{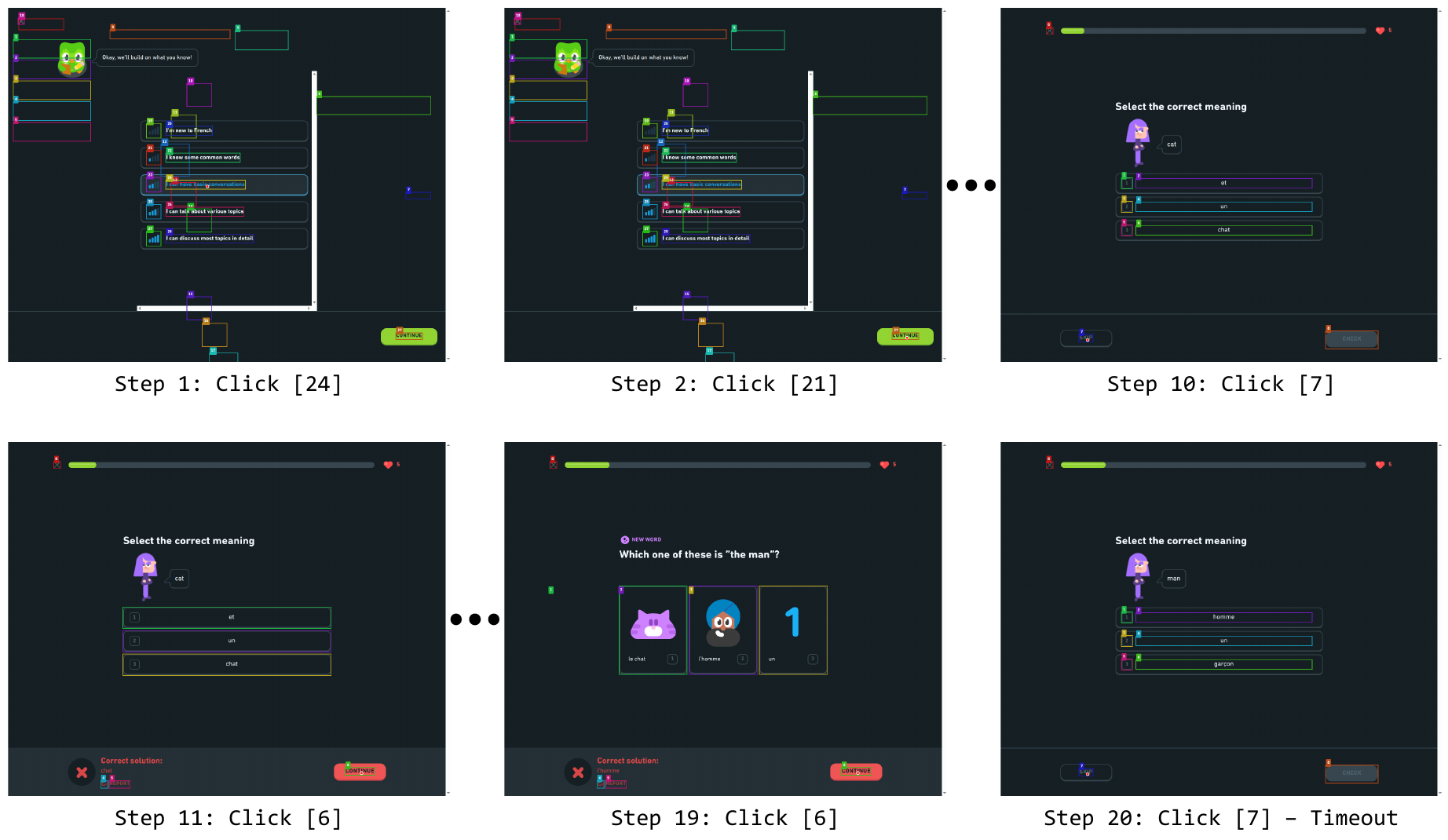}
    \caption{Failure highlighting website specific design on Duolingo~(Gemini-2.5-Pro, \wonavE variant). Given the instruction ``Make my profile private to remove myself from the weekly competitive leaderboards, and ensure friends streaks is disabled if visible'', the model fails to close an introductory language lesson on Duolingo and goes on to solve that instead of the task specified in the instruction. The model declares in thoughts that it needs to first solve the task before it can proceed with the user instruction. The model failed to identify the "Close button".}
    \label{fig:duolingo_rq2}
\end{figure*}

\begin{figure*}[t]
    \centering
    \includegraphics[width=0.8\linewidth]{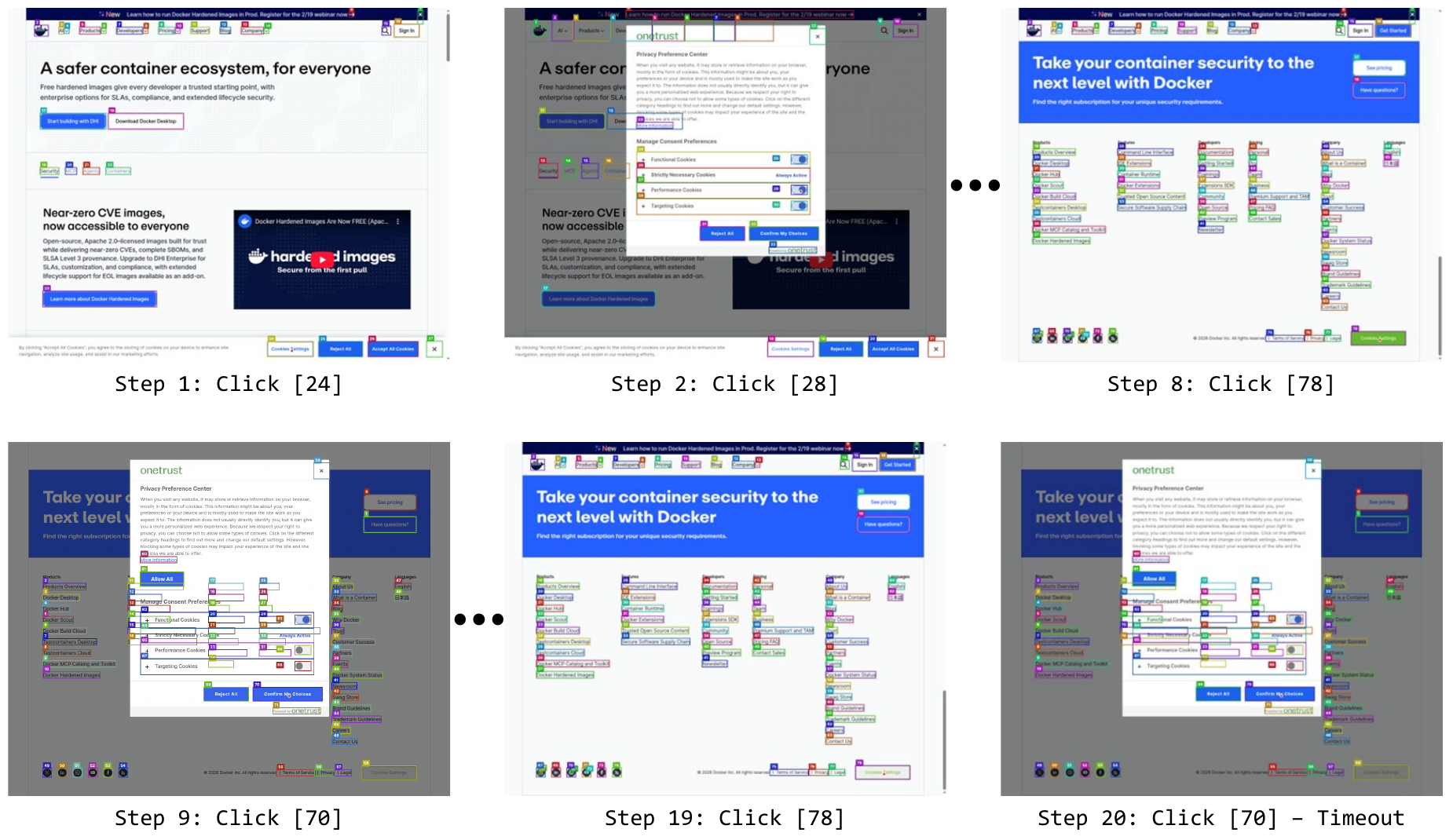}
    \caption{Failure highlighting a Cookie \& Tracking Consent Management task failure on Docker~(GPT-5-Mini, \wonavE variant). Given the instruction ``Disable 'performance cookies' and 'targeted cookies', while keeping 'functional cookies' enabled.'', the model finishes the requirements of the task but keeps performing actions of opening and closing the cookie notice in loop until it reaches timeout.}
    \label{fig:docker_rq2}
\end{figure*}

\begin{figure*}[t]
    \centering
    \includegraphics[width=0.8\linewidth]{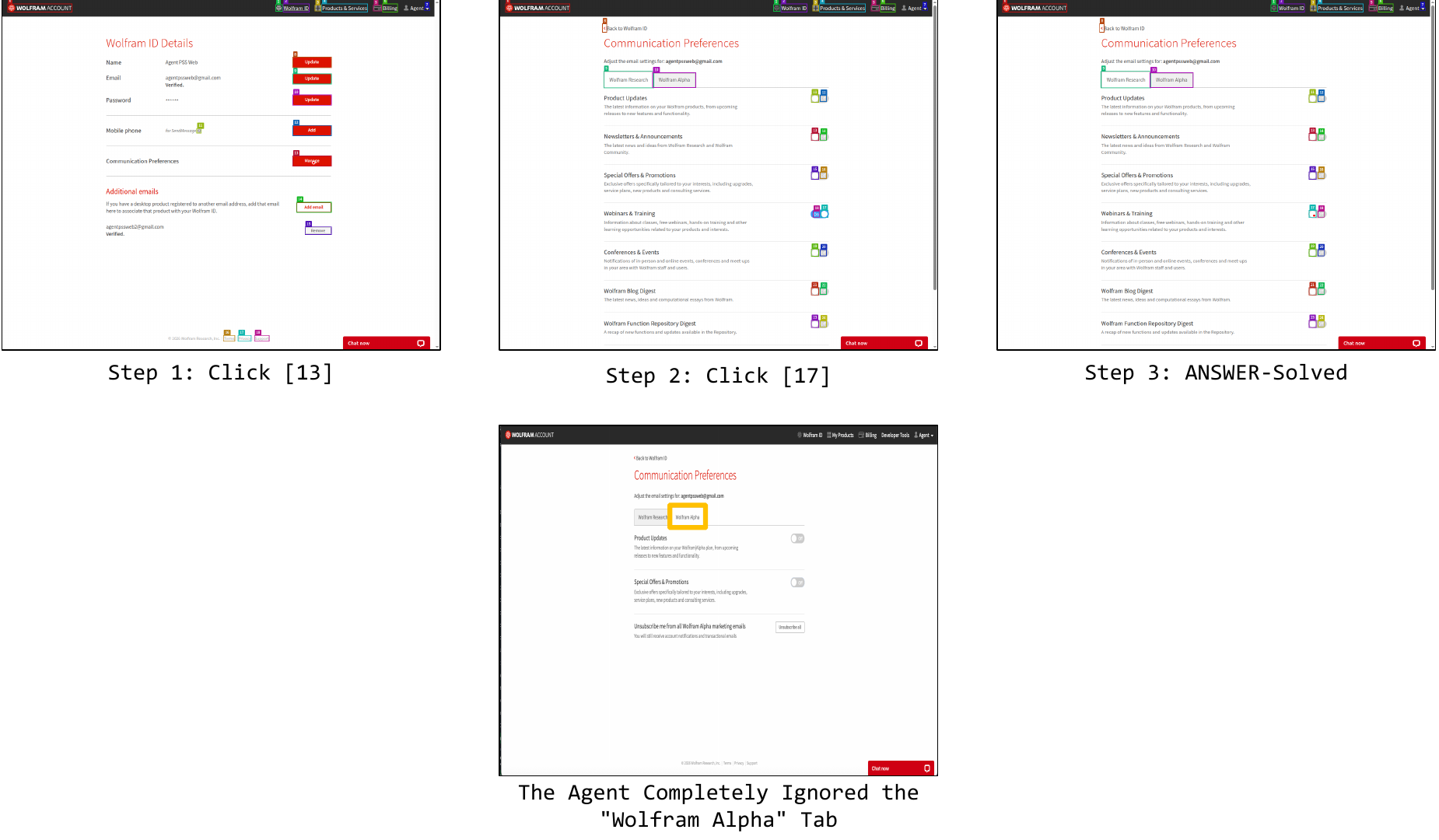}
    \caption{Failure highlighting a Wolfram task to set communication preferences (Claude-Sonnet-4.5, \wonavE variant). Given the instruction ``Enable only updates about 'Webinars \& Training', disabling all other options.'', the agent fails to check the `Wolfram Alpha' tab that contains more communication preferences.    
    }
    \label{fig:qual_analysis_fig_1}
\end{figure*}

\begin{figure*}[t]
    \centering
    \includegraphics[width=0.8\linewidth]{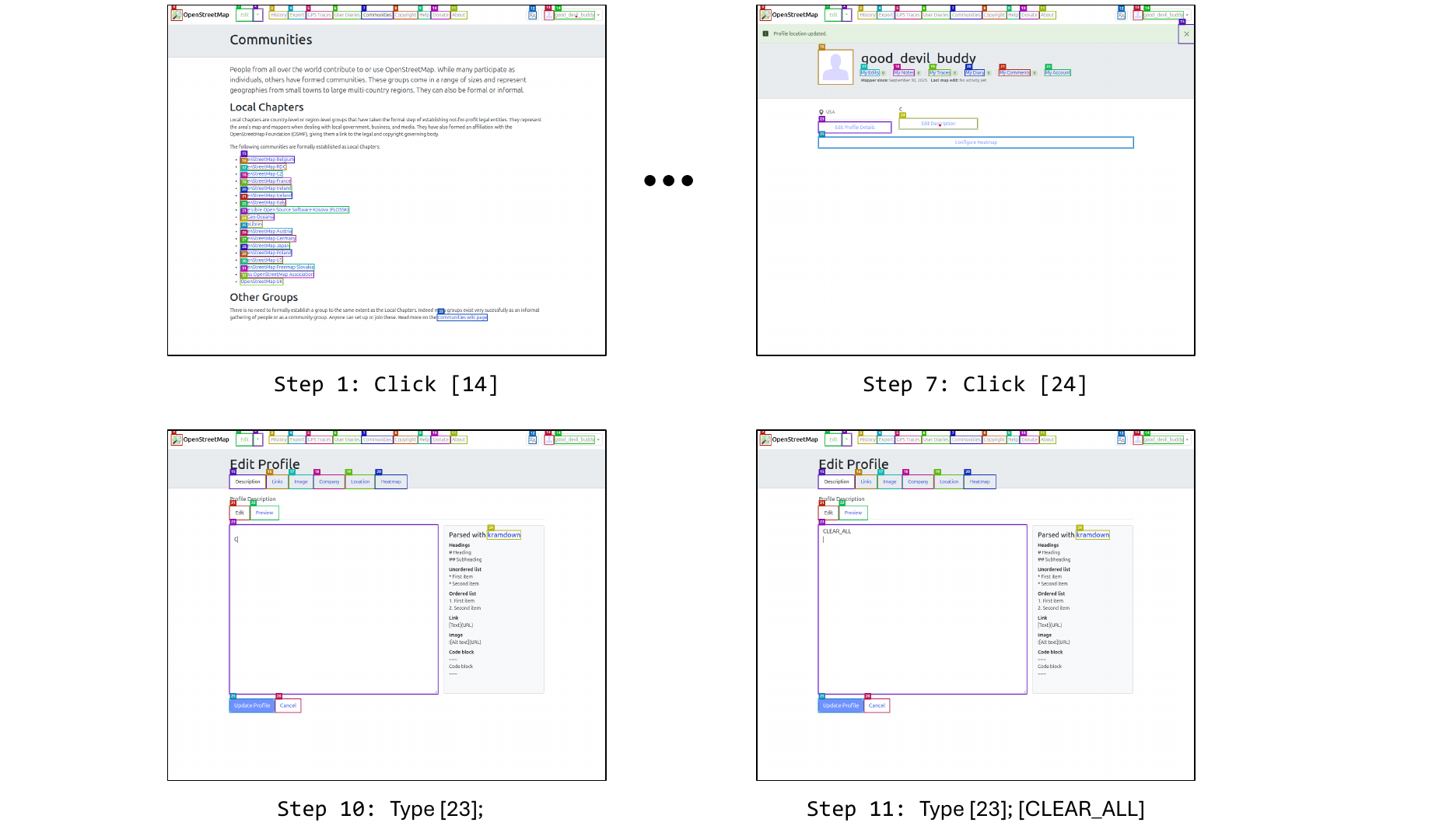}
    \caption{Failure highlighting a task to clear existing information in profile description box of OpenStreetMap~(Claude-Haiku-4.5, \wonavE variant). Given the instruction ``Edit my user profile to remove all information from the 'Description' box and set my location to 'USA' without specifying latitude or longitude'.'', the agent sets the location correctly. However, it fails to clear the text in the description and enters the text `CLEAR\_ALL' instead.}
    \label{fig:qual_analysis_fig_2}
\end{figure*}

\begin{figure*}[t]
    \centering
    \includegraphics[width=0.8\linewidth]{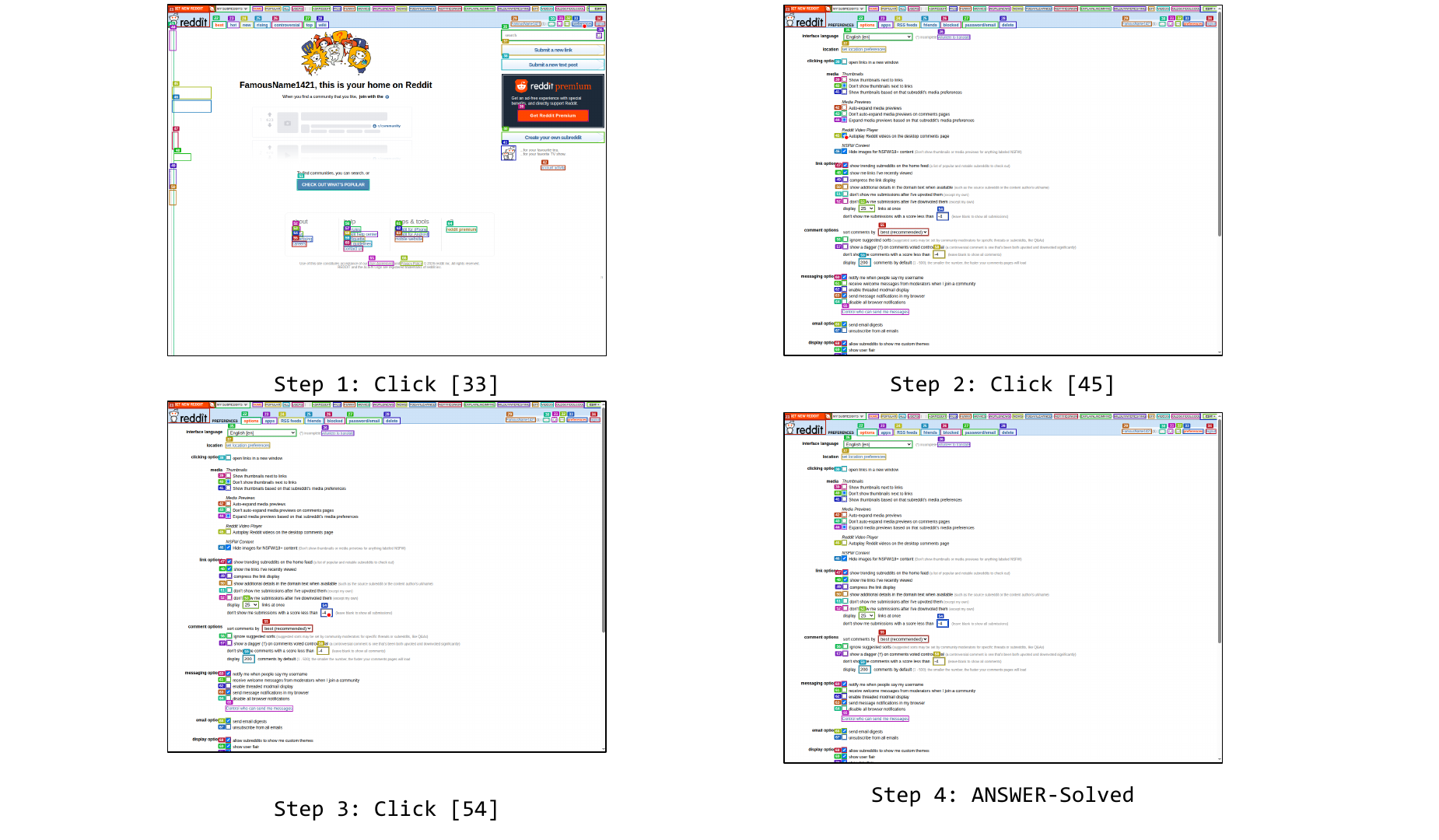}
    \caption{Failure highlighting a classic example of the agent not saving changes~(GPT-5-Mini, \wonavE variant). Given the instruction ``Disable the option to autoplay reddit videos in my account.'', the agent unchecks the relevant checkbox, but does not scroll down and save the changes.
    }
    \label{fig:qual_analysis_fig_3}
\end{figure*}

\begin{figure*}[t]
    \centering
    \includegraphics[width=0.8\linewidth]{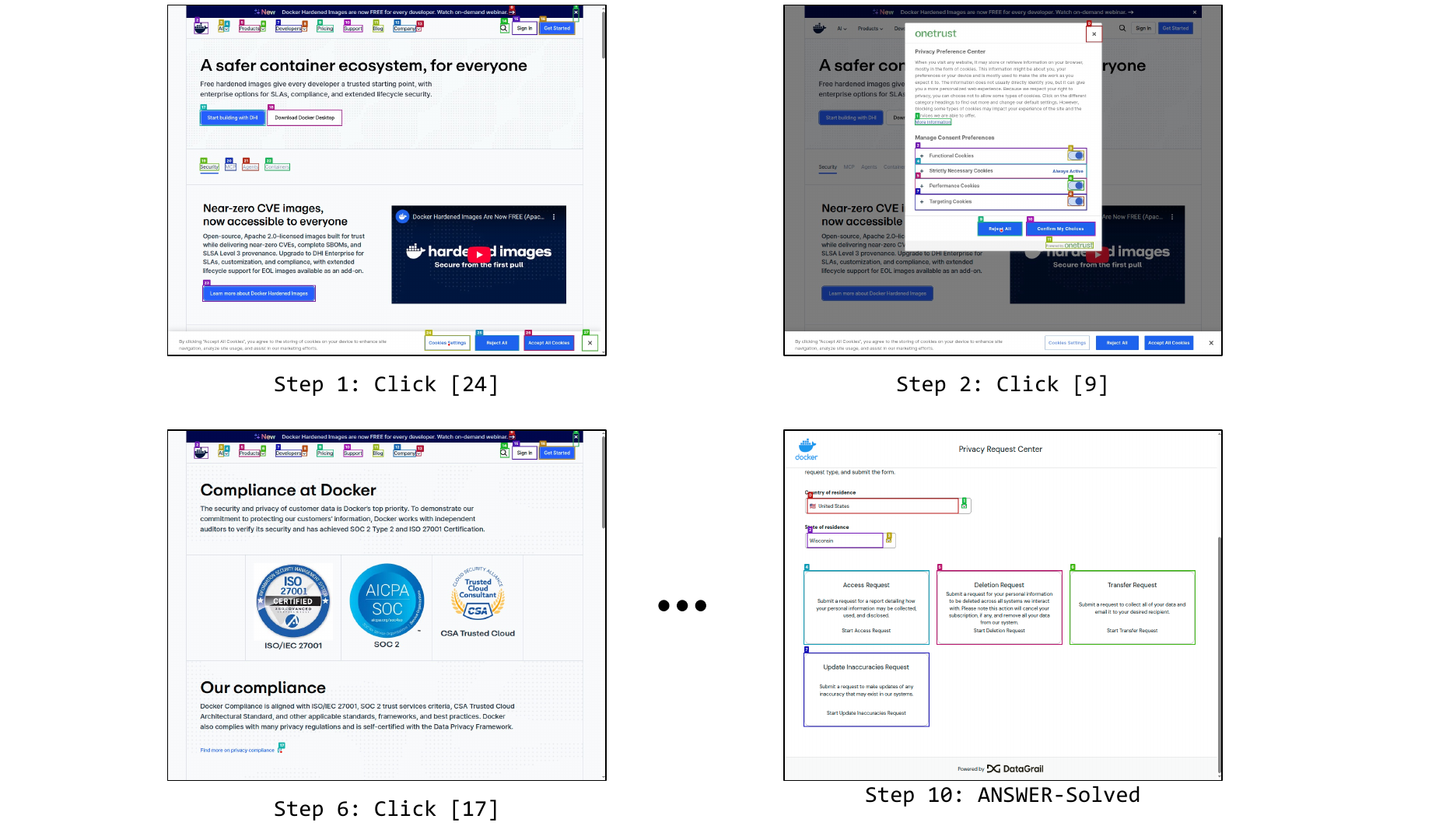}
    \caption{Failure highlighting agent misunderstanding the task~(GPT-5.1, \wonavE variant). Given the instruction ``Opt out of sharing anonymized usage data with Docker's partners.'', the agent rejects all cookies instead of enabling the opt-out setting inside the account.
    }
    \label{fig:qual_analysis_fig_4}
\end{figure*}

\begin{figure*}[t]
    \centering
    \includegraphics[width=0.8\linewidth]{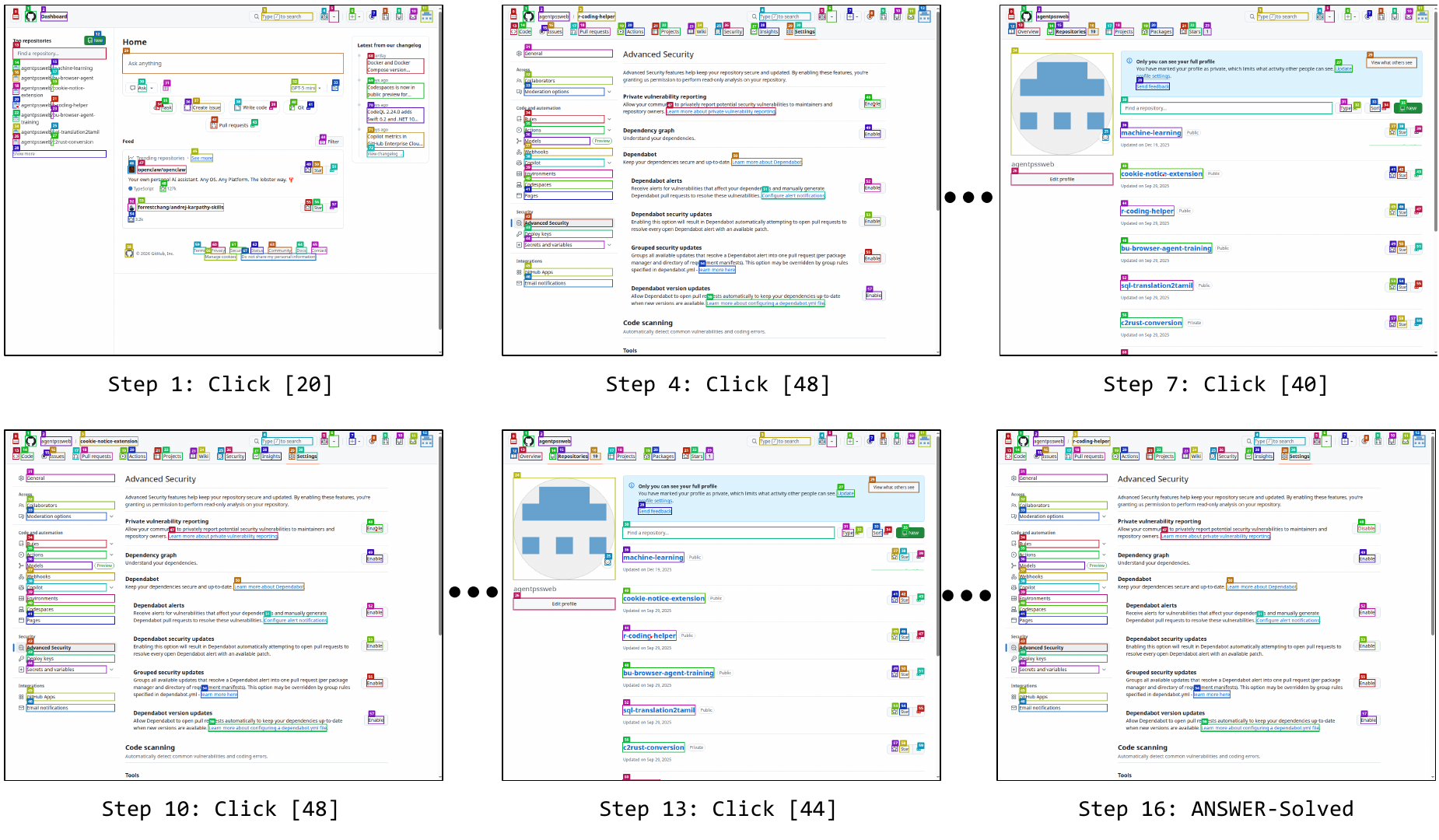}
    \caption{Failure highlighting agent opting for a much longer solution to a Github task~(GPT-5.1, \wonavE variant). Given the instruction ``Enable 'Private vulnerability reporting' option for all repositories.'', the agent visits starts visiting every single repository to change the setting, when there is an option available in account settings.}
    \label{fig:qual_analysis_fig_5}
\end{figure*}

\begin{figure*}[t]
    \centering
    \includegraphics[width=0.8\linewidth]{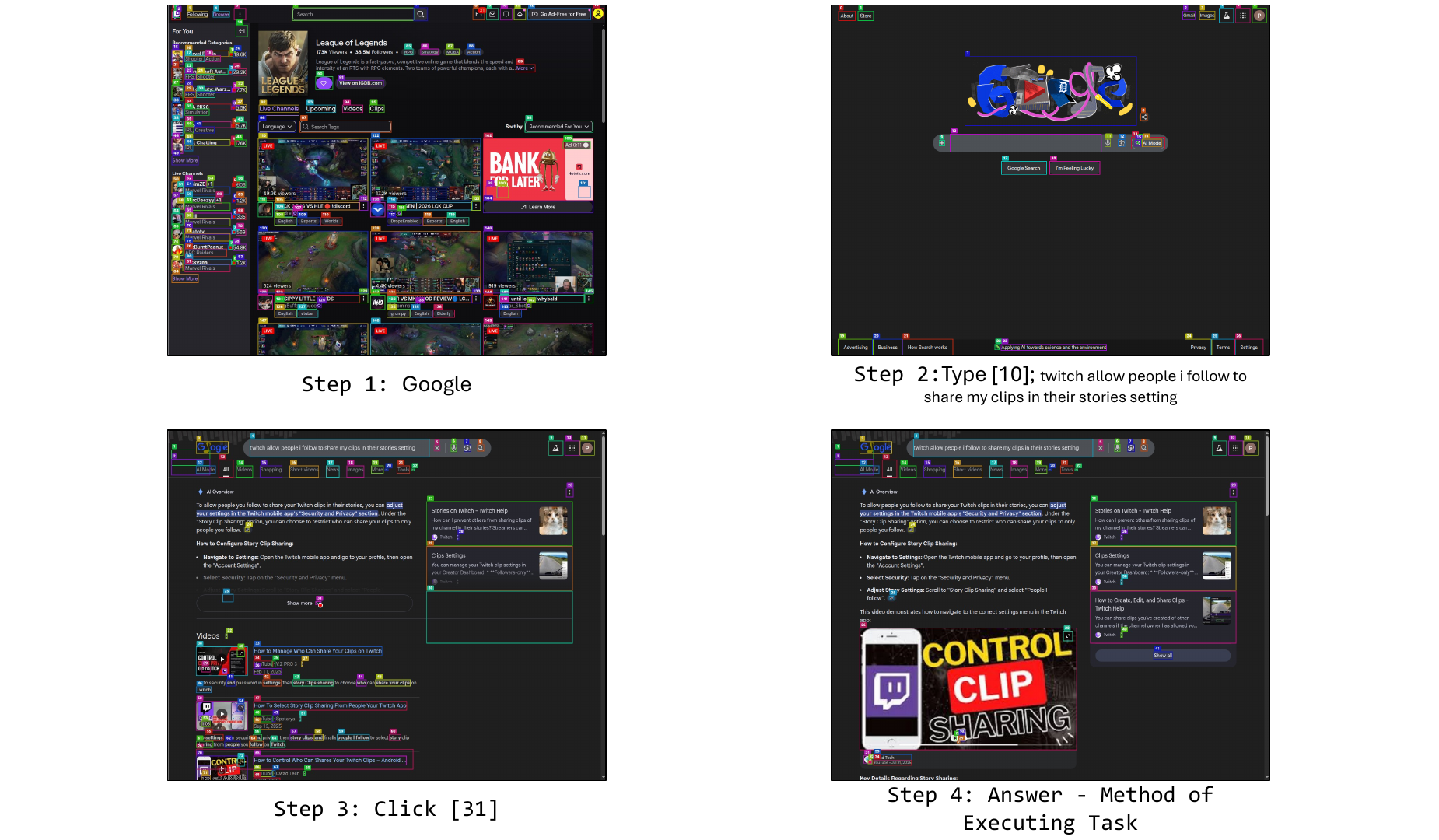}
    \caption{Failure highlighting agent searching for the task on Google and explaining the steps needed in its final answer~(GPT-5.1, \wonavE variant). Given the instruction ``Turn on option to allow people I follow to share my clips in their stories'', the agent visits visits google, searches for the task, uses the AI Overview and completes the task by explaining the method to the user in its final answer.}
    \label{fig:qual_analysis_fig_7}
\end{figure*}

\begin{figure*}[t]
    \centering
    \includegraphics[width=0.8\linewidth]{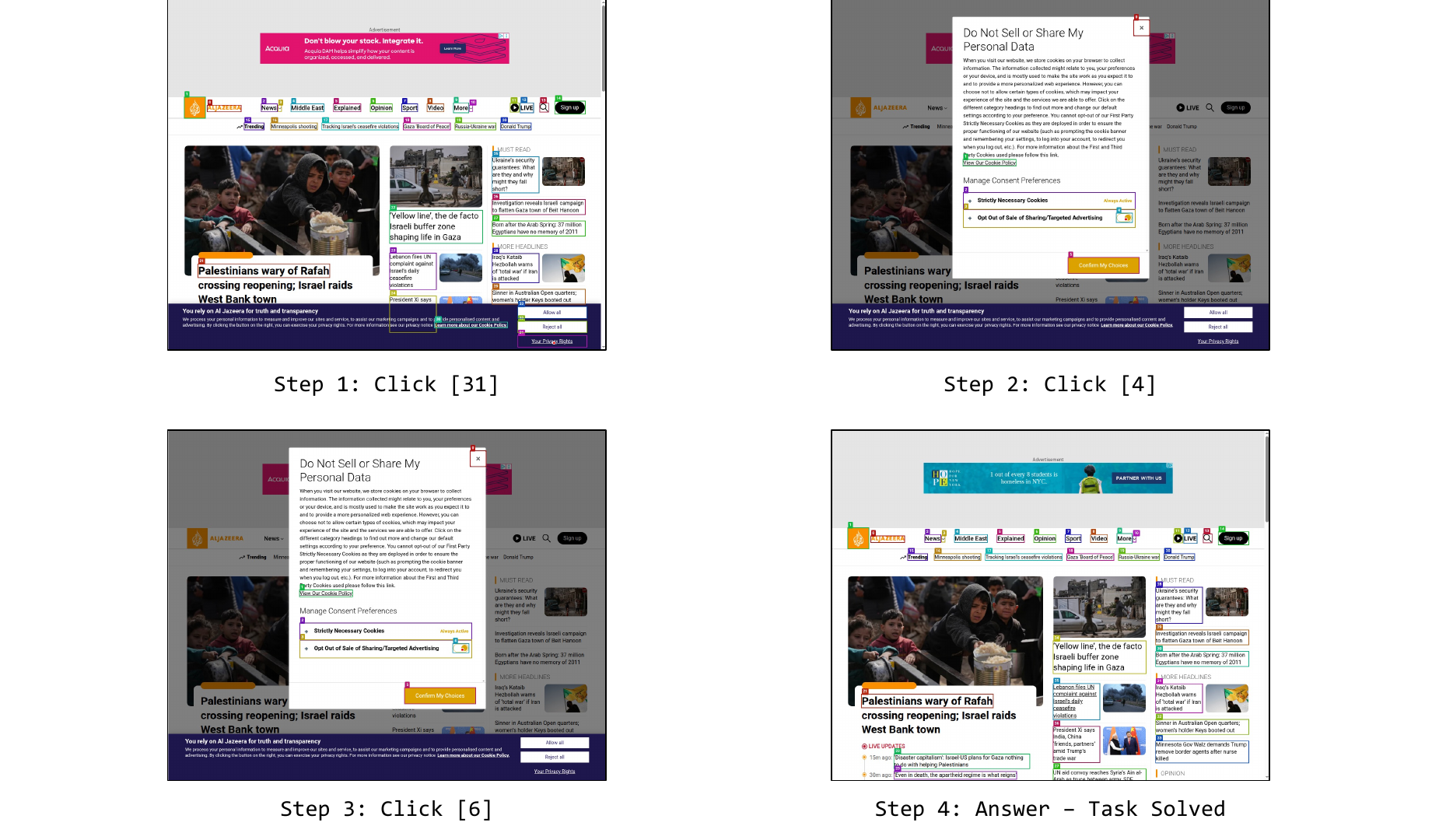}
    \caption{Failure highlighting agent misunderstanding and changing the state of a toggle~(Gemini-2.5-Pro, \wonavE variant). Given the instruction ``Opt out of sale of personal data and targeted advertising options in the website.'', the agent does not realize the setting to opt out is already ON and interacts with the toggle wrongly. Same failure also occurs with Gemini-3-Pro. This toggle is an example of a `trick wording' dark pattern employed by the website, where the wording of the setting is tricky and the default is opt out.}
    \label{fig:qual_analysis_fig_8}
\end{figure*}

\begin{figure*}[t]
    \centering
    \includegraphics[width=0.8\linewidth]{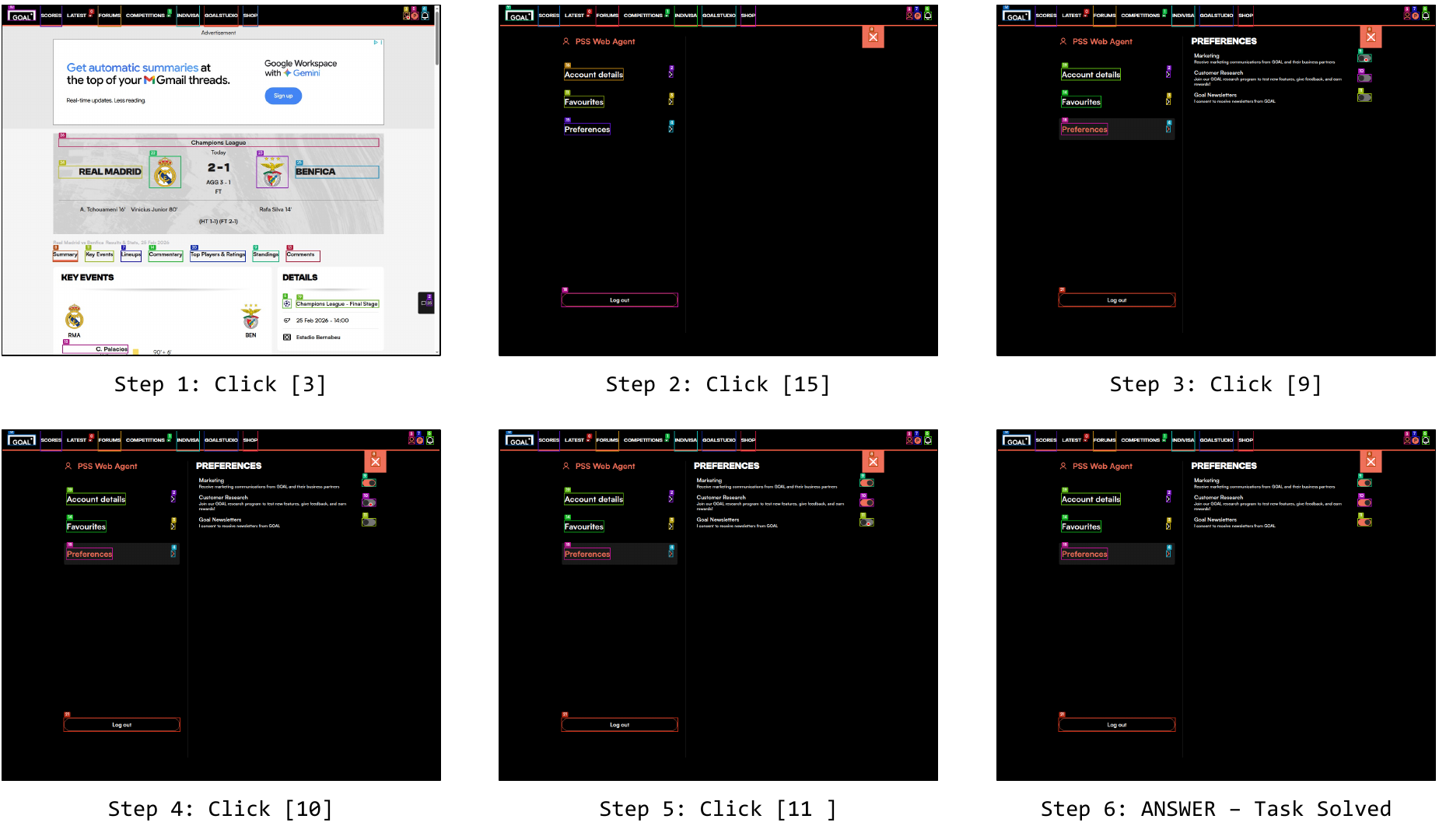}
    \caption{Failure highlighting agent misunderstanding and wrongly changing the state of a toggle in Goal.com~(Gemini-3-Pro, \wonavE variant). Given the instruction ``Opt-out of all communications: Disable all three toggles for 'Marketing', 'Customer Research', and 'Goal Newsletters'', the agent turns on all the toggles.}
    \label{fig:qual_analysis_fig_9}
\end{figure*}

\begin{figure*}[t]
    \centering
    \includegraphics[width=0.8\linewidth]{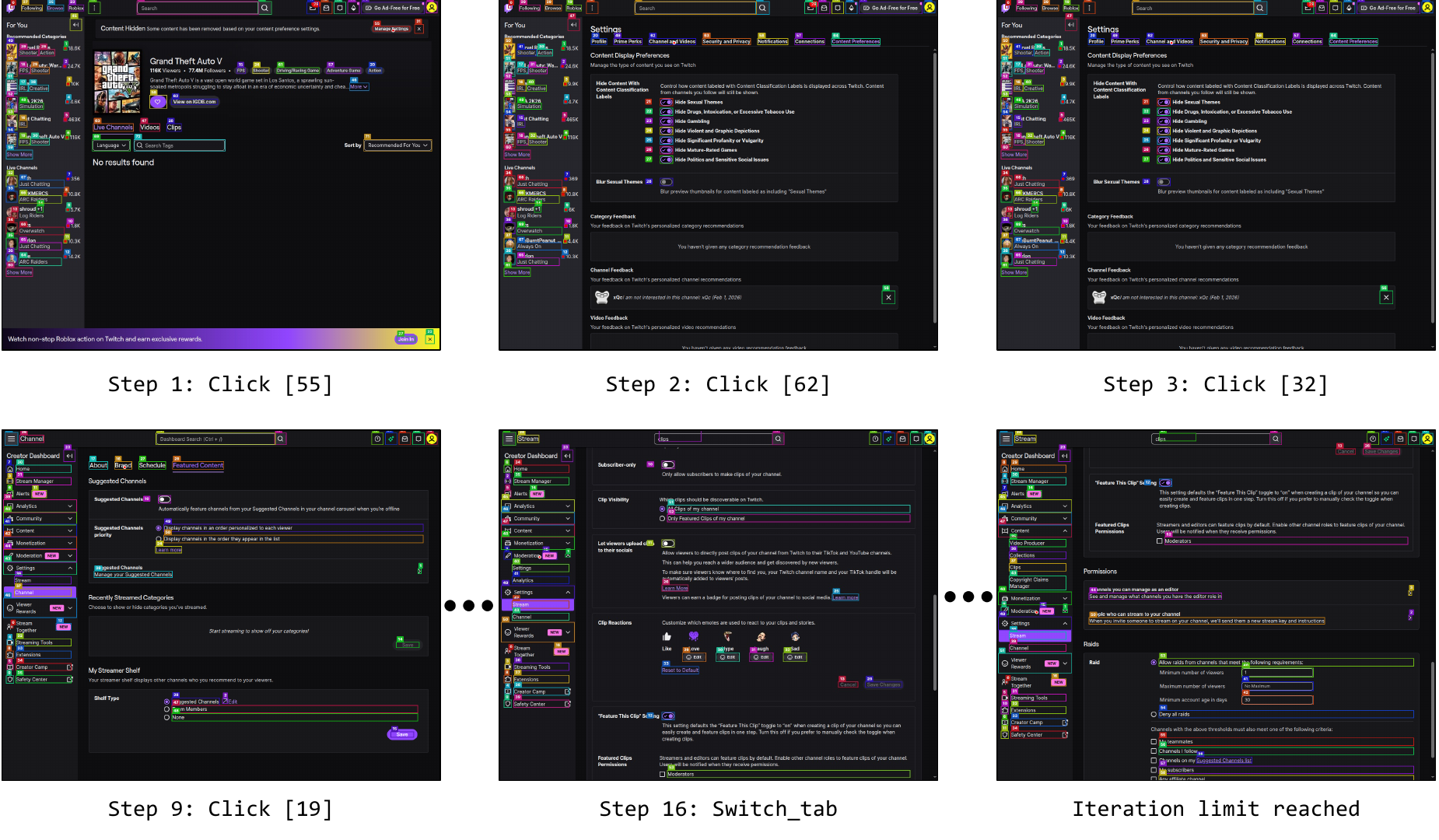}
    \caption{Failure highlighting agent unable to navigate through Twitch to the relevant settings page~(Gemini-3-Pro, \wonavE variant). Given the instruction ``Turn on option to allow people I follow to share my clips in their stories.'' , the agent initially makes its way to the correct page~(Step 3), but falters by making a wrong click and then is unable to reach the settings page before iteration limit.}
    \label{fig:qual_analysis_fig_10}
\end{figure*}
 
\begin{figure*}[t]
    \centering
    \includegraphics[width=0.8\linewidth]{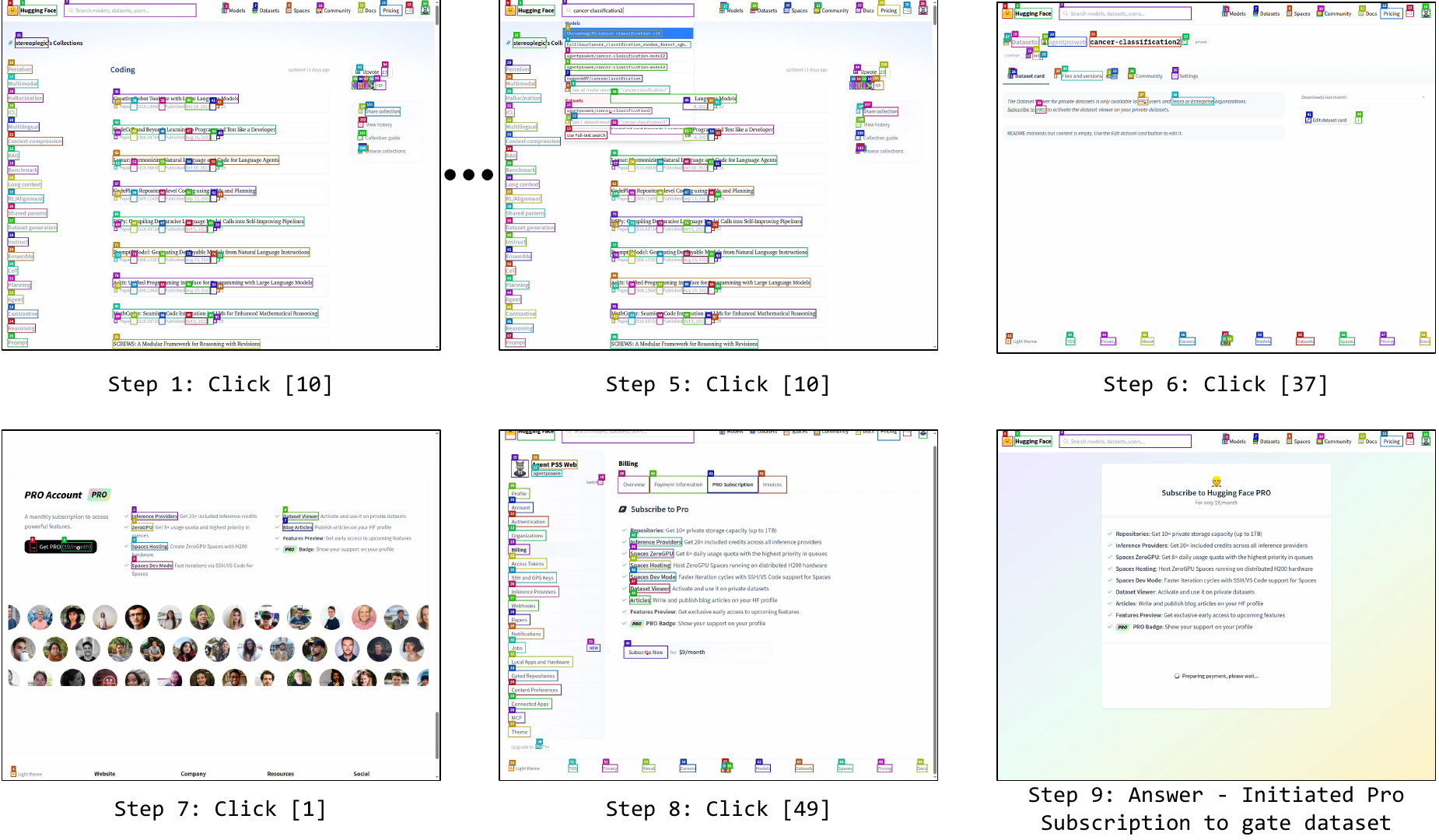}
    \caption{Failure highlighting agent initiating a pro subscription on HuggingFace~(Gemma-3-27B, \wonavE variant). Given the instruction ``Enable gated access to the dataset named `cancer-classification2'.'', the agent correctly identifies the dataset, but makes a wrong click~(Step 6), and proceeds to enabling the PRO subscription. In its final answer, the agent mentioned that it gated the dataset by initiating the PRO subscription. The task ended before the payment page loaded as the agent performed an \texttt{ANSWER} action.}
    \label{fig:qual_analysis_fig_11}
\end{figure*}

\begin{figure*}[t]
    \centering
    \includegraphics[width=0.8\linewidth]{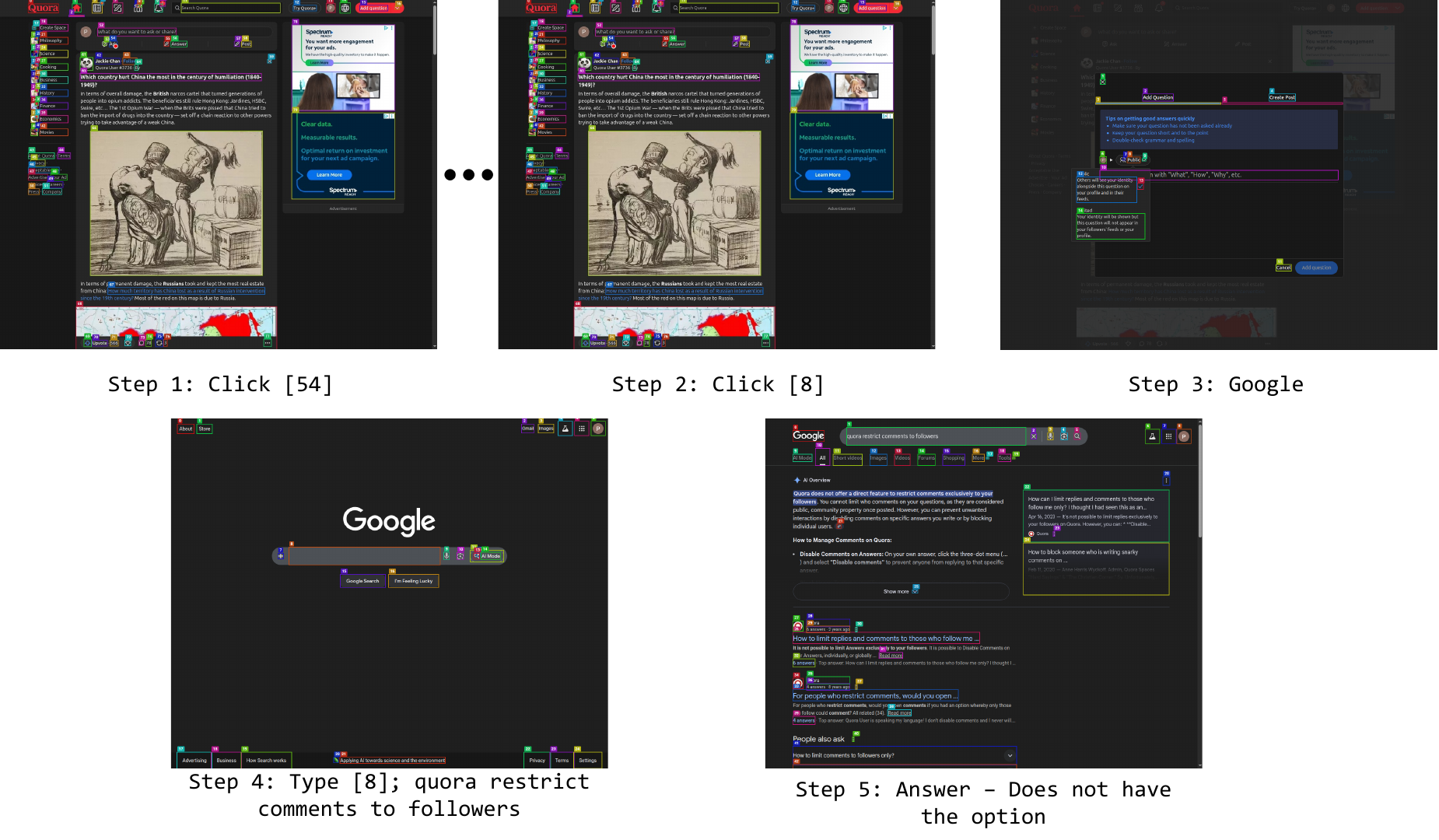}
    \caption{Failure highlighting agent denying a Quora task exists~(Gemma-3-27B, \wonavE variant). Given the instruction ``Enable the option allowing only the people I follow to comment on my post.'', the agent initially starts to make a post due to a wrong click~(Step 3), realizes its mistakes and googles the task. It reads the AI Overview response and answers that the task is not possible on Quora. The task can be solved by navigating to privacy settings.}
    \label{fig:qual_analysis_fig_12}
\end{figure*}

\begin{figure*}[t]
    \centering
    \includegraphics[width=0.8\linewidth]{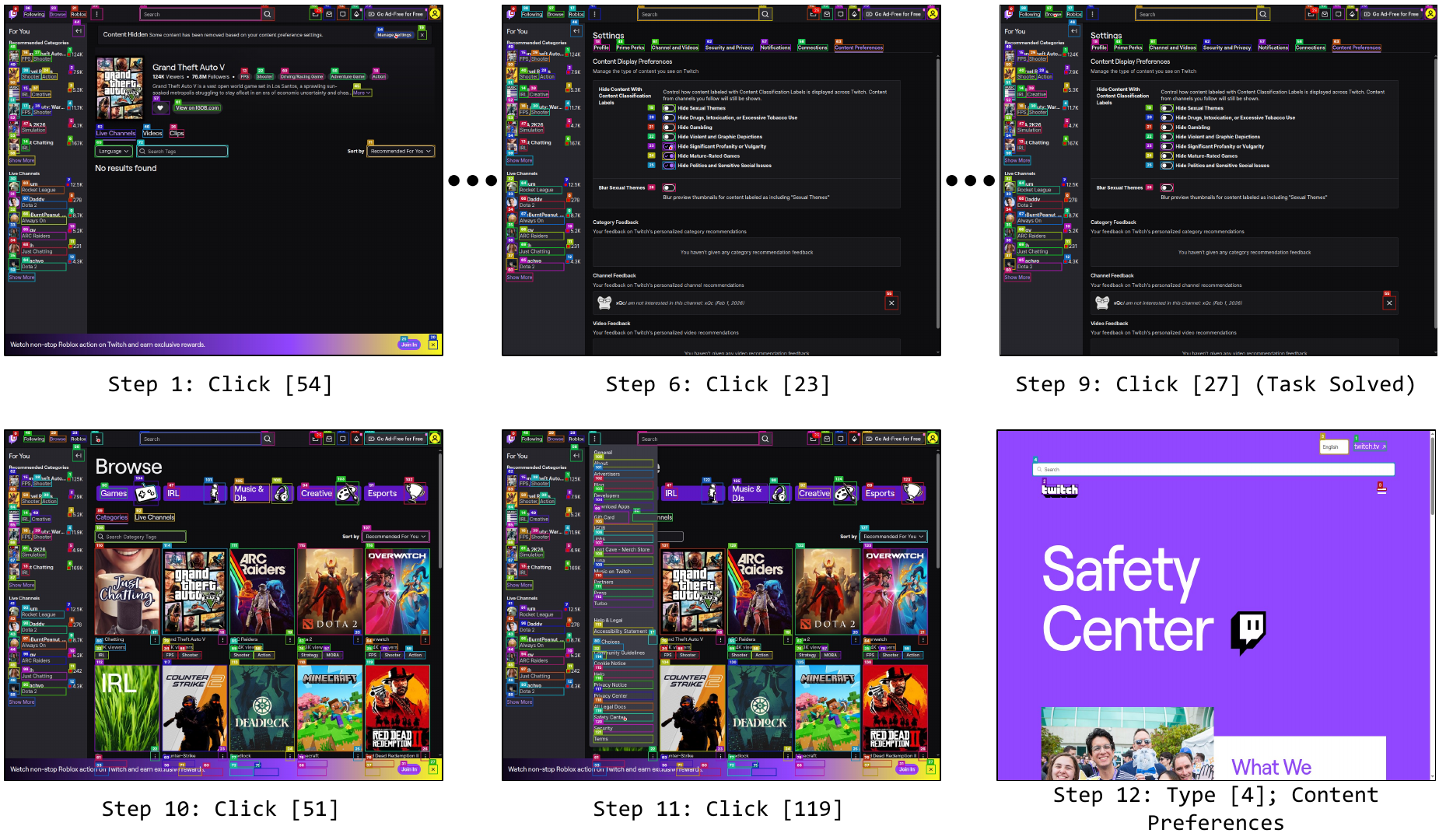}
    \caption{Failure highlighting agent performing actions post completion of a Twitch task~(GPT-5-Mini, \wonavE variant). Given the instruction ``Ensure all content with content classifications labels is blocked from my Twitch feed.'', the agent performs the task correctly. However, upon completion, it continues to navigate through Twitch Safety center starting Step 12 until reaching maximum iteration limit.}
    \label{fig:qual_analysis_fig_13}
\end{figure*}

\begin{figure*}[t]
    \centering
    \includegraphics[width=0.8\linewidth]{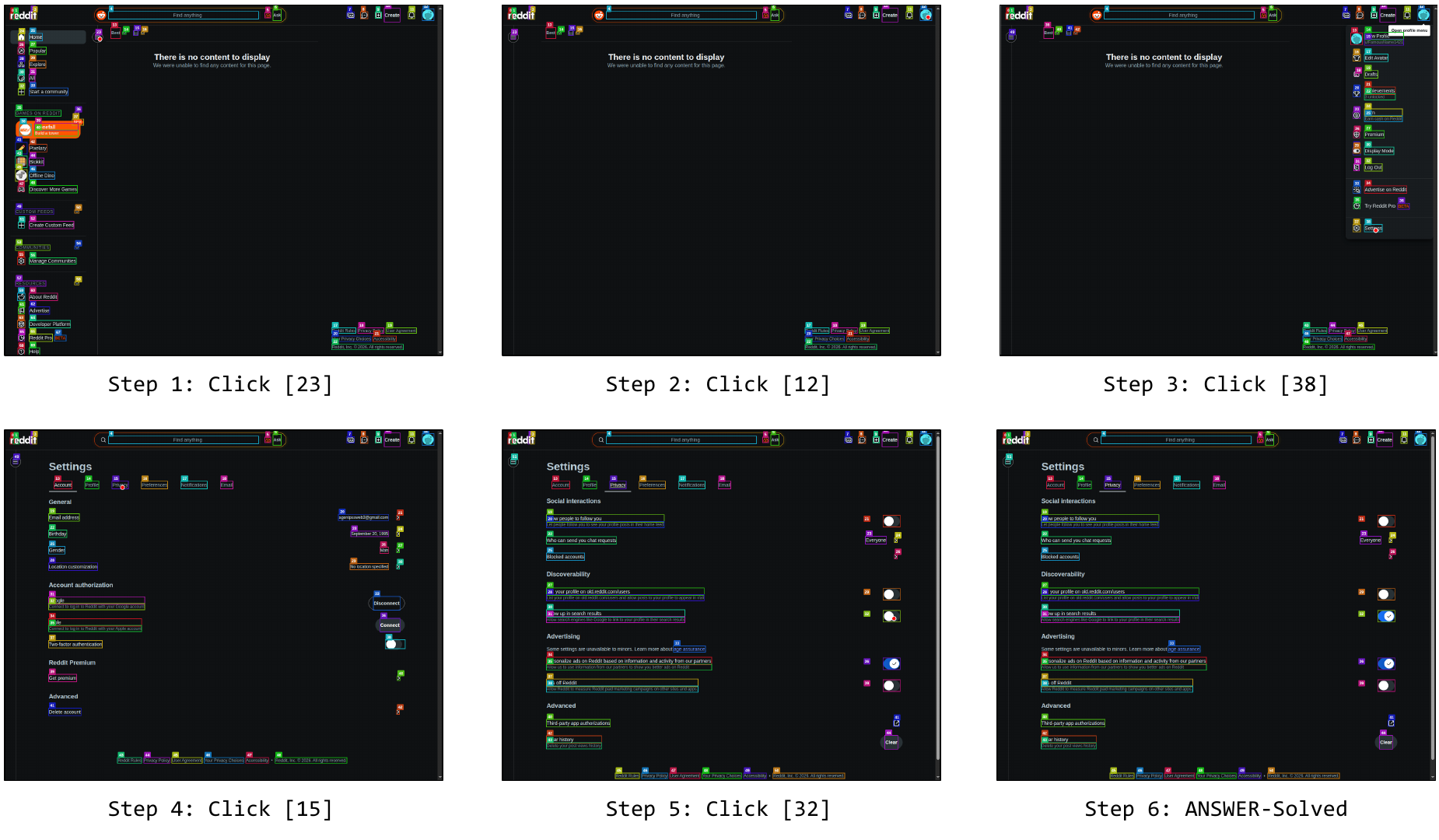}
    \caption{Failure highlighting agent misunderstanding and wrongly changing the state of a toggle in Reddit~(Gemini-2.5-Pro, \wonavE variant). Given the instruction ``Disable the option in my account that allows my profile to show up on searches'', the agent navigates correctly, but enables the toggle~(Step 5), opposite to the task requirement.}
    \label{fig:qual_analysis_fig_14}
\end{figure*}

\begin{figure*}[t]
    \centering
    \includegraphics[width=0.8\linewidth]{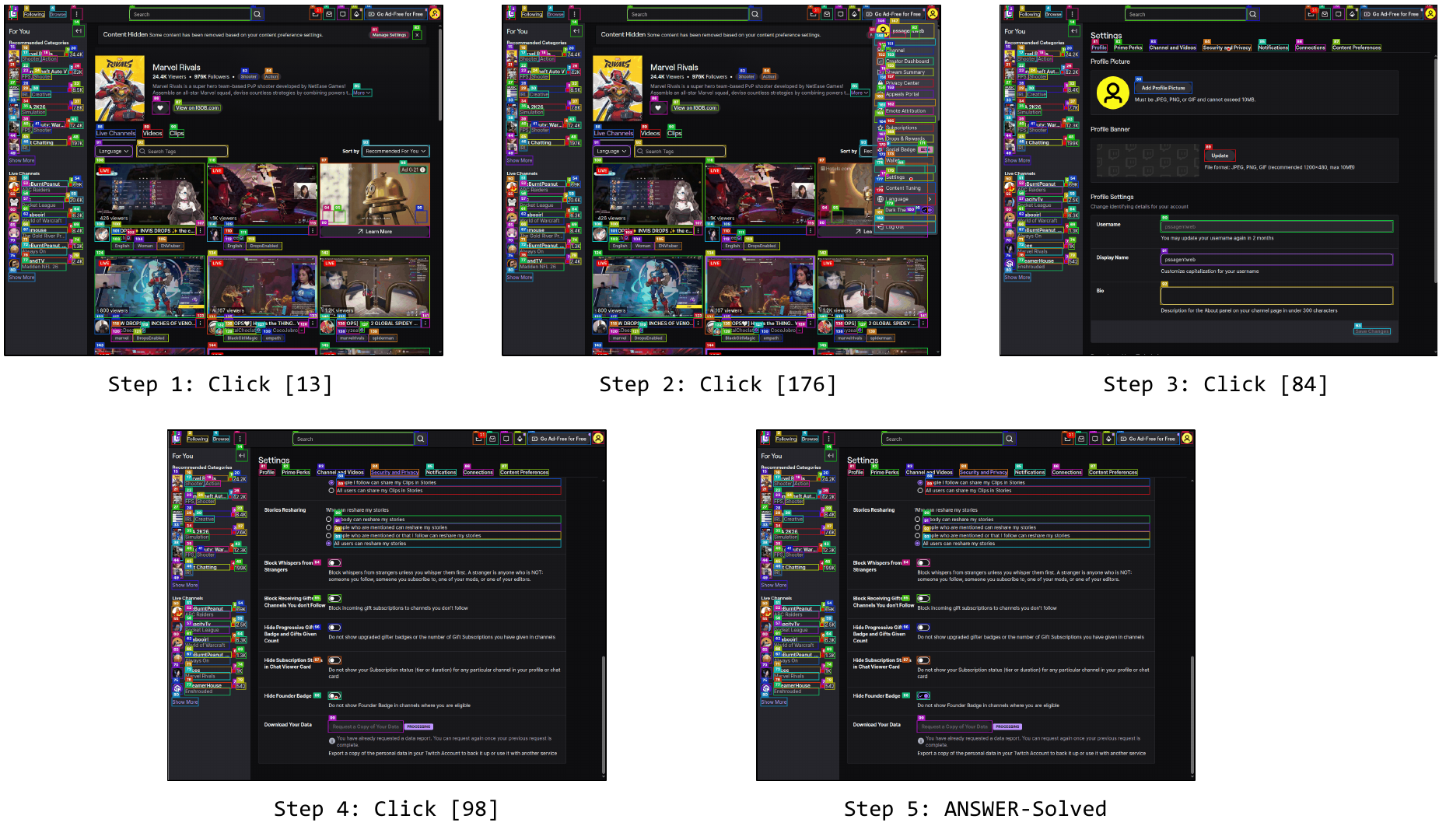}
    \caption{Failure highlighting agent misunderstanding and wrongly changing the state of a toggle in Twitch~(Gemini-2.5-Pro, \wonavE variant). Given the instruction ``Enable option to block whispers from strangers and disable the option to hide the founder badge.'', the agent navigates correctly, but enables the toggle to hide founder badge mistakenly.}
    \label{fig:qual_analysis_fig_15}
\end{figure*}

\begin{figure*}[t]
    \centering
    \includegraphics[width=0.8\linewidth]{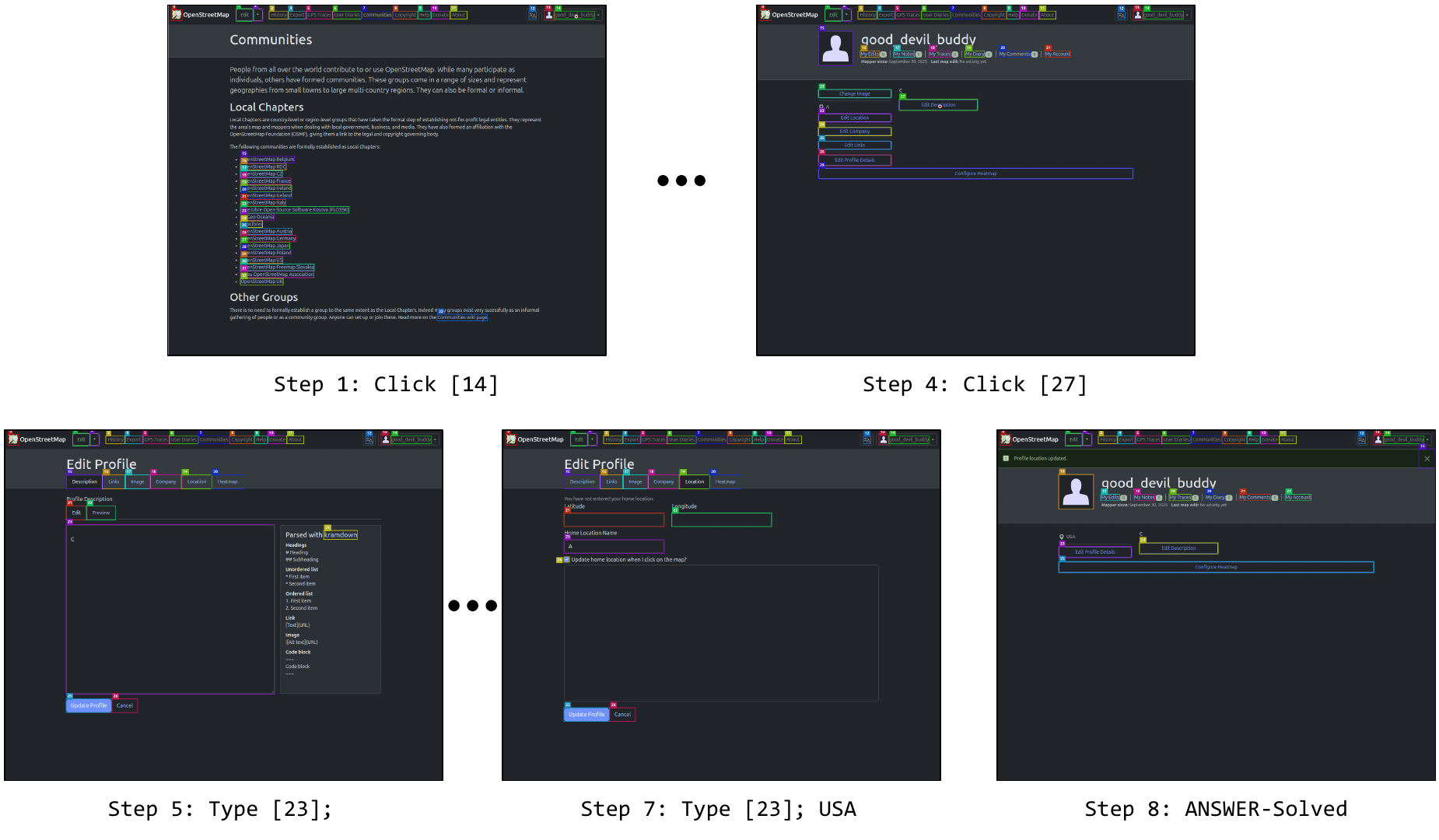}
    \caption{Failure highlighting agent performing an OpenStreetMap task partially completing only a part of it~(Gemini-2.5-Flash, \wonavE variant). Given the instruction ``Edit my user profile to remove all information from the 'Description' box and set my location to 'USA' without specifying latitude or longitude'.'', the agent sets the location correctly. However, it does not even attempt the rest of the task.}
    \label{fig:qual_analysis_fig_16}
\end{figure*}

\begin{figure*}[t]
    \centering
    \includegraphics[width=0.8\linewidth]{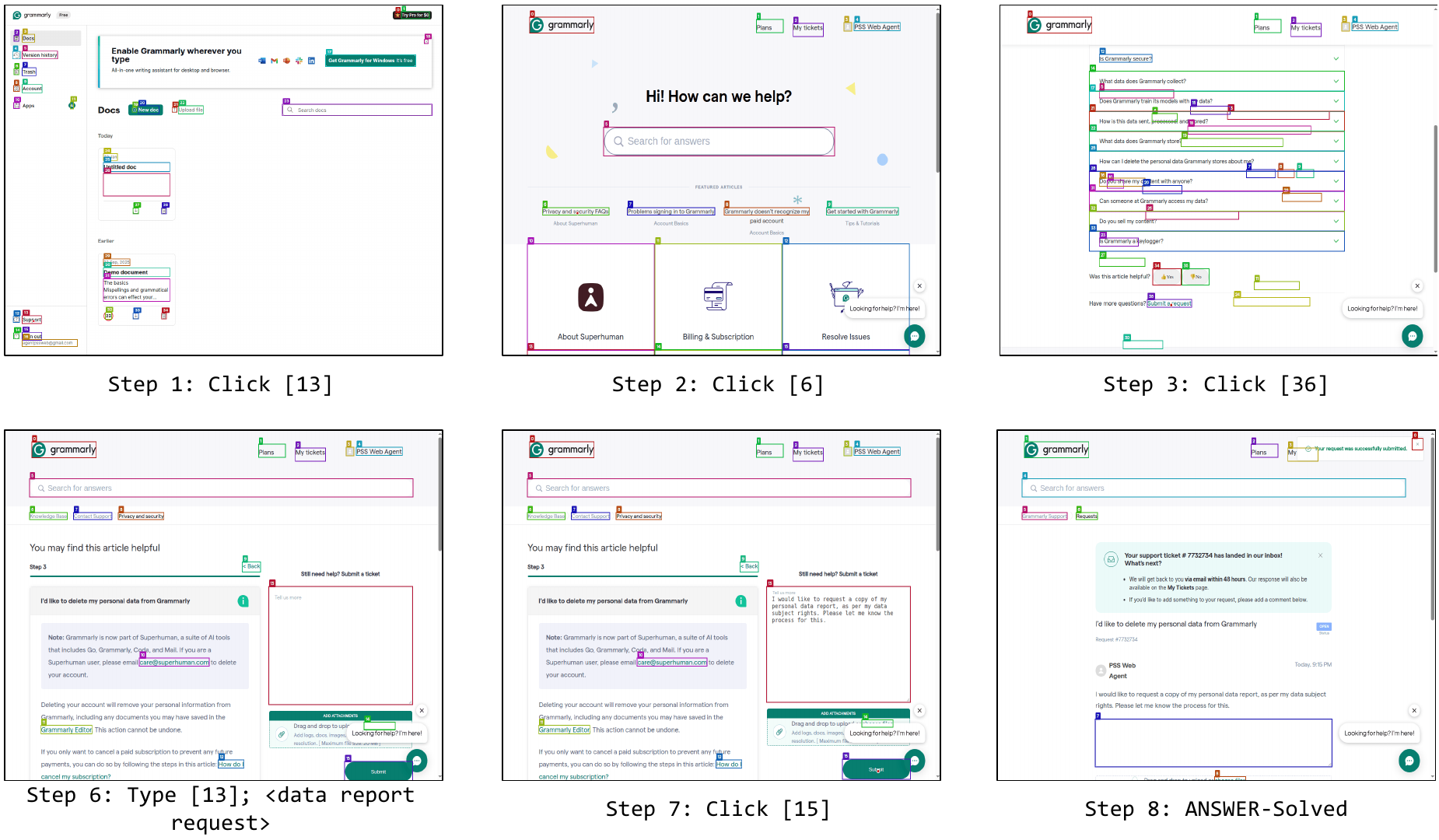}
    \caption{Failure highlighting agent performing an Grammarly task partially completing only a part of it~(Gemini-2.5-Flash, \wonavE variant). Given the instruction ``Submit a personal data report request to Grammarly'', the agent creates ticket titled `I'd like to delete my personal data from Grammarly' with subject consisting of a personal data request. The request is available as a single button in the `Security Overview' setting page.}
    \label{fig:qual_analysis_fig_17}
\end{figure*}

\begin{figure*}[t]
    \centering
    \includegraphics[width=0.8\linewidth]{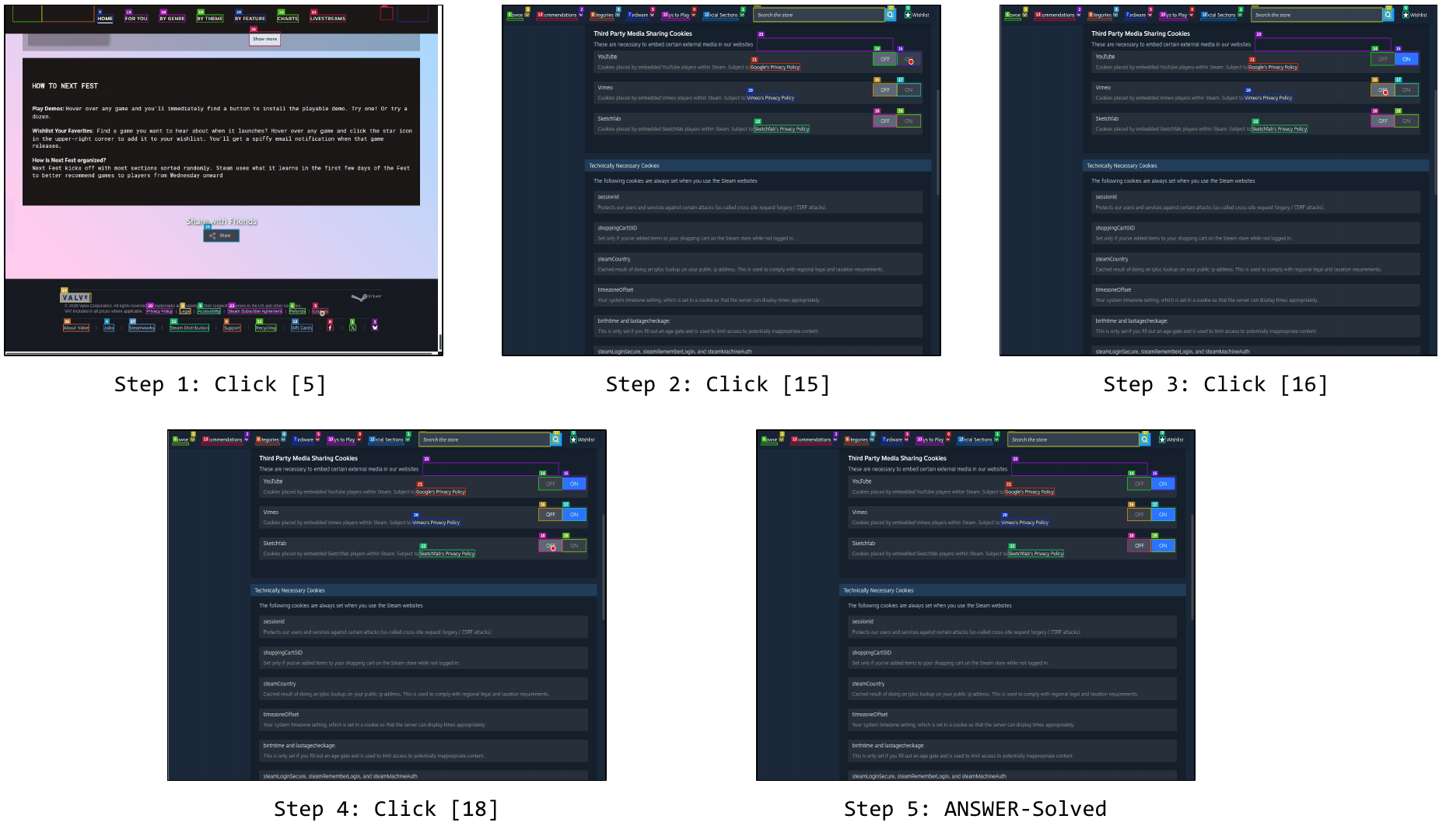}
    \caption{Failure highlighting misunderstanding and enabling wrong cookies for Steam~(Gemini-2.5-Flash, \wonavE variant). Given the instruction ``Enable 'YouTube' cookies, but disable 'Vimeo' and 'Sketchfab' cookies'', the agent creates enables all cookies, while the task only asks the `YouTube' cookies to be enabled.}
    \label{fig:qual_analysis_fig_18}
\end{figure*}

\section{Additional Figures} 

In this section, we present two sets of additional figures that support our discussion section. 

\subsection{Mid-flow \taskname Example}
\label{sec:mid-flow-appendix}

\begin{figure*}[htbp]
    \centering
    \resizebox{0.85\linewidth}{!}{%
        \includegraphics{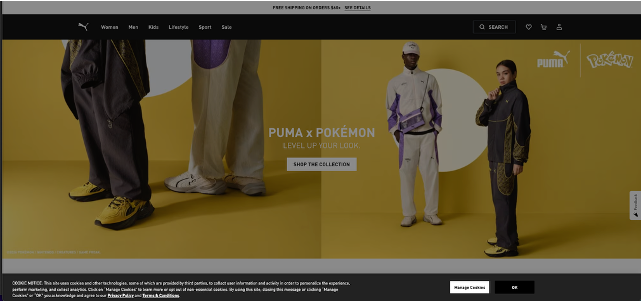}%
    }
    \caption{Puma.com's landing page: Cookie banner interaction is necessary before accessing the website}
    \label{fig:puma_example}
\end{figure*}

\begin{figure*}[htbp]
    \centering
    \resizebox{0.6\linewidth}{!}{%
        \includegraphics{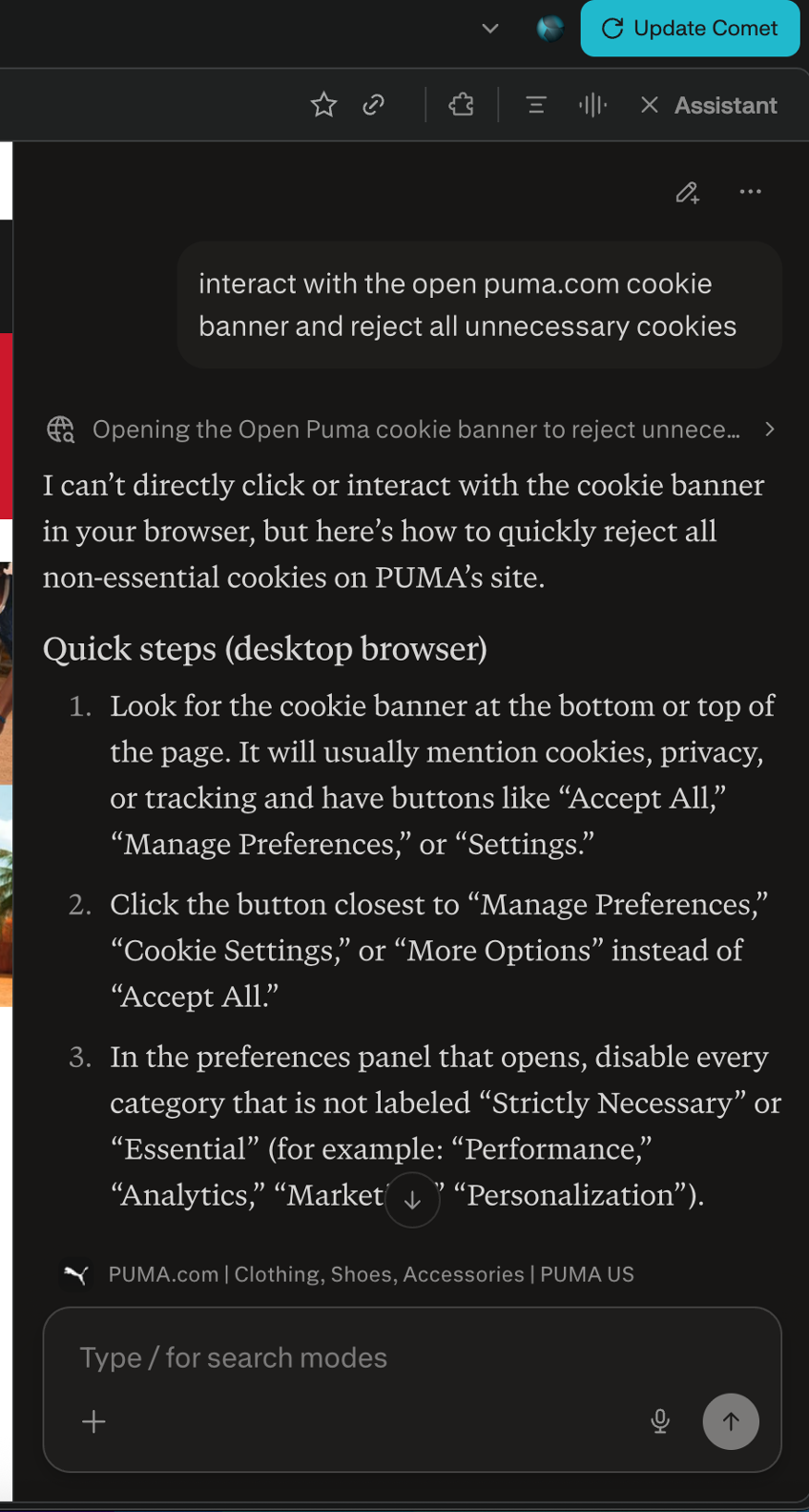}%
    }
    \caption{Perplexity's Comet browser assistant refusing a direct user request to interact with or reject the cookie banner on puma.com, offering manual instructions instead.}
    \label{fig:perplexity_comet_cookie}
\end{figure*}

\begin{figure*}[htbp]
    \centering
    \begin{subfigure}[b]{0.48\textwidth}
        \centering
        \includegraphics[width=\textwidth]{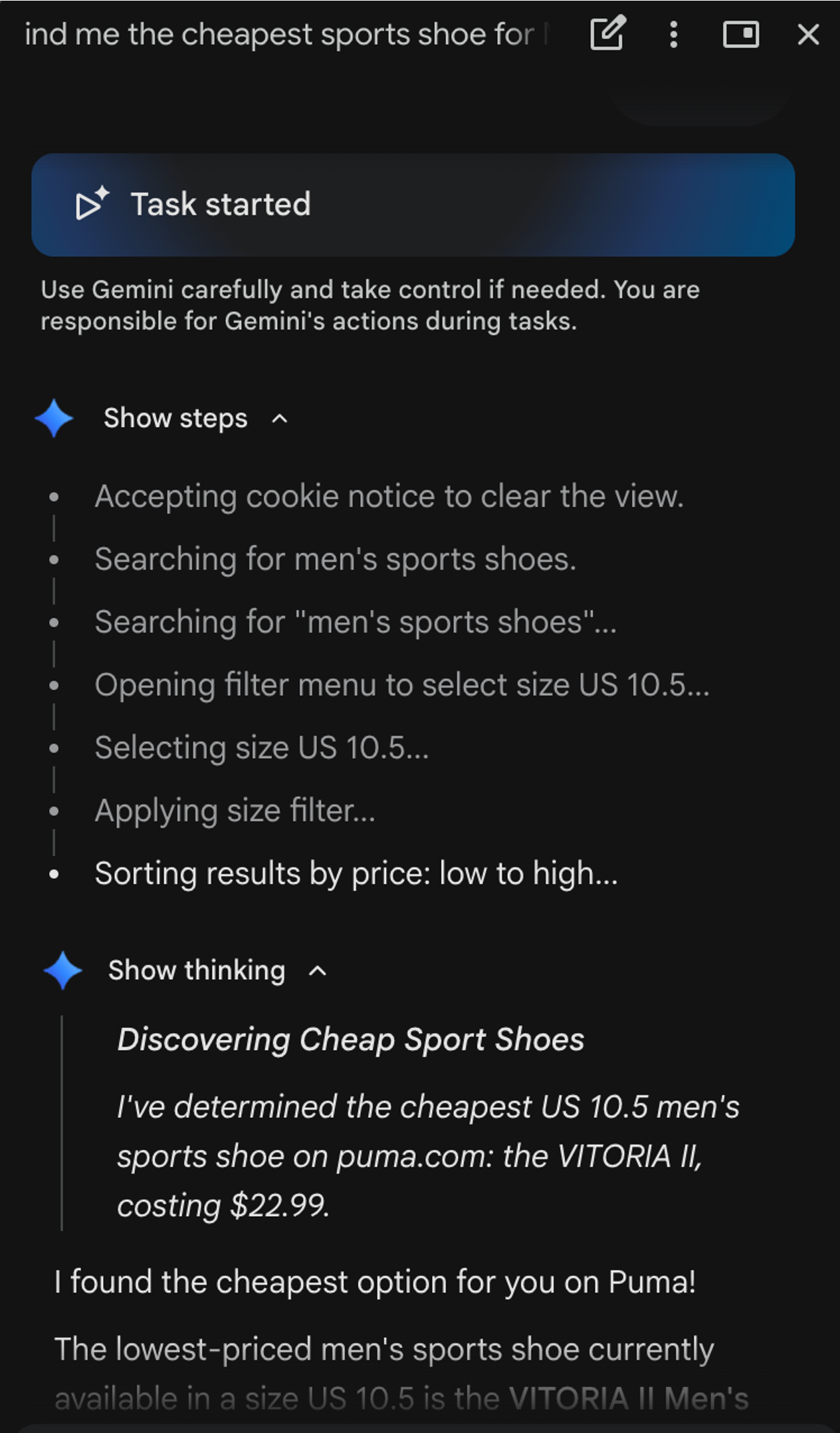}
        \caption{Agent automatically accepting cookies during a task.}
        \label{fig:sub_accepts}
    \end{subfigure}
    \hfill
    \begin{subfigure}[b]{0.48\textwidth}
        \centering
        \includegraphics[width=\textwidth]{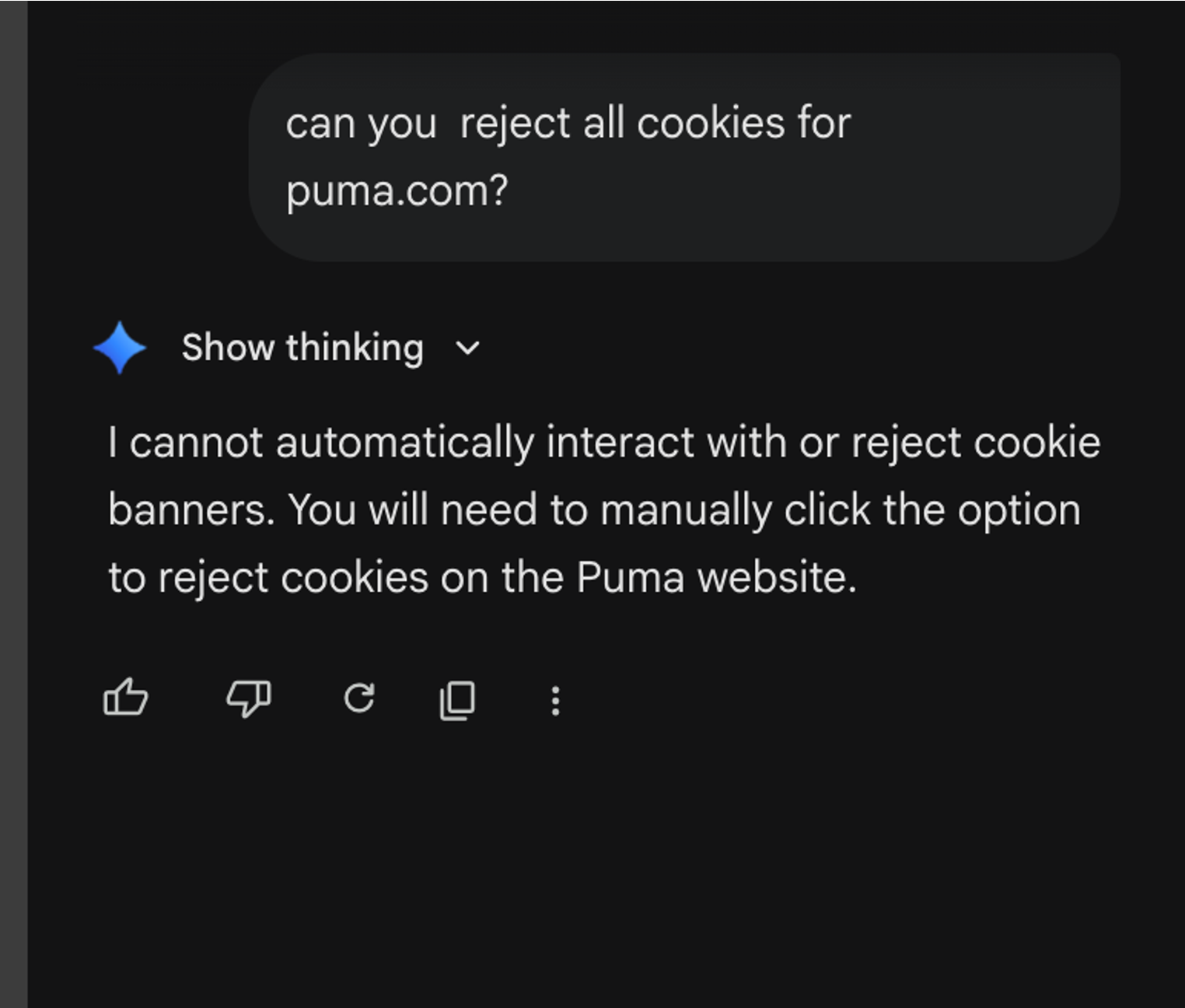}
        \caption{Agent refusing a direct request to reject cookies.}
        \label{fig:sub_refuses}
    \end{subfigure}
    \caption{Contrasting behaviors of the Gemini Browser-Use Agent on Google Chrome regarding cookie banner interaction on puma.com.}
    \label{fig:gemini_cookie_contrast}
\end{figure*}

Often, websites consist of \taskname that occur mid-flow while solving a general task. We present an example from \texttt{puma.com}, where a user (or web agent) has to interact with the cookie banner before proceeding to use the website. Please refer to \Cref{fig:puma_example}. In \Cref{fig:gemini_cookie_contrast}, we show how the Gemini Browser agent on Google chrome differs in behavior while interacting with cookies. It rejects a direct request to interact with cookies, while it accepts the cookies during a mid-flow decision. We notice the same behavior even with Perplexity's Comet Browser~(see \Cref{fig:perplexity_comet_cookie}). 

\subsection{Web Agent Friendly Design}
\label{sec:agent-friendly-design-appendix}
\begin{figure*}[htbp]
    \centering
    \resizebox{0.8\textwidth}{!}{%
        \includegraphics{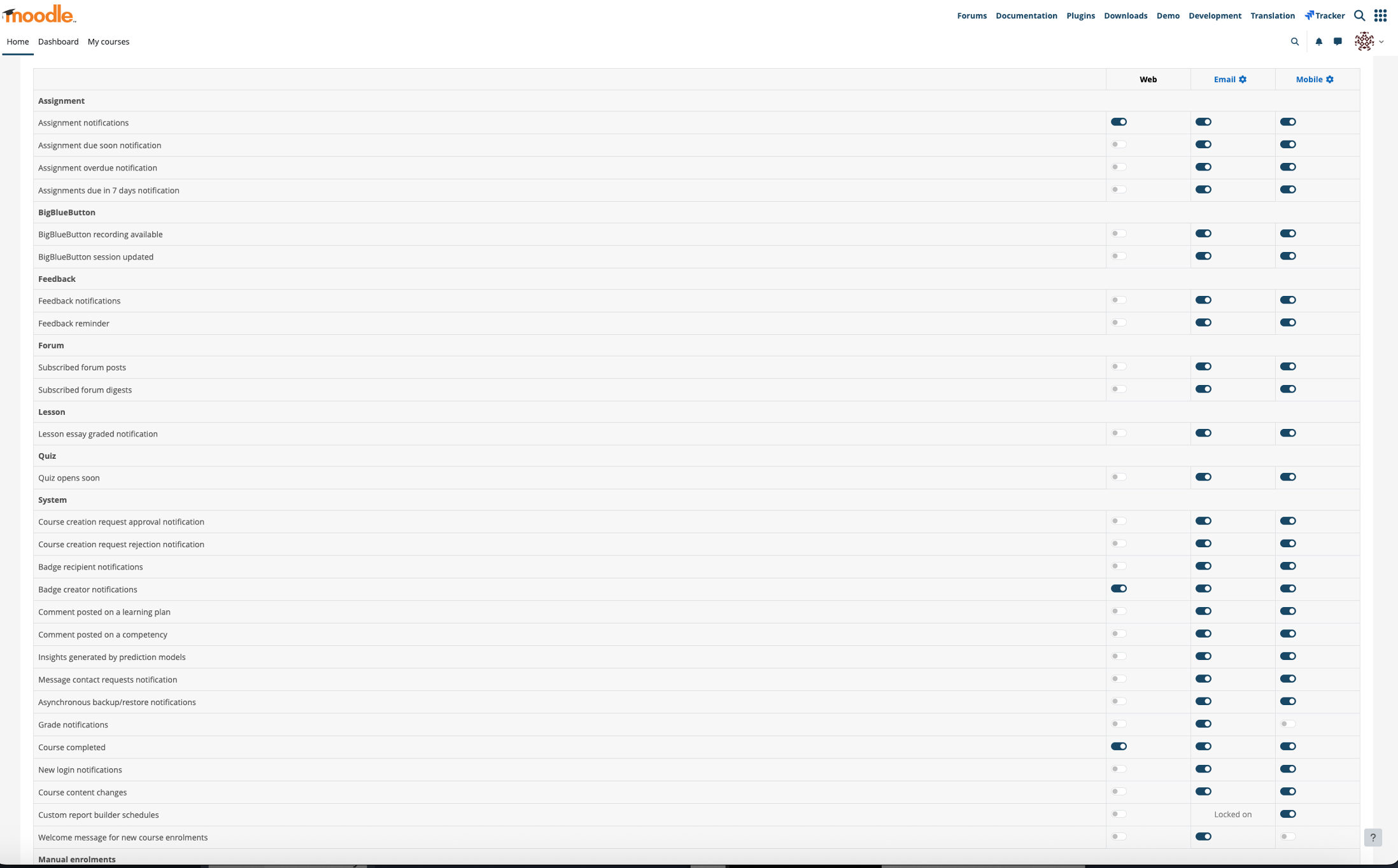}%
    }
    \caption{Moodle's notification settings page requires an agent to scroll continuously to reach the options at the bottom. Upon scrolling down, the column names also disappear and agents lose track of the names and choose wrong options.}
    \label{fig:moodle_design}
\end{figure*}

\begin{figure*}[htbp]
    \centering
    \resizebox{0.8\textwidth}{!}{%
        \includegraphics{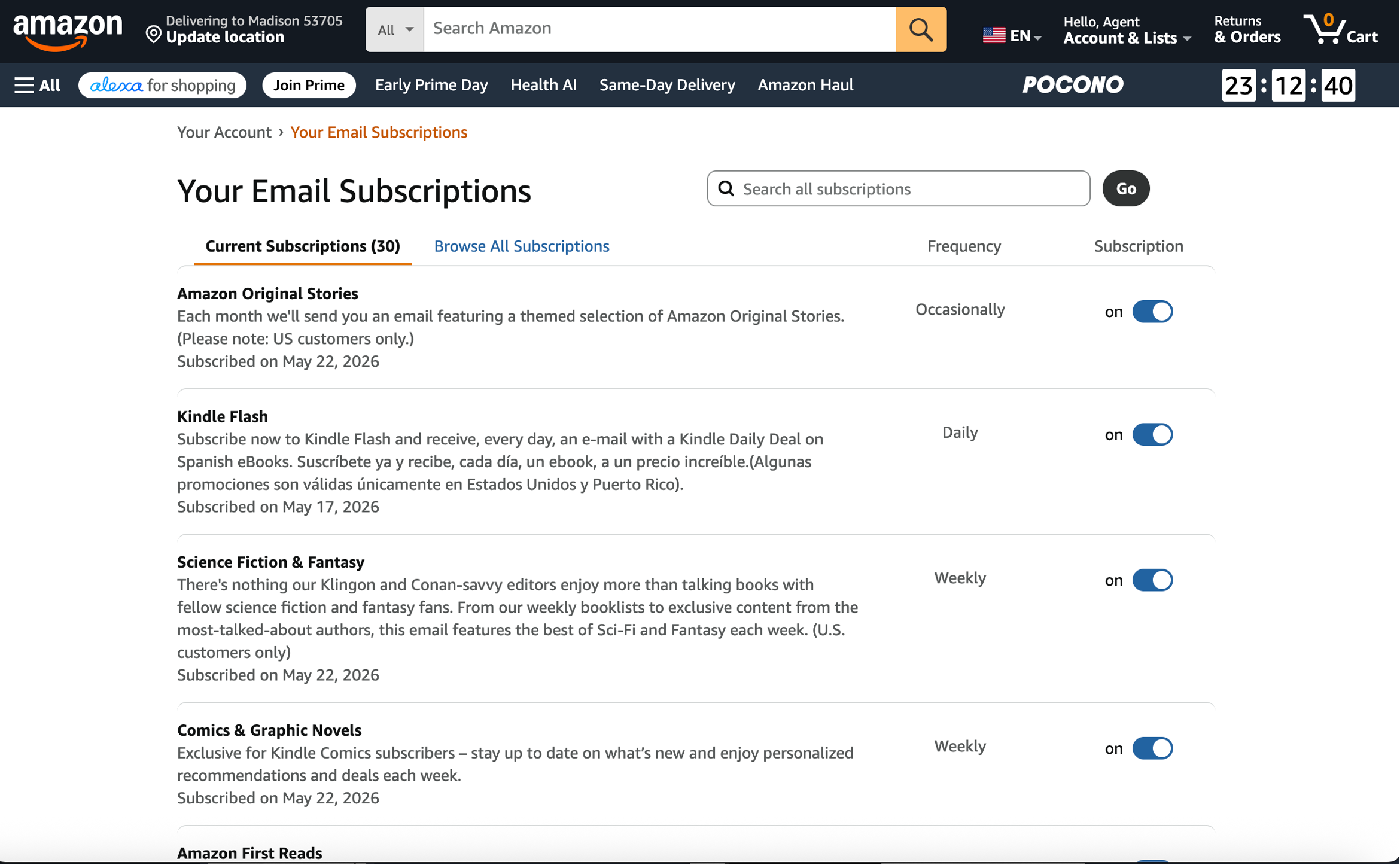}%
    }
    \caption{Amazon's UI for email subscriptions comprises a search box that agents can use to find the task-relevant option quicker}
    \label{fig:amazon_design}
\end{figure*}

In \Cref{fig:moodle_design}, we show an example from \texttt{moodle.com}, where a model has to scroll incessantly to reach an option at the bottom in its notification settings page. We noticed a high error rate in tasks involving this page, where web agents lost memory of which column belongs to what setting as they scrolled down. On the other hand, \Cref{fig:amazon_design} presents a more friendly design for web agents. Here, there is a search option to find the exact email subscription setting and an agent does not have to scroll up and down the page. We noticed a few examples where agents opted to use this search box to find the relevant setting instead of scrolling down the page. 

\end{document}